\documentclass[12pt]{article}

\usepackage{amssymb,amsmath,amsfonts,eurosym,geometry,ulem,graphicx,caption,color,setspace,sectsty,comment,caption,pdflscape,array,hyperref,indentfirst,accents,float,subcaption,algorithm,algorithmic,bbm,appendix,booktabs,dcolumn,placeins,afterpage,optidef, authblk}

\usepackage[hang,flushmargin]{footmisc} 

\newcommand*{\dt}[1]{%
   \accentset{\mbox{\large\bfseries .}}{#1}}

\newcolumntype{L}[1]{>{\raggedright\let\newline\\arraybackslash\hspace{0pt}}m{#1}}
\newcolumntype{C}[1]{>{\centering\let\newline\\arraybackslash\hspace{0pt}}m{#1}}
\newcolumntype{R}[1]{>{\raggedleft\let\newline\\arraybackslash\hspace{0pt}}m{#1}}

\newcolumntype{d}[1]{D..{#1}}

\usepackage[american]{babel}

\usepackage[
    backend=biber,
    style=bwl-FU,
  ]{biblatex}

\begin{document}

\begin{titlepage}
\title{Estimating the Currency Composition of \protect\\ Foreign Exchange Reserves}
\author{Matthew Ferranti\thanks{ferranti@g.harvard.edu. The author thanks Kenneth Rogoff, Neil Shephard, Jeffrey Frankel, Isaiah Andrews, Elie Tamer, Pol Antras, Antonio Coppola, and participants at the Harvard International Lunch for helpful comments. This paper is based upon work supported by a National Science Foundation Graduate Research Fellowship under Grant No. DGE1745303. The author declares no conflicts of interest.}}
\affil{Department of Economics, Harvard University}
\date{\today}
\maketitle
\begin{abstract}
Central banks manage about \$12 trillion in foreign exchange reserves, influencing global exchange rates and asset prices. However, some of the largest holders of reserves report minimal information about their currency composition, hindering empirical analysis. I describe a Hidden Markov Model to estimate the composition of a central bank's reserves by relating the fluctuation in the portfolio's valuation to the exchange rates of major reserve currencies. I apply the model to China and Singapore, two countries that collectively hold about \$3.4 trillion in reserves and conceal their composition. I find that both China's reserve composition likely resembles the global average, while Singapore probably holds fewer US dollars.
\end{abstract}


\setcounter{page}{0}
\thispagestyle{empty}
\end{titlepage}
\pagebreak \newpage

\doublespacing

\section{Introduction} \label{sec:introduction}

Although the US dollar's share of international reserves has fallen from 72\% in 2001 to 58\% in 2022, empirical studies of changes in reserve currency composition have been limited by a lack of country-level data. Some of the largest holders of reserves report minimal information about their currency composition, including China, which possesses approximately one quarter of global reserves. 

I describe a method to estimate the foreign exchange reserve currency composition of countries that do not publicly report this information, regardless of the countries' exchange rate regimes. First, I decompose changes in reserves into two components, the rate of return and the fluctuation due to exchange rates. Then, I implement a Hidden Markov Model to estimate the currency composition of the portfolio. The model's structure provides sufficient flexibility to fit most countries. 

A Hidden Markov Model is a method for estimating an unknown time series by exploiting its relationship to one or more known time series. Intuitively, the Hidden Markov Model infers the currency composition of the reserve portfolio from the co-movement between the fluctuation in the portfolio's valuation and the exchange rates of major reserve currencies, adjusting for assumed returns on the reserve holdings. I apply the Hidden Markov Model to estimate the foreign exchange currency composition of China and Singapore, which do not self-report their currency composition. I validate my Hidden Markov Model using the self-reported currency shares of Brazil and Switzerland, two countries that collectively hold more than \$1 trillion in reserves.

China's secrecy accomplishes two objectives. First, it is more difficult for market participants to front-run China's trades if China conceals its portfolio composition. For example, in May 2010, China issued a rare denial\footnote{wsj.com/articles/SB10001424052748704269204575269993992429022} of a Financial Times article\footnote{ft.com/content/7049ad6e-68ea-11df-910b-00144feab49a} reporting that China might reduce its holdings of Eurozone debt. The euro had dropped 1.5\% against the dollar following the publication of the Financial Times article, likely in anticipation of a Chinese sale of euro-denominated assets. Second, \cite{prasad} points out that the composition of China's reserves is considered politically sensitive because the portion of reserves invested in US dollar securities helps to finance deficits of the United States in spite of disagreements between China and the United States on various political and economic issues.

The conventional wisdom among financial market participants is that China has been reducing its holdings of US dollar assets and diversifying into other reserve currencies. China occasionally reinforces this perception. For example, a September 2020 article\footnote{globaltimes.cn/content/1199833.shtml} in the Global Times (a state-run newspaper) stated that China will continue reducing its US debt holdings, and China might accelerate the pace of reductions if relations between the US and China worsen. Nevertheless, I estimate that China held approximately 60\% of its reserves in US dollar assets as of Q3 2022 (interquartile range 10\%), close to the global average of 58\%, as reported by the IMF.

If the price elasticity of demand for government debt is low, then the distribution of sovereign debt across central banks is important because the reserve management policies of central banks could affect equilibrium exchange rates and the prices of government bonds. \cite{gabaix} develop this hypothesis in relation to equity markets. The impact of the May 2010 Financial Times report on the exchange rate of the euro suggests that China holds significant influence over prices in the foreign exchange and sovereign debt markets. Certainly, the trading activities of central banks can generate significant profits or losses in the short-term. \cite{jpepaper} finds that several central banks lost billions of dollars during the 1970s by trading against exchange rate movements. 

The paper is structured as follows. Section \ref{sec:literature} provides an overview of related literature in international finance, forensic economics, and macroeconometrics. Section \ref{sec:accounting} derives a portfolio accounting identity, which forms the basis for the Hidden Markov Model described in the following Section \ref{sec:model}. Section \ref{sec:data} details data sources. Section \ref{sec:results} reports estimates for China and Singapore, and Section \ref{sec:benchmarking} validates the model with estimates for Brazil and Switzerland. Finally, Section \ref{sec:conclusion} concludes.

\afterpage{\FloatBarrier}
\section{Related Literature} \label{sec:literature}

Approximately 58\% of world reserves are held in dollar-denominated assets, even though the US economy comprises only one quarter of world GDP. To explain this phenomenon, \cite{aerpnp} propose a theoretical model relating the currency denomination of reserves to the funding choices of banks and the invoicing of international trade. \cite{gopinathwp} document that dollars comprise about 62\% of the foreign currency local liabilities of non-US banks, and the dollar’s share as an invoicing currency for imported goods is approximately 4.7 times the share of US goods in imports.

Lack of country-level data has limited empirical research into determinants of reserve currency composition. While the vast majority of central banks report their currency composition confidentially to the IMF, the IMF almost never makes these data available to researchers. The most recent paper using confidential IMF data is \cite{mathieson}, which explained currency composition using trade flows, financial flows, and trade pegs. Some researchers have used aggregate IMF data instead of confidential country-level data. For example, \cite{irfcl} track combined "big four" reserve currency holdings (US dollar, euro, British pound, and yen) before and after the 2008 financial crisis. \cite{frankel} also use aggregate IMF data to suggest that output and trade, capital controls and the depth of financial markets, confidence in the value of the currency, and network externalities explain reserve currency shares.

A newer body of literature gathers publicly available information as a substitute for IMF confidential data. However, these papers cannot handle countries such as China, which report minimal information about their reserve composition. By scraping central bank websites, \cite{itorecent} obtained the currency composition of 58 countries; they relate the dollar and euro share to the currency zone weights and the currency denomination of trade and debt. \cite{weidner} examine the currency composition of 36 Eurozone countries, and show that the euro share among those countries declined following the 2010 euro crisis. \cite{rmbreserves} identify countries that began investing in renminbi-denominated assets for the first time via a combination of news articles and conversations with central banks. They show that a country's decision to invest in renminbi assets is influenced by the degree of political alignment between that country and China.

This paper also relates to the forensic economics literature. By combining several official data sources, I can infer the hidden composition of the reserve portfolio that is most likely to have generated the observed data. \cite{forensic} provides an overview of common forensic methods. \cite{armstrade} apply forensic techniques in an international trade setting by detecting evidence of illegal arms trade in the stock prices of weapons exporters.

This paper contributes to the literature that applies Hidden Markov Models to topics in finance. \cite{markovfinance} compile several papers applying Hidden Markov Models to option pricing, hedging costs, trend-following trading, the term structure of interest rates, and currency markets. More recently, \cite{hmmcrypto} apply a Hidden Markov Model to detect regime changes in the evolution of Bitcoin prices.

Lastly, I implement particle filtering to sample from my Hidden Markov Model. Particle filtering is a well-known procedure in financial econometrics and macroeconometrics. The technique is presented in \cite{durbin} and \cite{chopin}. \cite{schorf} describe filtering methods for solving DSGE models. There are several more sophisticated versions of particle filters, such as the auxiliary particle filter proposed by \cite{shephard}, the bootstrap particle filter discussed by \cite{jesusquintana}, and the tempered particle filter suggested by \cite{schorf2}. However, these techniques are not necessary to sample from my model.

\afterpage{\FloatBarrier}
\section{Portfolio Accounting} \label{sec:accounting}

As in \cite{sheng}, I decompose the foreign exchange portfolio of a central bank into its components. Suppose a central bank holds assets denominated in $N$ different currencies. Let $W_t$ be the US dollar value of the portfolio at market exchange rates at time $t$. Let $e^i_t$ be the exchange rate in US dollars per unit of currency $i$, and let $x^i_t$ be the quantity of currency $i$ assets owned by the central bank. Then the portfolio of the central bank can be written as:
\begin{equation} \label{cbportfolio}
W_t = \sum^N_{i=1} e^i_tx^i_t
\end{equation}
This portfolio will evolve over time due to purchases or withdrawals, interest earnings, and capital gains or losses. Let $C_t$ represent net purchases of reserves measured in US dollars, $\pi_t$ represent net profit from trading (henceforth assumed to be zero), and $r_t^i$ represent the rate of return on investments in currency $i$ measured in local currency.\footnote{This paper excludes gold reserves from the estimation process, because most countries self-report their gold holdings, so estimating the gold share of reserves is not useful.} Then the central bank portfolio evolves according to the following budget constraint:
\begin{equation} \label{budget}
W_t = C_t + \pi_t + \sum^N_{i=1} (1+r^i_{t-1})e^i_tx^i_{t-1}
\end{equation}
I obtain the following identity by subtracting equation (\ref{cbportfolio}) at time $t-1$ from equation (\ref{budget}), and dividing by equation \ref{cbportfolio} at time $t-1$:
\begin{equation} \label{decomp}
\dt{W_t} = \underbrace{\dt{C_t}}_\text{net purchase rate} + \underbrace{R_{t-1}}_\text{rate of return} + \underbrace{\sum^N_{i=1} \beta^i_{t-1}(1+r^i_{t-1})\dt{e^i_t}}_\text{rate of change due to currency fluctuation}
\end{equation}
where:

$\dt{W_t} = (W_t - W_{t-1})/W_{t-1}$ is the growth rate of foreign exchange reserves;

$\dt{C_t} = (C_t - C_{t-1})/W_{t-1}$ is the net purchase rate of foreign exchange reserves;

$\dt{e^i_t} = (e^i_t - e^i_{t-1})/e^i_{t-1}$ is the growth rate of the exchange rate; and

$\beta^i_{t-1} = x^i_{t-1}e^i_{t-1}/W_{t-1}$ is the value share of reserve assets denominated in currency $i$. 

Equation (\ref{decomp}) decomposes the rate of change of foreign exchange reserves into three components. The first component $\dt{C_t}$ is the net purchase rate of reserves. The second component $R_{t-1} = \sum^N_{i=1} \beta^i_{t-1}r^i_{t-1}$ is the rate of return on reserves, comprised of the weighted average of the interest rates earned in each currency. The third component $\sum^N_{i=1} \beta^i_{t-1}(1+r^i_{t-1})\dt{e^i_t}$ is the rate of change of reserves due to currency fluctuation. Because $\beta^i_{t-1}$ is a value share, it can change as exchange rates fluctuate or as the country adjusts the quantity of currency $i$ within its portfolio through trading.

\afterpage{\FloatBarrier}
\section{Hidden Markov Model} \label{sec:model}

\afterpage{\FloatBarrier}
\subsection{Observation Equation} 

Equation \ref{decomp} suggests a specification in which the coefficients $\beta^i_t$ represent the currency shares, and $\epsilon_t / \sigma_{obs, t}$ is standard Laplace (double exponential) distributed:
\begin{equation} \label{firstobservation}
y_t = \dt{W_t} - \dt{C_t}  =  \sum^N_{i=1} \beta^i_{t-1}(1+r^i_{t-1})\dt{e^i_t} + \sum^N_{i=1} \beta^i_{t-1}r^i_{t-1} + \epsilon_t \; \; 
\end{equation}

For any given country, I observe the left side of the equation: $\dt{W_t}$ and $\dt{C_t}$ are available from the central bank or IMF. 

The error term $\epsilon_t$ accounts for accounting irregularities, measurement error, and model misspecification. Drawing inspiration from \cite{levy}, who explore a class of heavy-tailed state space models, I choose the heavy-tailed Laplace distribution to model $\epsilon_t$. As discussed in \cite{laplace}, the Laplace distribution accommodates outliers in financial data better than the Normal distribution, preventing my model from mis-tracking during periods of elevated volatility in exchange rates, asset prices, or net flows into the portfolio. With finite moments of any order, the Laplace distribution is a useful compromise between the Normal distribution and the Cauchy distribution, an extremely heavy-tailed distribution with undefined moments. For Brazil, I present a comparison of the Laplace, Normal, and Cauchy distributions to illustrate the robustness of the Laplace approach. To illustrate the potential presence of accounting irregularities in the data, \cite{sheng} identifies a \$15 billion discrepancy between China's net purchases and total reserves in January 2002 that cannot be explained by exchange rate or asset price fluctuation. 

Model misspecification error captured by $\epsilon_t$ is primarily derived from necessity of making assumptions about the rate of return on the central bank's investments, because countries that do not publish the currency composition of their reserves typically do not publish the quarterly rate of return on their reserves. Accordingly, I assume that a central bank invests its reserves primarily into short- and intermediate-term bonds, earning a rate of return comprised of capital gains from zero-coupon 2-year, 5-year, or 7-year sovereign bonds of each reserve currency in the portfolio. I assume the central bank maintains a constant maturity in its investment portfolio by purchasing the zero-coupon bonds at the start of the quarter and selling them at the end of the quarter. The IMF \cite{imfreservereport} indicate that a number of central banks are concerned about reporting quarterly negative returns; maintaining a short or intermediate duration limits their interest rate risk. 

A portion of the error $\epsilon_t$ also comes from a lack of knowledge of the precise timing of reserve purchases or withdrawals due to the quarterly frequency of the data. For example, a large purchase of reserves in the middle of the quarter would generally earn a rate of return that differs from the return on reserves that were invested at the beginning of the quarter. In order to capture the temporal variation in the error term, I allow $\sigma_{obs, t}$ to vary over time in proportion to an index of global exchange rate volatility. For each quarter, I compute the standard deviation of the daily change in the exchange rate of the IMF Special Drawing Right (SDR) against the US dollar, as a proxy for global reserve currency exchange rate volatility. Then, I re-scale this series such that its average is equal to half the interquartile range of $\lVert \dt{W_t} - C_t/W_{t-1} \rVert$, the non-purchase rate of change in the reserve portfolio. I prefer the interquartile range to the standard deviation so that the process of computing $\sigma_{obs, t}$ is more resistant to outliers. 

Because I set $\sigma_{obs, t}$ based on the volatility of the SDR, the model adjusts estimated currency shares to match new data more closely during time periods when exchange rate volatility is low, so that the US dollar value of portfolio purchases and withdrawals does not greatly depend on the timing of the flows within the quarter. However, the model does benefit from uncorrelated exchange rate movement across reserve currencies, which is necessary for identification. Two currencies with the same exchange rate histories are indistinguishable from each other (assuming sovereign bond yields for those two currencies are similar). Exchange rate volatility of major reserve currencies has trended lower since the Great Recession, as noted in \cite{exchcorr}.

Countries that possess a larger quantity of reserves relative to GDP may invest their reserves more aggressively, because those countries can afford to bear more risk. In particular, some central banks hold corporate bonds and even equities. In order to account for the possibility that the central bank could hold equities, or investments whose performance is correlated with equities, I estimate the equity share of reserves through the following optimization procedure:

\begin{mini}|l| 
  {x_t}{\sum^{i=\text{min}(t+10,T)}_{i=\text{max}(t-10,0)} \left[y_t - x\left(\sum^N_{i=1} \beta^i_{t-1}(1+r^i_{eq,t-1})\dt{e^i_t} + \sum^N_{i=1} \beta^i_{t-1}r^i_{eq,t-1}\right) \right.}{}{}
  \breakObjective{ \left. - (1-x) \left( \sum^N_{i=1} \beta^i_{t-1}(1+r^i_{bd,t-1})\dt{e^i_t} + \sum^N_{i=1} \beta^i_{t-1}r^i_{bd,t-1}\right)\right]^2/\sigma_{obs,t}^2}
  {\label{equityoptimization}}{}
  \addConstraint{0 \leq x_t }{ \leq 1. \quad}
 \end{mini}

Here, $x_t$ is the estimated equity share of reserves. $r^i_{eq,t}$ and $r^i_{bd,t}$ are the local currency equity and bond returns of reserve currency issuers, respectively. Because $\beta^i_t$ are unknown at this stage -- the currency shares of reserves have not yet been estimated -- I substitute the world average reserve currency shares from the IMF Currency Composition of Official Foreign Exchange Reserves (COFER) dataset. Using a rolling window spanning 20 quarters helps smooth the estimated equity share and prevents overfitting. To the extent that the investment returns of a central bank differ from my assumptions, that misspecification error is contained within $\epsilon_t$. I conduct a sensitivity analysis using 2-year, 5-year, 7-year, and 10-year sovereign bonds in the case of China.

After estimating the equity share of reserves, I split the return portion of equation (\ref{firstobservation}) into the returns from equities and bonds. I assume that the equities and bonds have the same currency composition. Because central banks hold a low equity share, it is not feasible to estimate separate currency compositions for bonds and the equities.

\begin{equation}
\begin{aligned} \label{observation}
y_t &= \; x\left(\sum^N_{i=1} \beta^i_{t-1}(1+r^i_{eq,t-1})\dt{e^i_t} + \sum^N_{i=1} \beta^i_{t-1}r^i_{eq,t-1}\right) \\ &+ (1-x)\left(\sum^N_{i=1} \beta^i_{t-1}(1+r^i_{bd,t-1})\dt{e^i_t} + \sum^N_{i=1} \beta^i_{t-1}r^i_{bd,t-1}\right) + \epsilon_t \; \; \mbox{(observation equation)}
\end{aligned}
\end{equation}

The observation volatility $\sigma_{obs, t}$ for China is illustrated in Figure \ref{obsstdev}. Further discussion about the importance of $\epsilon_t$ can be found in Appendix \ref{optimappendix}.

\begin{figure}[h]
 \centering
    \includegraphics[scale = 1.1]{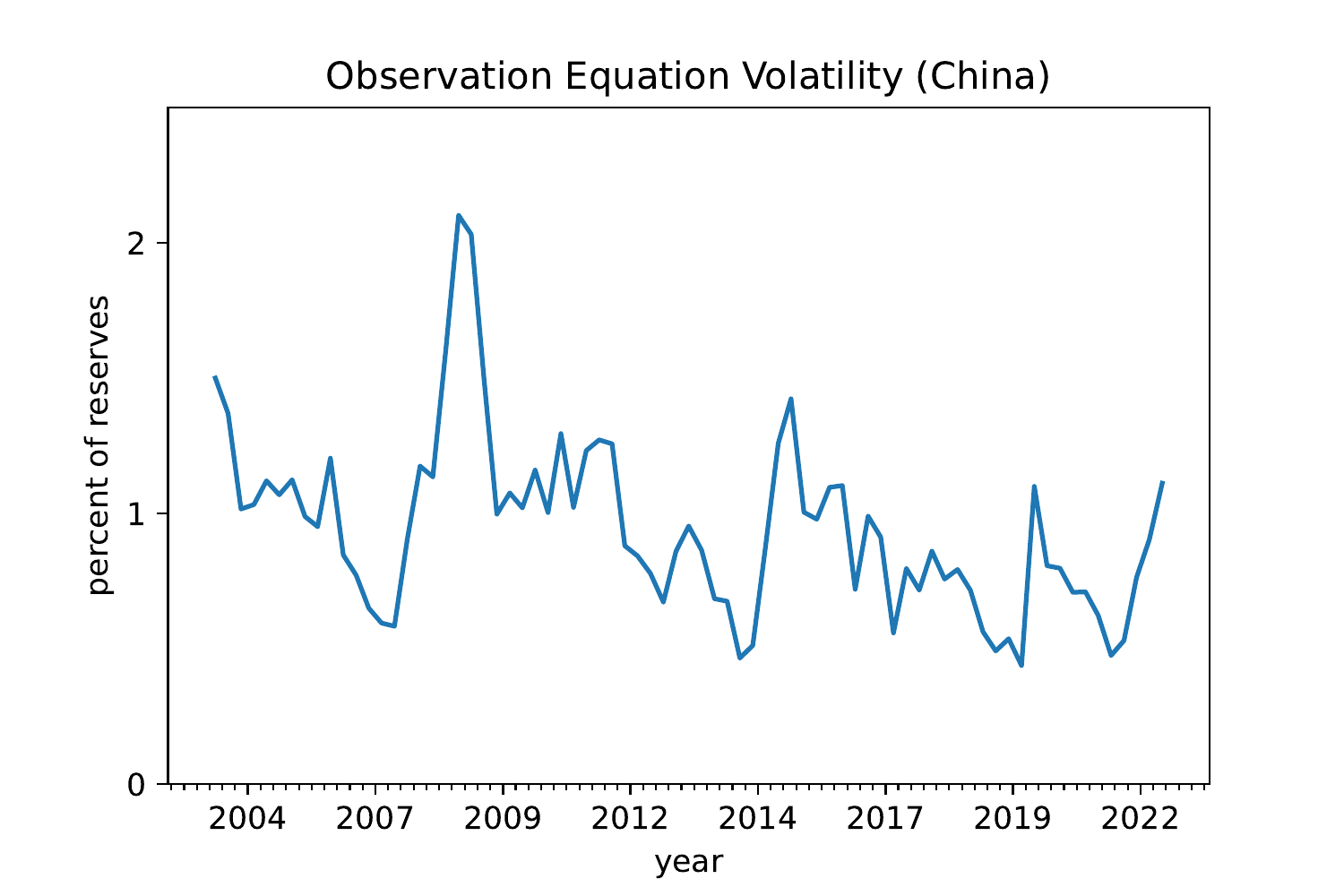}
    \vspace{-1em}
    \begin{flushright}Source: SAFE, author's calculations \end{flushright}
    \caption{The error $\sigma_{obs, t}$ in the observation equation.}
    \label{obsstdev}
\end{figure}

\afterpage{\FloatBarrier}
\subsection{State Space Equation} 

Previous empirical research, such as \cite{itorecent} and \cite{mathieson}, has consistently found that reserve currency shares are highly auto-correlated. Consequently, I adopt the following time series structure for the reserve currency shares:
\begingroup\belowdisplayskip=\belowdisplayshortskip
\begin{equation} \label{statespaceeq}
\vec{\beta_{t+1}} | \vec{\beta_t} \sim \text{Dirichlet} \; (\alpha_{t} \beta^{usd}_{t},\alpha_{t} \beta^{euro}_{t},\ldots, \alpha_{t} \beta^N_t) \; \; \mbox{(state space equation)}
\end{equation}
\endgroup
\begin{equation} \label{alpha}
\alpha_{t} | \beta^{usd}_t = \frac{\beta^{usd}_{t} - (\beta^{usd}_{t})^2 - \gamma}{\gamma} \; \; \mbox{(Dirichlet variance scale factor)}
\end{equation}
The Dirichlet distribution is commonly used for modeling compositional time series, in which the vector of dependent variables at each time point is a set of proportions that sum to one. For example, \cite{dirichletts} model multivariate dependent variables using the Dirichlet distribution and an ARMA process for the Dirichlet parameters. This paper differs from the compositional time series literature in that I model the unobserved independent variable as Dirichlet distributed, instead of the dependent variable, which is scalar. 

In this application, I prevent the currency shares from becoming too small by ensuring that the mean of the distribution for drawing the next period's shares is at least 1\% for all shares. Specifically, for the purpose of computing $\vec{\beta_{t+1}}$, I set $\beta^i_t = 0.01$ in the Dirichlet distribution if $\beta^i_t < 0.01$. Therefore, $\beta^i_t$ may fall below 1\%, but $\mathbb{E} (\beta^i_t) >= 0.01$ for all $i, t$.

The Dirichlet distribution is useful for several reasons. First, equation (\ref{statespaceeq}) is a martingale, because the currency shares at time $t+1$ are drawn from a distribution with a mean equal to the shares at time $t$:
\begin{equation} \label{expectationstatespace}
\mathbb{E} \left( \beta^i_{t+1} \right) = \frac{\alpha_t \beta^i_{t}}{\sum^N_{i=1} \alpha_t \beta^i_{t}} = \beta^i_{t}
\end{equation}
Second, the Dirichlet distribution constrains the currency shares, such that $0 \leq \beta^i_t \leq 1$ for all $i,t$, and $\sum^N_{i=1} \beta^i_t = 1$ for all $t$. Central banks generally do not hold significant leveraged or short positions, so these constraints are reasonable.

Third, the Dirichlet distribution provides an intuitive variance-covariance structure:
\begingroup\belowdisplayskip=\belowdisplayshortskip
\begin{equation} \label{varstatespace}
Var \left( \beta^i_{t+1} \right) = \frac{\beta^i_t(1-\beta^i_t)}{\alpha_t+1}
\end{equation}
\endgroup
\begin{equation} \label{covstatespace}
Cov \left( \beta^i_{t+1}, \beta^j_{t+1} \right) = -\frac{\beta^i_t \beta^j_t}{\alpha_t + 1} \; \; \mbox{for i $\neq$ j}
\end{equation}
The scale factor $\alpha_t$ controls the variance of the Dirichlet distribution. It is helpful to fix the variance of the US dollar share in order to provide a reference point for estimating the remaining variances and covariances. Substituting equation (\ref{alpha}) into equation (\ref{varstatespace}) reveals that $Var \left( \beta^{usd}_{t} \right) = \gamma$. Setting $\gamma = 0.015^2$ ensures that a one-standard deviation innovation in the state space is equal to a 1.5 percentage point change in the US dollar share. As a comparison, \cite{chinafx} found that quarterly changes in China's US dollar share were on the order of 0.5 percentage points. My choice of a higher standard deviation enables the model to track larger changes over time and weakens the influence of the Bayesian prior on my estimates. As a consequence of this assumption, the variance of the US dollar share is constant over time, but the variance of all other currency shares can rise or fall according to equation (\ref{varstatespace}).

As illustrated in equation (\ref{varstatespace}), the variance of currency shares other than the US dollar depends on the size of the share: for shares less than 50\%, the variance falls as the share falls, but for shares greater than 50\%, the variance falls as the share rises. The intuition can be illustrated in a two-currency example. Consider two portfolios, one split evenly between dollars and euro, and the other divided 75\% dollars and 25\% euro. If the exchange rate of the euro depreciates 10\% against the US dollar, and the portfolios are not rebalanced, the first portfolio will now be allocated 52.6\% in dollars and 47.4\% in euro, while the second portfolio will be allocated 76.9\% in dollars and 23.1\% in euro. In the first portfolio initially containing 50\% euro, the euro share declined 2.6 percentage points, whereas in the second portfolio initially containing 25\% euro, the euro share declined only 1.9 percentage points. Conversely, the US dollar share rose 2.6 percentage points in the first portfolio initially containing 50\% US dollars, but rose 1.9 percentage points in the second portfolio initially containing 75\% US dollars.

Because of this variance relationship, the currency share distributions will generally not be symmetric: they will be skewed right if less than 50\%, and skewed left if less than 50\%. This effect will be more pronounced the farther the currency share is from 50\%. As a result, the model assigns more probability to the possibility of a significant increase (reduction) in a small (large) currency share, rather than vice versa. It is reasonable for the distribution of a small share to be skewed right because the share cannot fall below 0\%, so a small share can increase more than it can decrease.

The covariance between any two shares is negative and increasing in either share. The intuition is that a change in one currency share must be offset by an opposite change in at least one other currency share. The larger the product of the two shares, the more likely that changes in one of the shares will be offset by corresponding changes in the other.

The covariance structure provided by the Dirichlet distribution obviates the need to estimate or choose many covariance parameters. Both the covariances and variances are functions of the shares and the $alpha_t$ scale parameter.

The observation equation (\ref{observation}), observation volatility $\sigma_{obs, t}$, state space equation (\ref{statespaceeq}), and state space variance parameter (\ref{alpha}) collectively comprise the Hidden Markov Model to estimate the unknown reserve currency shares.

\afterpage{\FloatBarrier}
\subsection{Priors} I employ a Bayesian approach to solving this Hidden Markov Model. Bayesian inference requires specifying a prior, which can augment inferences drawn on smaller samples. Choosing a reasonable starting point is necessary to initialize the model.

I implement a Dirichlet prior for the currency shares at $t=0$, the beginning of 2004. For each country, I select the prior based on either its own self-reported currency shares or the aggregate world currency shares of allocated reserves in 2004 as reported by the IMF COFER dataset. Country-specific priors are available in Appendix \ref{appendixparams}, and a sensitivity analysis to the variance of the prior in the case of China is available in Appendix \ref{appendixprior}. The prior becomes less influential as time passes, so that my 2022 estimates are not especially sensitive to the prior.

\afterpage{\FloatBarrier}
\subsection{Computation} I solve this Hidden Markov Model numerically using a particle filter, a technique commonly employed for solving state-space models as described in the related literature section \ref{sec:literature}. Particle filtering produces a discrete approximation of the state space $p(\vec{\beta_t} \mid y_1, ... y_t)$ for each $t= 1, 2, ... T$ by predicting, reweighting, and resampling a set of $N_F$ filter particles to generate estimates for the next time period. Specifically, for each particle $i = 1, 2, ... N_F$:
\begingroup\belowdisplayskip=\belowdisplayshortskip
\begin{equation} \label{predict}
\vec{\beta^i_{t}} \sim p(\vec{\beta_{t}} \mid \vec{\beta_{t-1}}) \; \; \mbox{(predict)}
\end{equation}
\endgroup
\begin{equation} \label{reweight}
w^i_{t} \propto p(y_{t} \mid \vec{\beta^i_{t}}) \; \; \mbox{(reweight)}
\end{equation} 
\begin{equation} \label{resample}
p(\vec{\beta_t} \mid y_1, ... y_t) \sim (\vec{\beta_{t}}, w_{t}) \; \; \mbox{(resample)}
\end{equation}
Pseudo-code for the particle filter algorithm is available in Appendix \ref{appendixalgorithms}. The estimates below are generated with $N_F = 10,000$ particles.

\afterpage{\FloatBarrier}
\section{Data} \label{sec:data}
\begin{figure}[h]
 \centering
    \includegraphics[scale = 1.1]{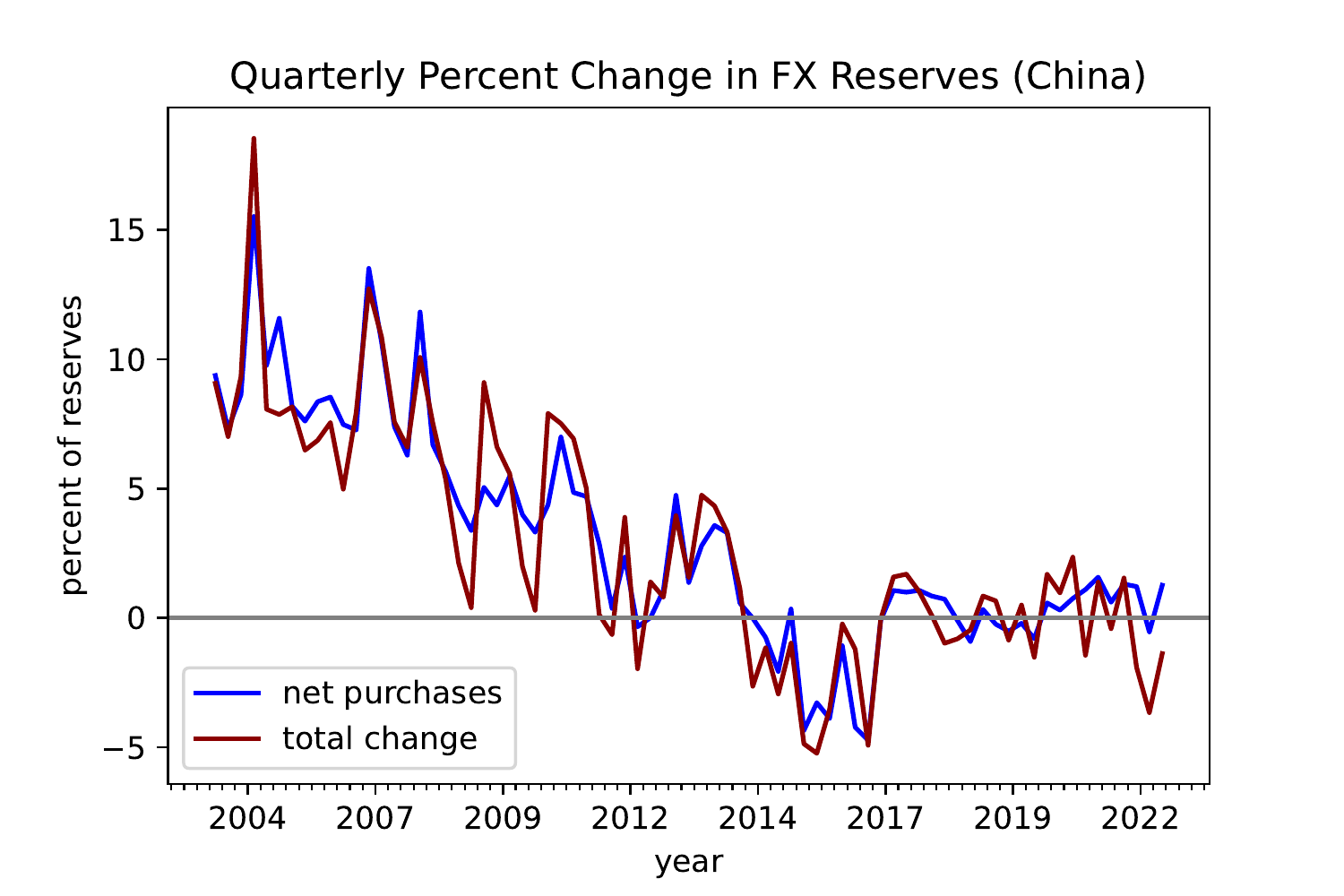}
    \caption{The total rate of change, and the purchase rate, of China's foreign exchange reserves. The difference between these lines represents the non-purchase rate of change, comprised of investment returns and currency fluctuation.}
    \vspace{-1em}
    \begin{flushright}Source: State Administration for Foreign Exchange \end{flushright}
    \label{reschange}
\end{figure}

I obtain the total size of a central bank's foreign exchange reserves (excluding gold) from the central bank's website, annual reports, or the IMF International Reserves and Foreign Currency Liquidity (IRFCL) database. I obtain the quarterly change in reserves from the IMF's Balance of Payments database and the central bank's published balance of payments statistics. Figure \ref{reschange} plots these series for China. I obtain local currency equity total return indices, net of withholding tax, from MSCI. I use 2-year, 5-year, or 7-year zero-coupon sovereign bonds to approximate the fixed income rate of return on an investment in a particular currency. I obtain sovereign bond yields from FactSet and investing.com, from which I calculate the quarterly total returns using the discounted cash flow equation. Sovereign bond returns have been similar across advanced economies over the last 15 years, so identification in the model mostly derives from uncorrelated exchange rate movement rather than differences in the assumed fixed income rate of return across currencies.

\afterpage{\FloatBarrier}
\section{Results} \label{sec:results}

In the estimation process, I use 7-year sovereign bonds for China and Singapore, which resembles the market capitalization-weighted duration of the global bond market.  A sensitivity analysis in which I compare results using 2-year, 5-year, 7-year and 10-year sovereign bonds for China is available in Appendix \ref{appendixchinasensitivity}. 

\afterpage{\FloatBarrier}
\subsection{China}

China possesses about one quarter of global foreign exchange reserves, a portfolio of approximately \$3 trillion accumulated from years of pursuing an export-oriented growth strategy. China's reserves briefly reached \$4 trillion in 2014, before sharp capital outflows forced China to intervene in support of its exchange rate. Since the end of 2016, China has implemented a managed floating exchange rate, targeting a basket of 24 currencies. The weightings change annually based on the currency denomination of China's trade, but top four currencies are currently the US dollar, the euro, the yen, and the won. 

\cite{sheng} is the first paper to produce empirical estimates of the currency composition of China's reserves, focusing on the period 2000-2007, during which China likely added euros to their portfolio. \cite{sheng} computes monthly net reserve purchases by assuming that China maintains a fixed exchange rate, which China abandoned in 2005. My approach differs in several respects. First, I do not make assumptions about the exchange rate regime of the central bank. I use quarterly balance of payments data, rather than monthly central bank balance sheet data, to compute net reserve purchases. Second, my model can accommodate a larger number of reserve currencies, because my model does not contain a variance-switching component, which adds significant computational overhead. Variance-switching may have been more important in the mid-2000s, when countries were adding euros to their reserves. However, the ability to track more currencies simultaneously is now more important, because reserve portfolios have become more diversified across currencies since the mid-2000s, as highlighted by \cite{stealth}. Third, I use the Dirichlet distribution to implement an appropriate covariance structure for the currency shares, which constrains the shares to be nonnegative and sum to one. This additional structure may help the model track the currency shares. Fourth, I implement a procedure to account for potential central bank investment in equities. Lastly, I employ particle filtering rather than Markov Chain Monte Carlo (MCMC) to solve the model. As a sequential Monte Carlo method, the particle filter estimates the state $p(\vec{\beta_t} \mid y_1, ... y_t)$ at time $t$ using only observations concurrent or prior to time $t$. By contrast, MCMC produces smoothed estimates of the state at time $t$ that incorporate information from the entire time series, including future observations. Updating filtered estimates with a new observation at time $t+1$ does not affect previous estimates. However, incorporating a new observation into MCMC requires re-estimating the entire time series, and may alter past estimates of the shares. I prefer the filter approach in this application. Overall, my model can be applied to a wider set of countries than that of \cite{sheng}, but my model is not designed to capture sudden, sharp shifts in portfolio composition.

In recent years, China's State Administration of Foreign Exchange (SAFE) has disclosed limited information about its holdings. In its 2018 Annual Report, SAFE revealed that the US dollar share of its reserve portfolio was 79\% in 1995, and this proportion declined to 58\% in 2014. \cite{prasad} notes that SAFE's statement appears targeted to make two points: that China is reducing the share of dollars in its reserves, and that the share of dollars it held in 2014 was below the global average for reserve portfolios. In subsequent annual reports, SAFE disclosed its US dollar share for 2015, 2016, and 2017. China has not disclosed the shares of any other currencies in its reserves.

\begin{figure}[h]
\begin{subfigure}{1.0\textwidth}
\includegraphics[width=\linewidth]{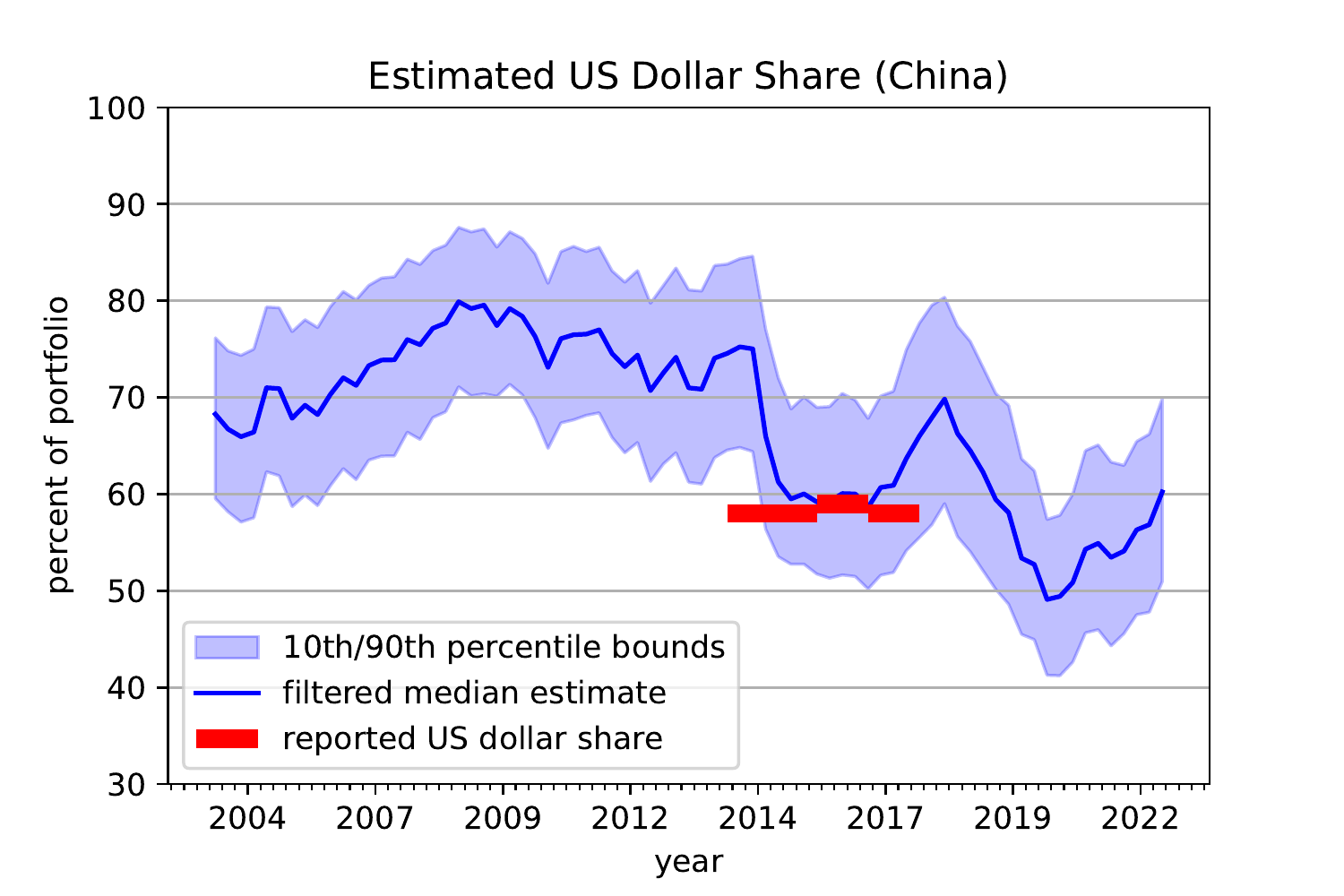}
\end{subfigure}\hspace*{\fill}

\begin{subfigure}{1.0\textwidth}
\includegraphics[width=\linewidth]{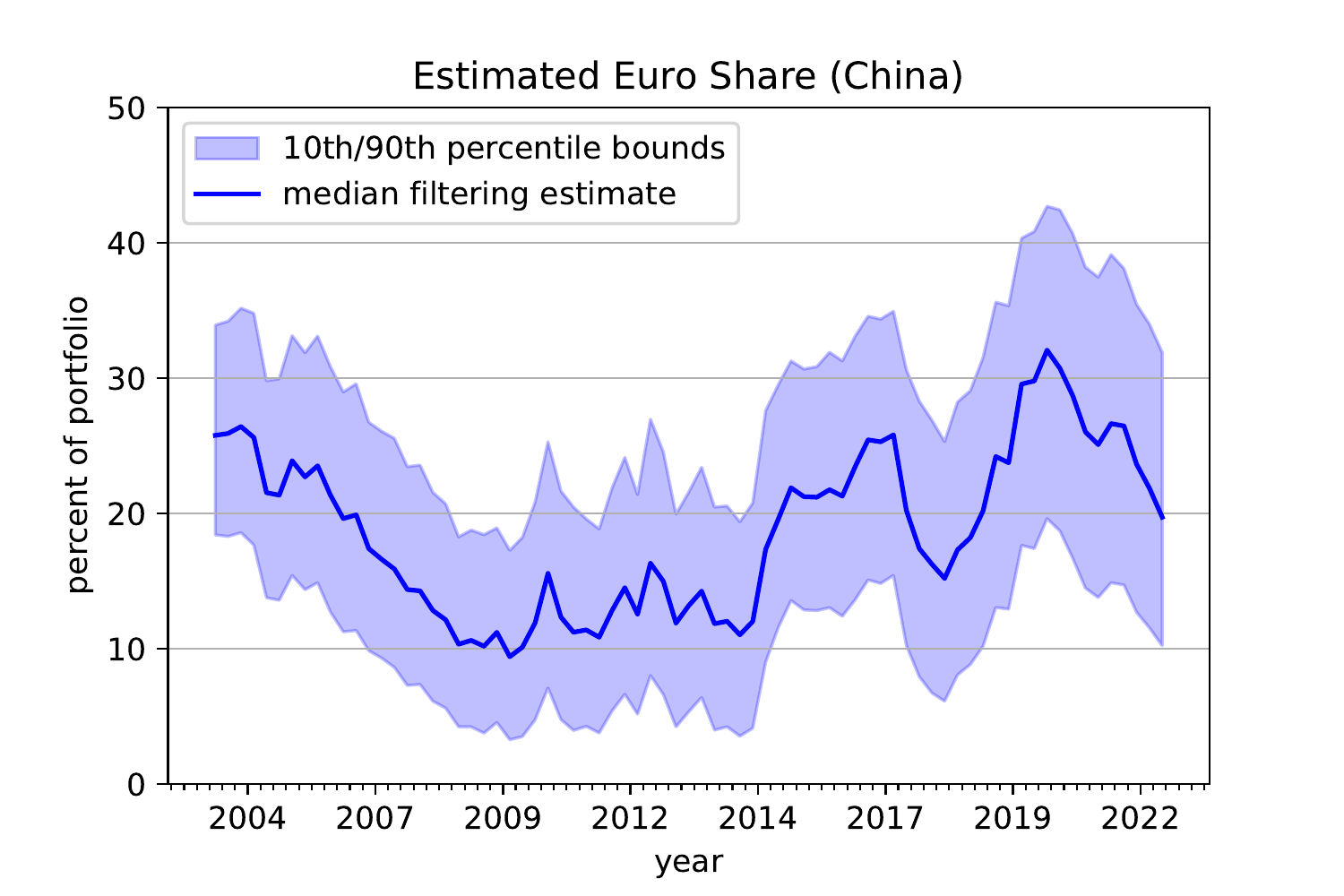}
\end{subfigure}\hspace*{\fill}

\caption{Filter estimates of China's currency shares.}

\label{chinafilter}

\end{figure}

\begin{figure}[h]\ContinuedFloat
\begin{subfigure}{1.0\textwidth}
\includegraphics[width=\linewidth]{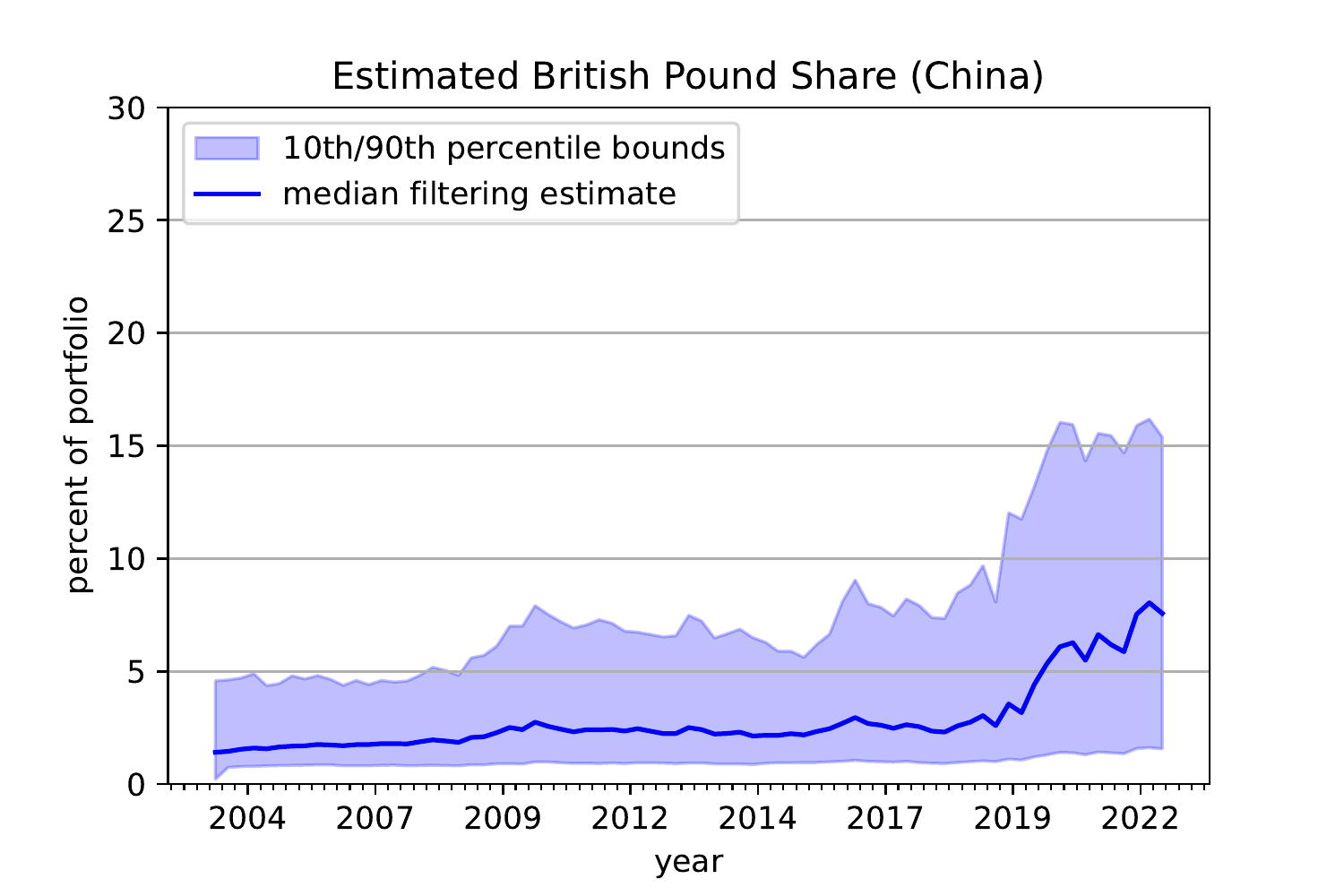}
\end{subfigure}\hspace*{\fill}

\begin{subfigure}{1.0\textwidth}
\includegraphics[width=\linewidth]{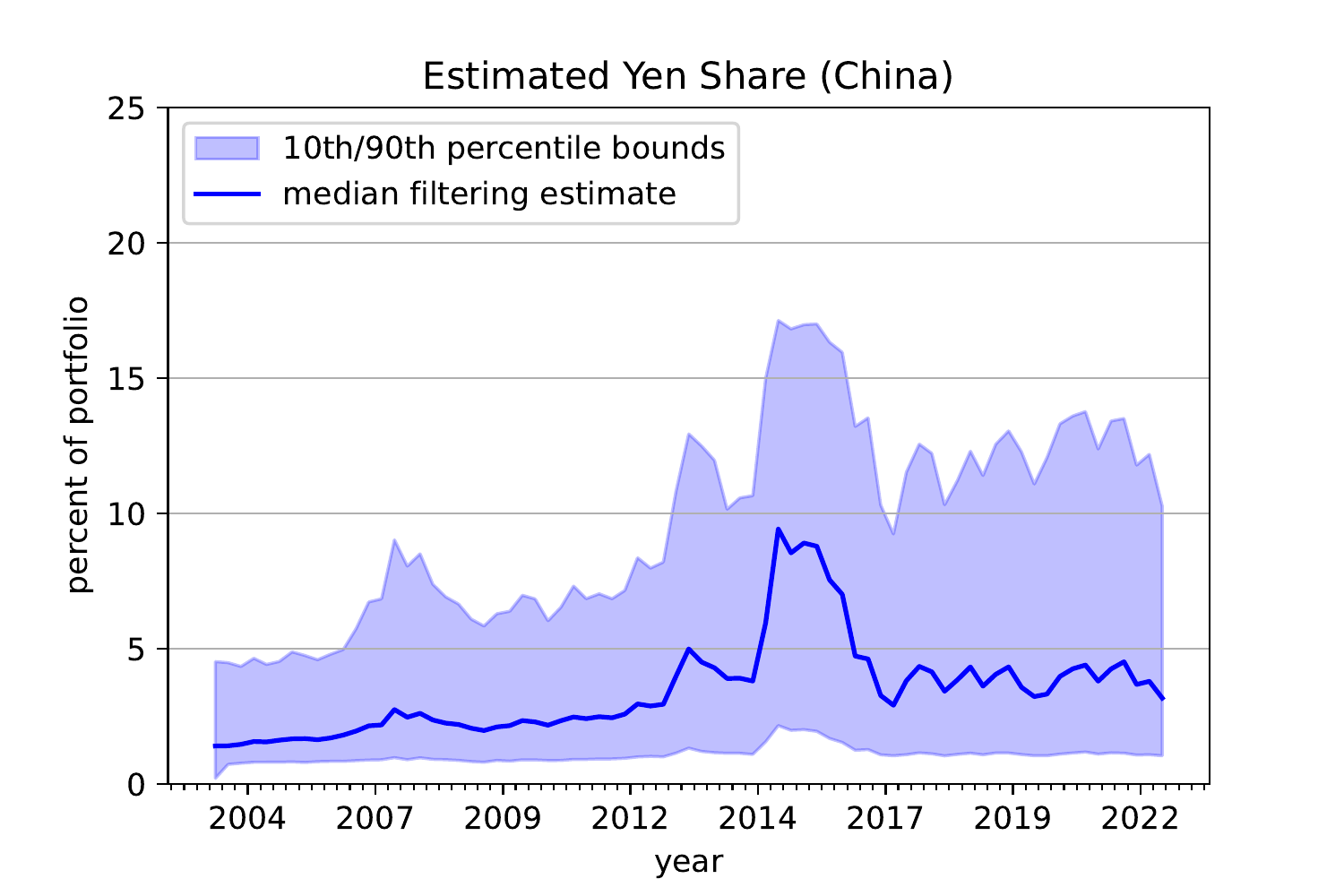}
\end{subfigure}\hspace*{\fill}

\caption{Filter estimates of China's currency shares.}

\label{chinafilter}

\end{figure}

\begin{figure}[h]\ContinuedFloat
\begin{subfigure}{1.0\textwidth}
\includegraphics[width=\linewidth]{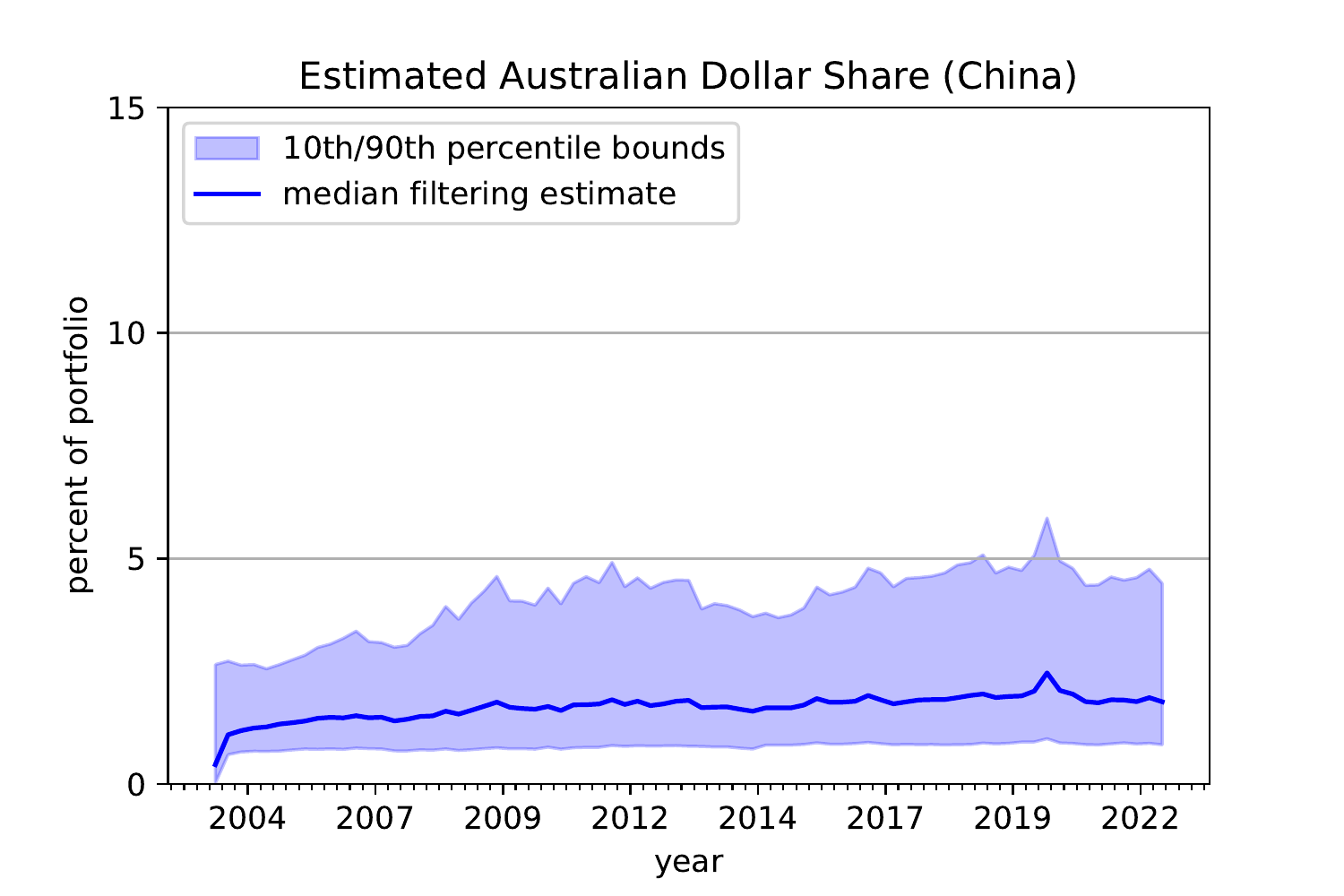}
\end{subfigure}\hspace*{\fill}

\begin{subfigure}{1.0\textwidth}
\includegraphics[width=\linewidth]{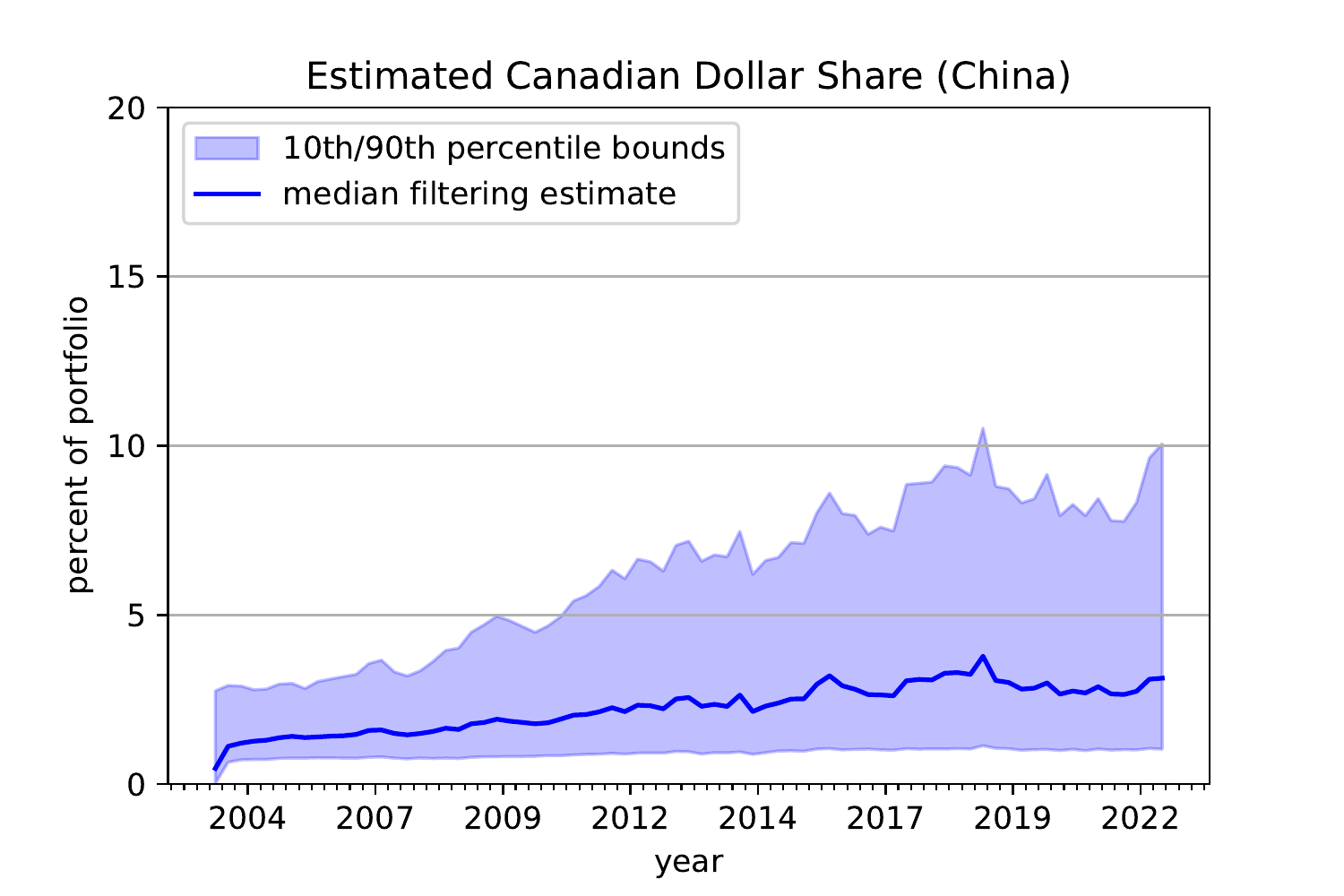}
\end{subfigure}\hspace*{\fill}

\caption{Filter estimates of China's currency shares.}

\label{chinafilter}

\end{figure}

Using the self-reported US dollar shares from SAFE, I can benchmark the performance of my model, while estimating China's reserve composition up to Q3 2022, as illustrated in Figure \ref{chinafilter}. The model includes the six major reserve currencies: the US dollar, euro, pound, yen, Canadian dollar, and Australian dollar. The filter estimates in Figure \ref{chinafilter} match the self-reported shares well. 

\begin{figure}[h]
\begin{subfigure}{1.0\textwidth}
\includegraphics[width=\linewidth]{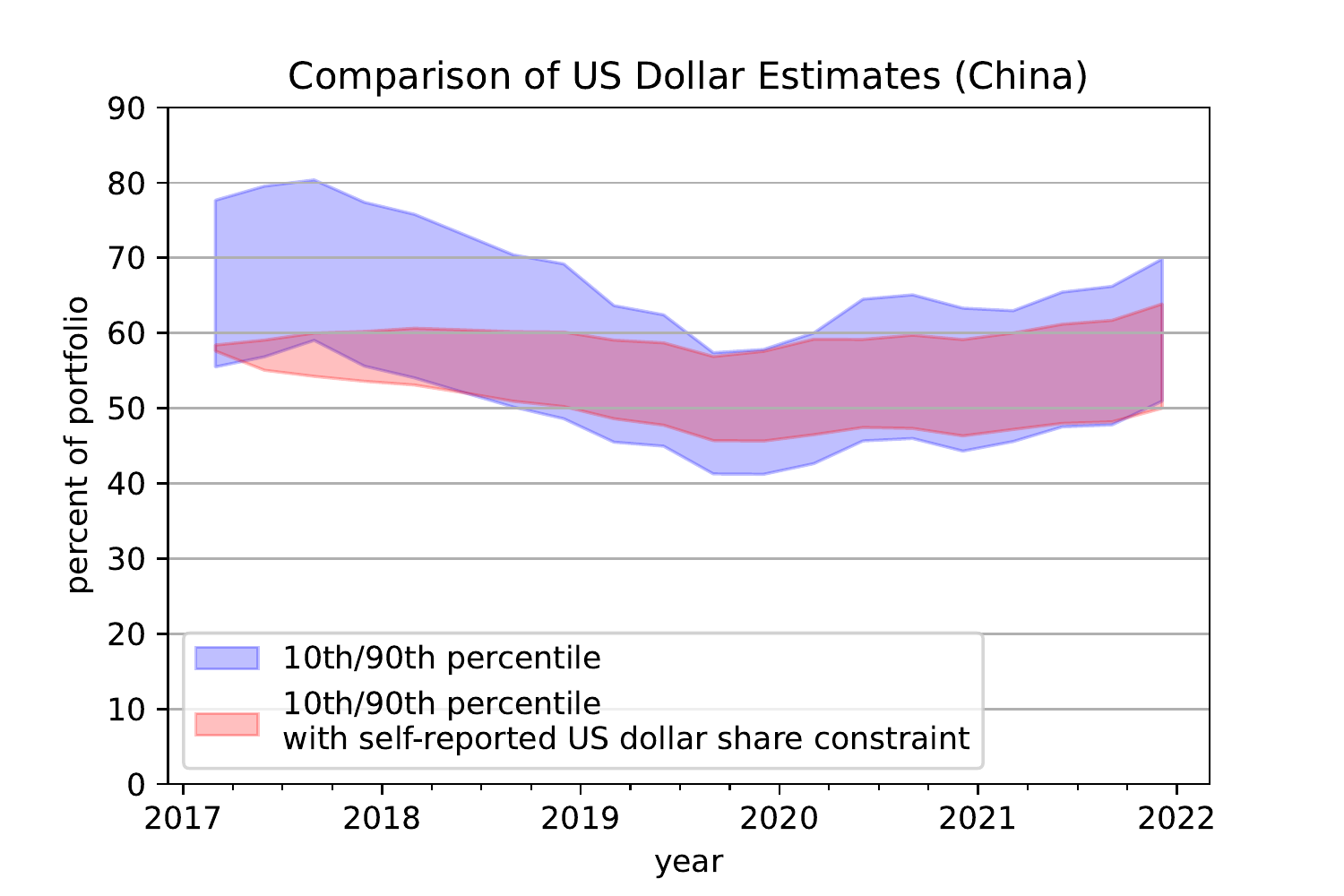}
\end{subfigure}\hspace*{\fill}

\begin{subfigure}{1.0\textwidth}
\includegraphics[width=\linewidth]{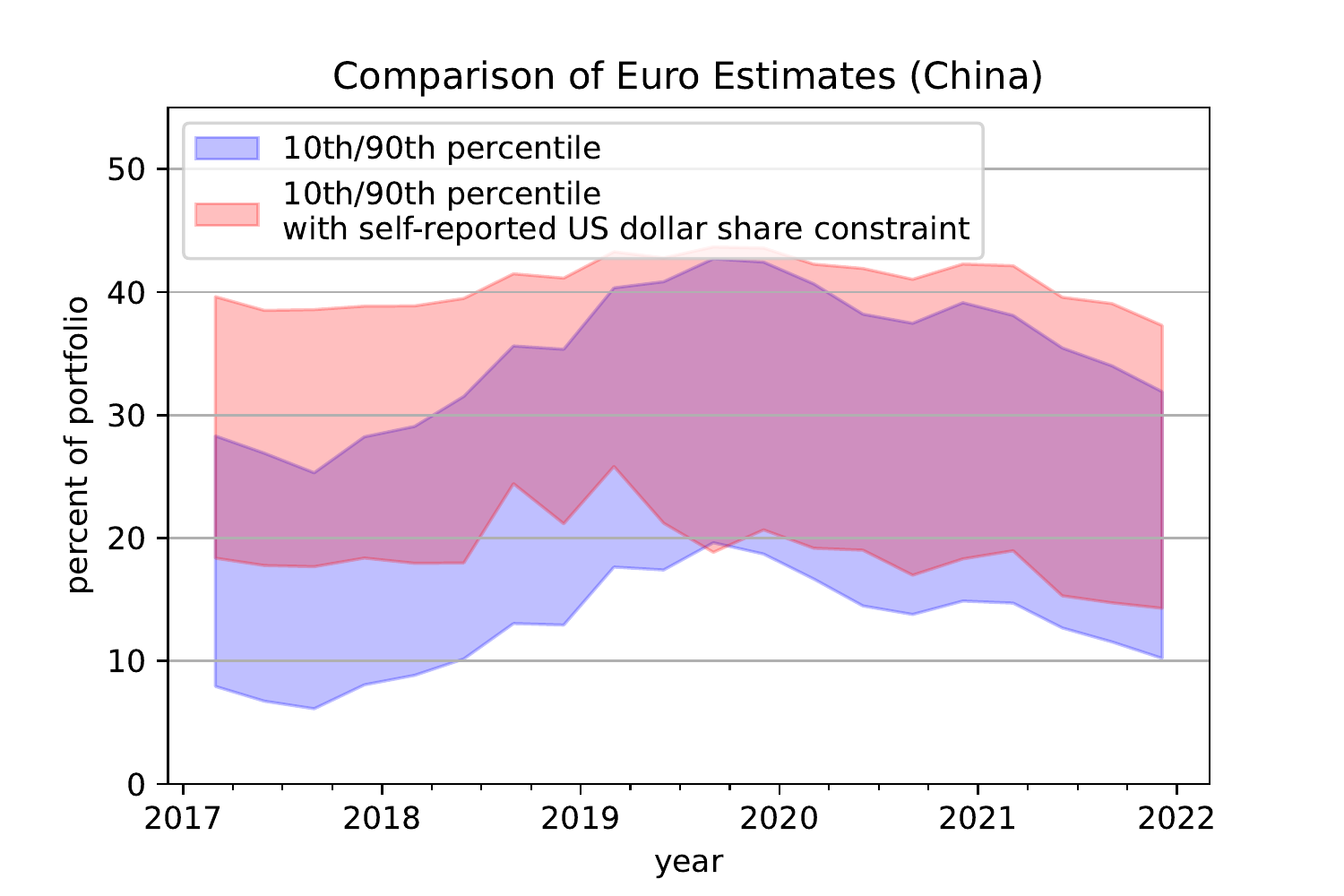}
\end{subfigure}\hspace*{\fill}

\caption{Filter estimates of China's currency shares, constrained using China's self-reported 2016 US dollar share.}

\label{chinafilterconstrained}

\end{figure}

In Figure \ref{chinafilterconstrained}, I present alternative filter estimates that begin in 2017, using a Dirichlet prior that constrains the 2017 US dollar share to be within 0.5\% of the China's self-reported share. Incorporating China's self-reported information into the model reduces the uncertainty in the estimates for several quarters, but the uncertainty interval around the estimates gradually widens over time, approaching the width of the uncertainty interval for the unconstrained estimates after about five years. The 2017 US dollar share constraint reduces the Q3 2022 median US dollar share estimate by 3.0 percentage points to 57.2\%, and increases the Q3 2022 median euro share estimate by 5.5 percentage points to 25.2\%.

\begin{figure}[h]
\begin{subfigure}{1.0\textwidth}
\includegraphics[width=\linewidth]{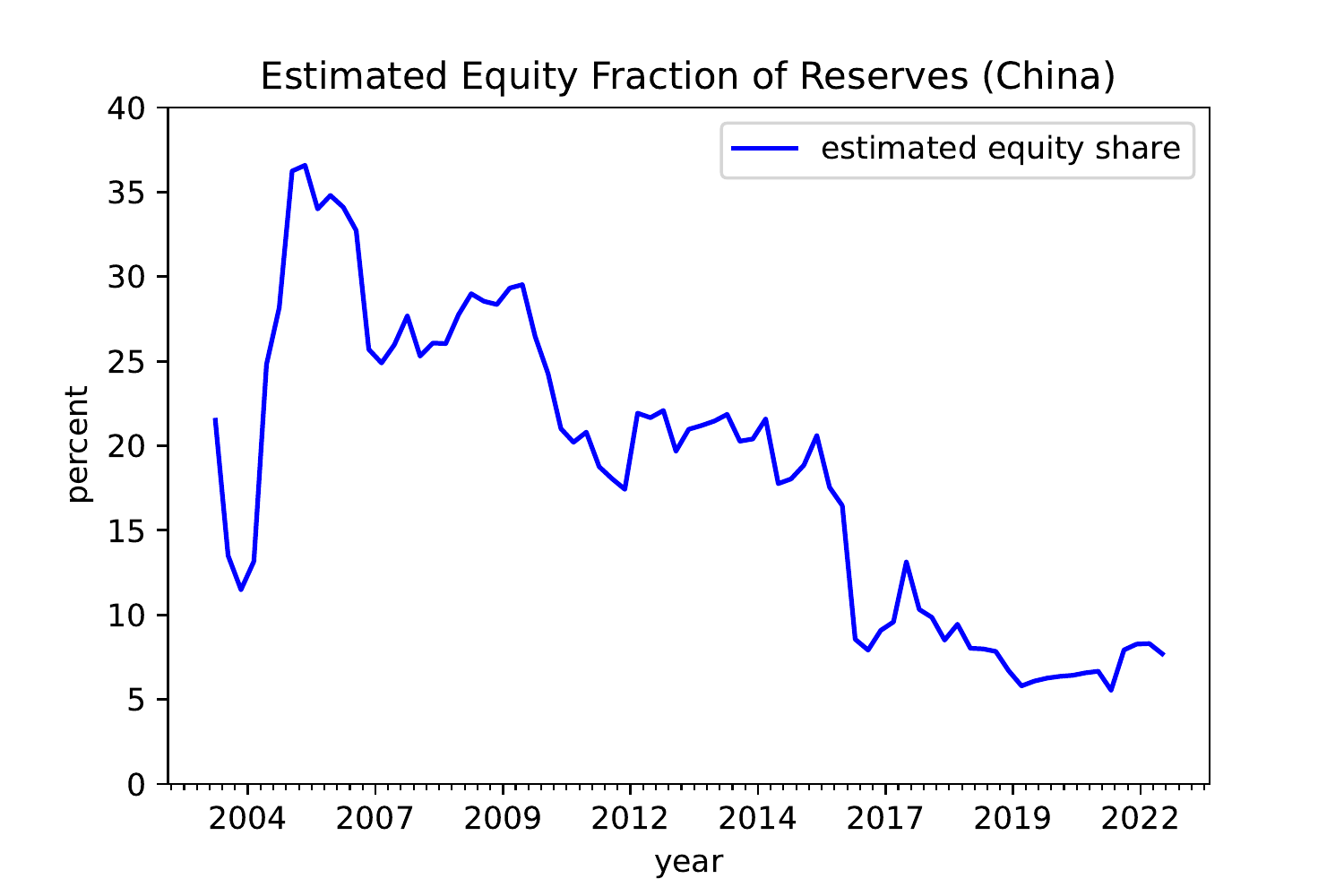}
\end{subfigure}\hspace*{\fill}

\caption{The estimated equity share of China's reserves.}

\label{chinaequity}

\end{figure}

Figure \ref{chinaequity} illustrates the estimated equity share of China's reserves, based on the optimization procedure (\ref{equityoptimization}). The elevated equity share during the financial crisis suggests that China may have owned some investments (such as mortgage backed securities) that were correlated with the performance of the stock market around that time.

\begin{figure}[h]
\begin{subfigure}{1.0\textwidth}
\includegraphics[width=\linewidth]{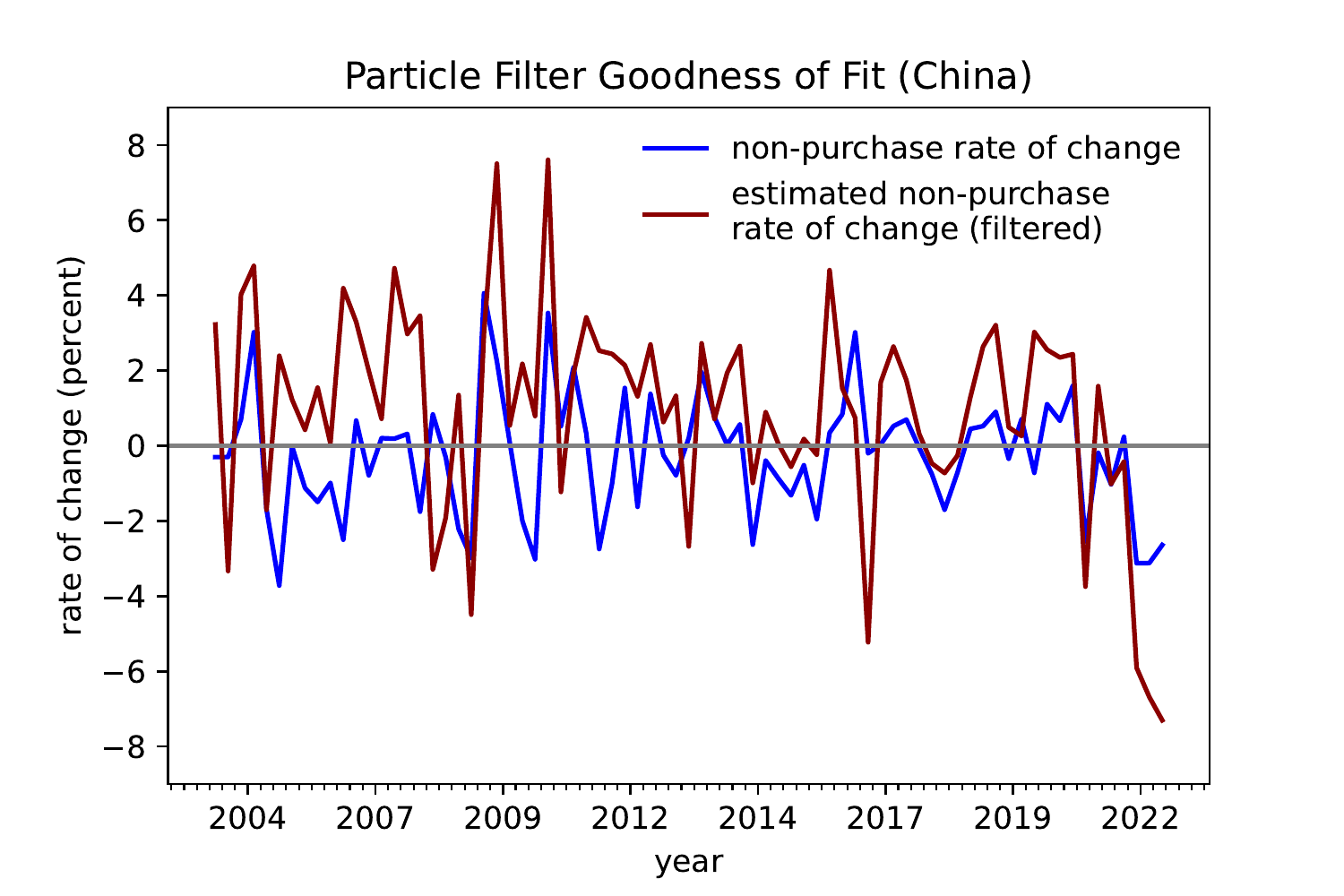}
\end{subfigure}\hspace*{\fill}

\caption{Median particle filter estimates, compared with the observed non-purchase rate of change of reserves.}

\label{chinagoodness}

\end{figure}

Figure \ref{chinagoodness} compares the median particle filter estimate to the non-purchase growth rate of reserves. This graphic illustrates how closely the estimated currency shares match the data. Curiously, during 2022, China's reserves outperformed both the global equity and bond markets. 

{\FloatBarrier}
\subsubsection{Comparison to Other Estimates}

\cite{chinafx} estimated China's reserve currency composition as of Q1 2015, using a least-squares optimization method rather than a Markov-based approach. Their method does not account for the time series structure of the currency shares. Nevertheless, their estimates largely agree with my own. A comparison with \cite{chinafx} can be found in Table \ref{optimcomp}. Note that the currency shares do not sum to one because calculating the median share does not preserve the sum constraint due to the skewness of the currency share distributions, and because \cite{chinafx} also estimated a 0.09\% allocation to the Swiss franc, a minor reserve currency which I omitted from my model.

\begin{table}[hbt!]
\caption{Comparison of Estimates with Shi and Nie (2017), as of Q1 2015} 
\label{optimcomp}
\centering 
\begin{tabular}{l *{6}{D..{-1}}} 
\toprule
 & \multicolumn{1}{c@{}}{USD} & \multicolumn{1}{c@{}}{EUR} & \multicolumn{1}{c@{}}{GBP} & \multicolumn{1}{c@{}}{JPY} & \multicolumn{1}{c@{}}{CAD} & \multicolumn{1}{c@{}}{AUD} \\ [0.5ex] 
\midrule
Shi and Nie (2017) & 63.6\% & 19.6\% & 4.9\% & 3.1\% & 2.2\% & 2.0\% \\ [0.5ex]
Hidden Markov Model (median) & 61.3\% & 19.6\% & 2.2\% & 9.4\% & 1.7\% & 2.4\% \\ [1ex] 
\bottomrule
\end{tabular}
\end{table}

\begin{figure}[h]
 \centering
    \includegraphics[scale = 1.1]{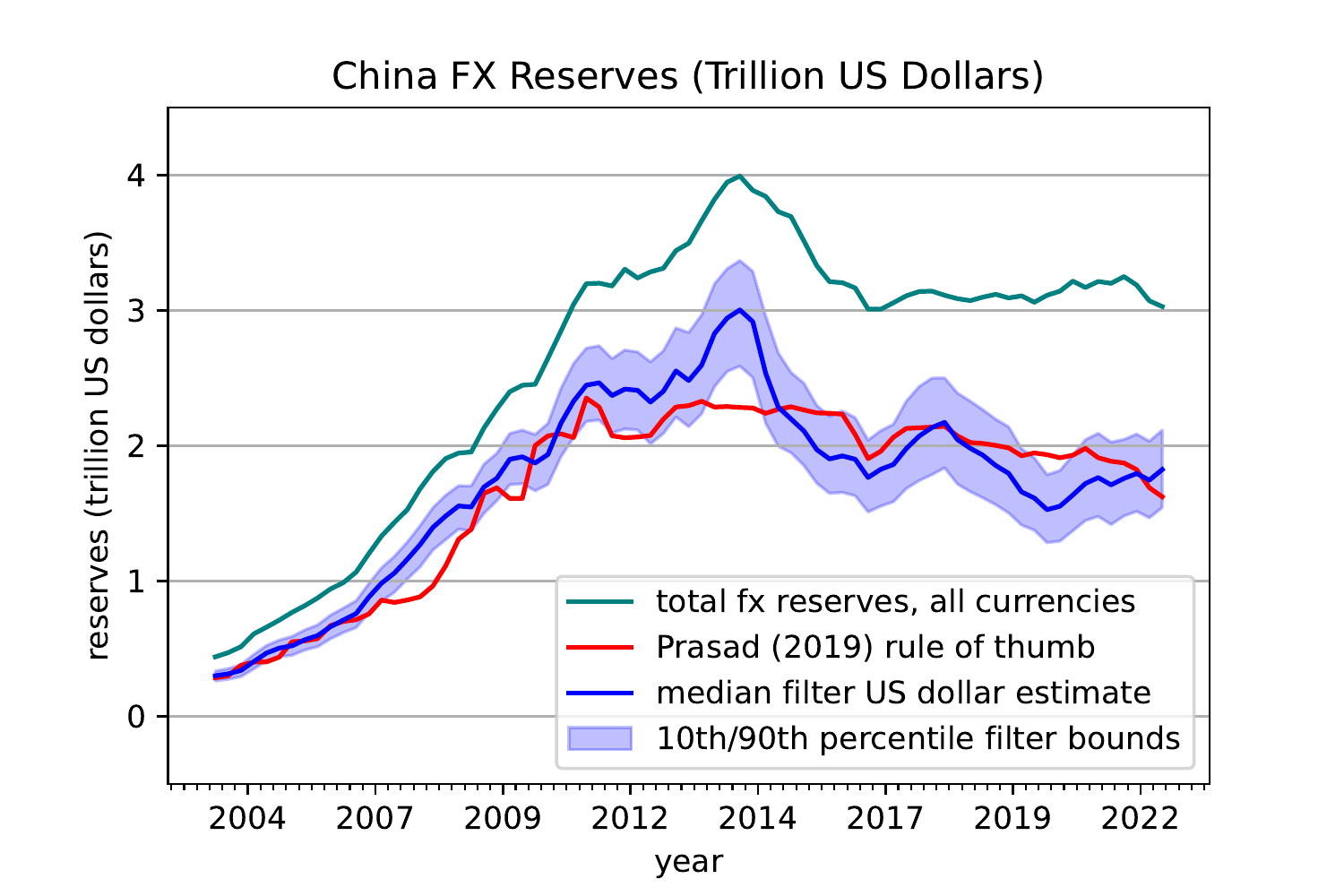}
    \caption{A comparison of the model estimates to the Prasad (2019) rule of thumb.}
    \label{tic}
\end{figure}

My model does not use any data from the Treasury International Capital (TIC) database.\footnote{For other estimates up to 2019 using TIC data, see: cfr.org/blog/lessons-phase-one-trade-war-china} The TIC database has some significant shortcomings. First, TIC measures the aggregate US Treasury holdings of all entities domiciled in mainland China, not just China's foreign exchange reserves. Second, TIC attributes assets to the country where the assets are held in custody, rather than to the beneficial owner. For example, if a Chinese entity buys a Treasury bond using an account in Germany, the asset is recorded against Germany rather than China. Finally, sub-annual TIC data prior to 2011 is constructed by adjusting annual custodial data using monthly transaction data. The transaction data records the location of the financial institution that executed the trade, rather than the location of the beneficiary of the trade. \cite{chinabook} points out that China appears to be aware of the TIC database and may strategically purchase US dollar securities using custodial accounts in other countries to evade scrutiny of its dollar holdings. 

Despite the limitations of the TIC dataset, financial media outlets often endow small changes in TIC holdings with great significance.\footnote{For example: reuters.com/article/us-china-economy-treasury-idUSKBN25V179} Consequently, it is informative to compare estimates from my Hidden Markov Model to estimates based on TIC data.

\cite{prasad} proposes a rule of thumb to estimate the dollar share of China's reserves based on TIC data and China's self-reported 2014 US dollar share. Prasad hypothesizes that the ratio of China's total dollar holdings to China's Treasury holdings reported by TIC has remained roughly constant over time. In 2014, that ratio was about 1.8. Figure \ref{tic} compares the Hidden Markov Model to this rule of thumb estimate. This rule of thumb matches the model's estimates reasonably well.

\afterpage
\FloatBarrier

\subsection{Singapore}

Although Singapore's foreign exchange reserves are much smaller than China's, they are large relative to the size of Singapore's economy. Foreign exchange reserves as of Q4 2022 were \$279 billion, compared to 2021 GDP of \$397 billion. Indeed, the Monetary Authority of Singapore occasionally transfers excess reserves into the Government Investment Corporation, Singapore's sovereign wealth fund, to be invested into riskier assets. Singapore follows a managed floating exchange rate regime, but does not disclose its currency basket. I am unable to find any previous estimates of Singapore's currency composition in the literature. 

My results are available in Figure \ref{sgresults}, based on seven reserve currencies (including the Chinese renminbi). I set a more diffuse prior in the case of Singapore, reflecting the lack of public information about the country's investment policies. Notably, Singapore appears to hold fewer US dollars, and more euros and British pounds, compared to world averages. Singapore's equity share of reserves is similar to that of China.

\begin{figure}[h]
\begin{subfigure}{1.0\textwidth}
\includegraphics[width=\linewidth]{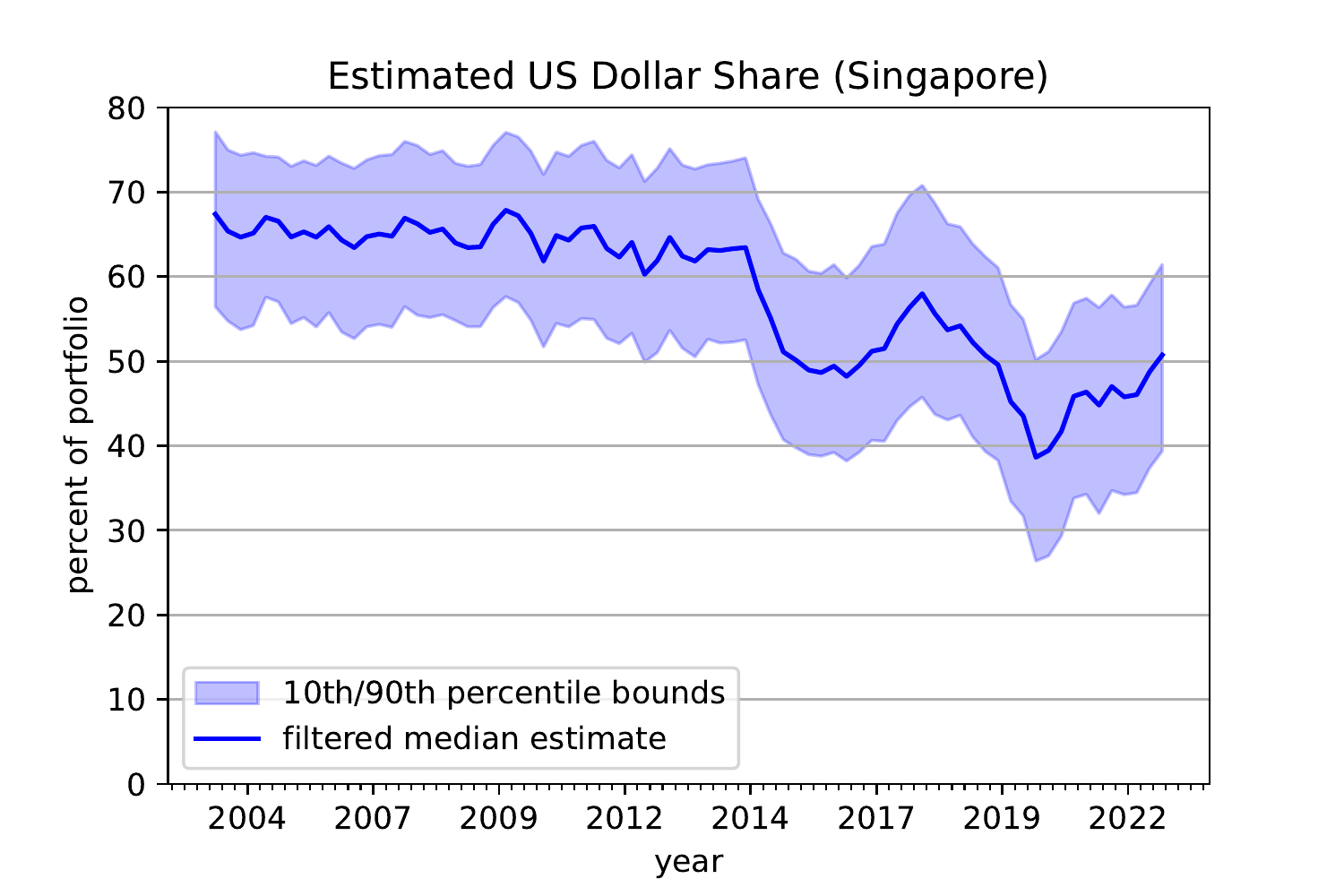}
\end{subfigure}\hspace*{\fill}

\begin{subfigure}{1.0\textwidth}
\includegraphics[width=\linewidth]{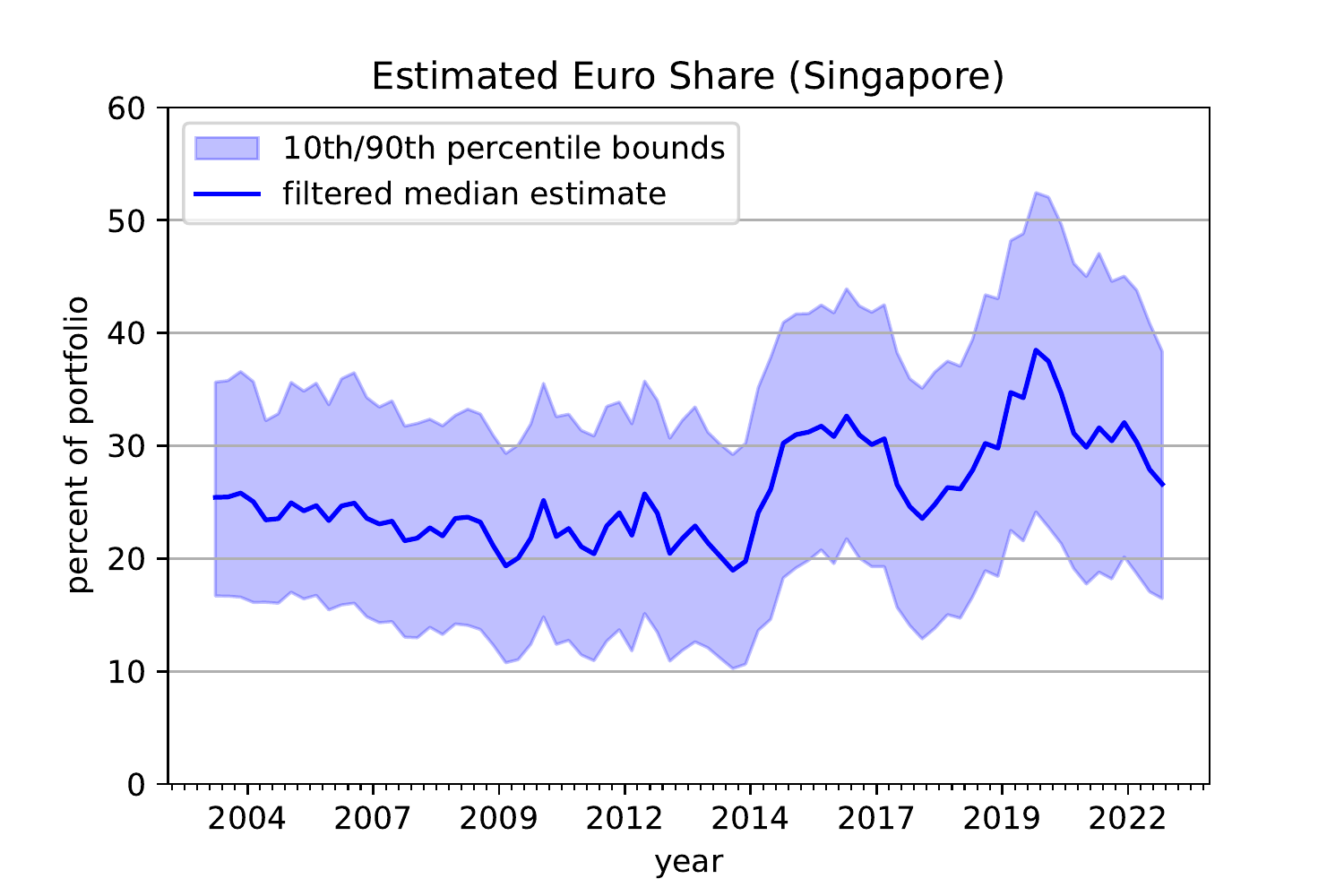}
\end{subfigure}\hspace*{\fill}

\caption{Filter estimates of Singapore's currency shares.}

\label{sgresults}
\end{figure}

\begin{figure}[h]\ContinuedFloat
\begin{subfigure}{1.0\textwidth}
\includegraphics[width=\linewidth]{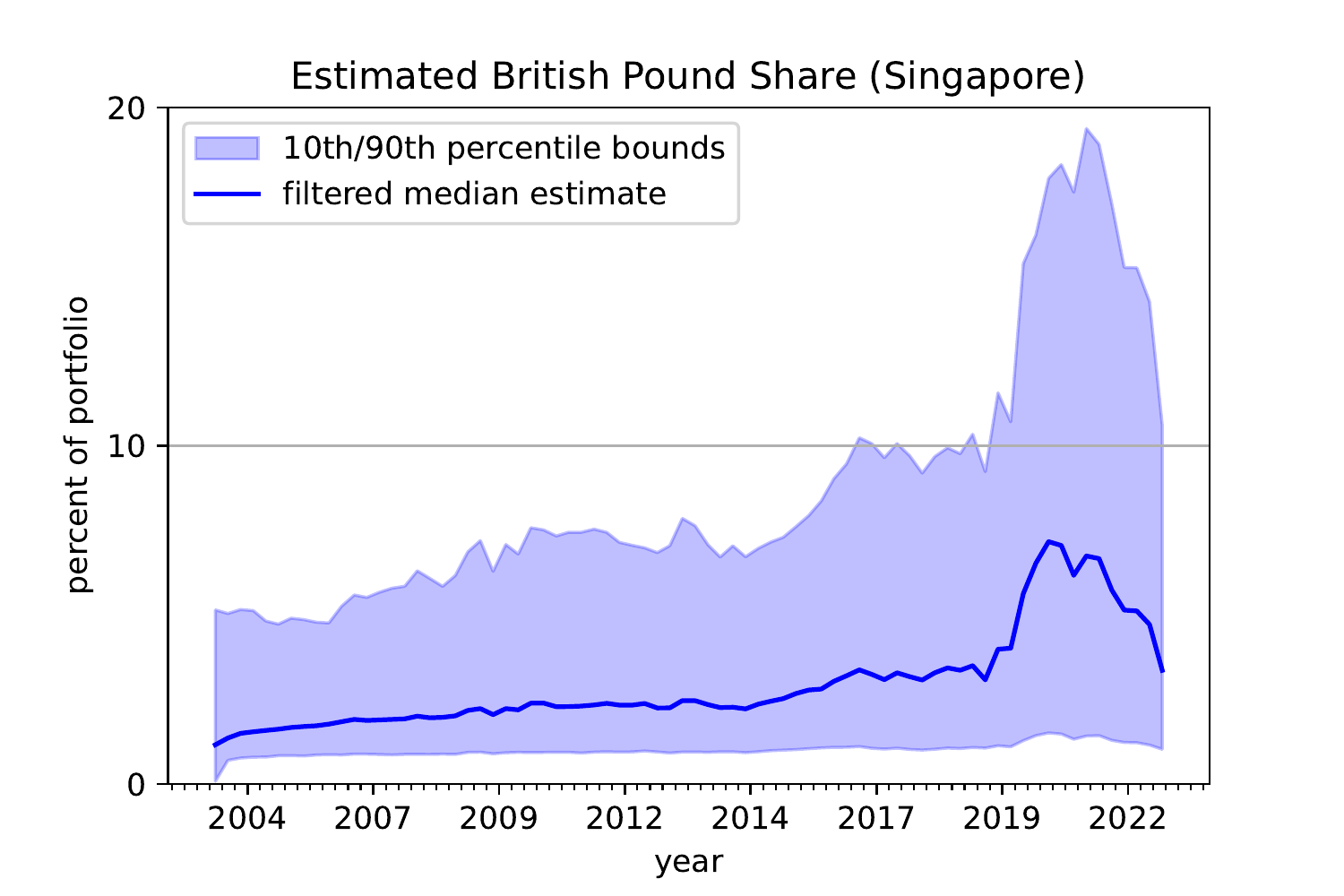}
\end{subfigure}\hspace*{\fill}

\begin{subfigure}{1.0\textwidth}
\includegraphics[width=\linewidth]{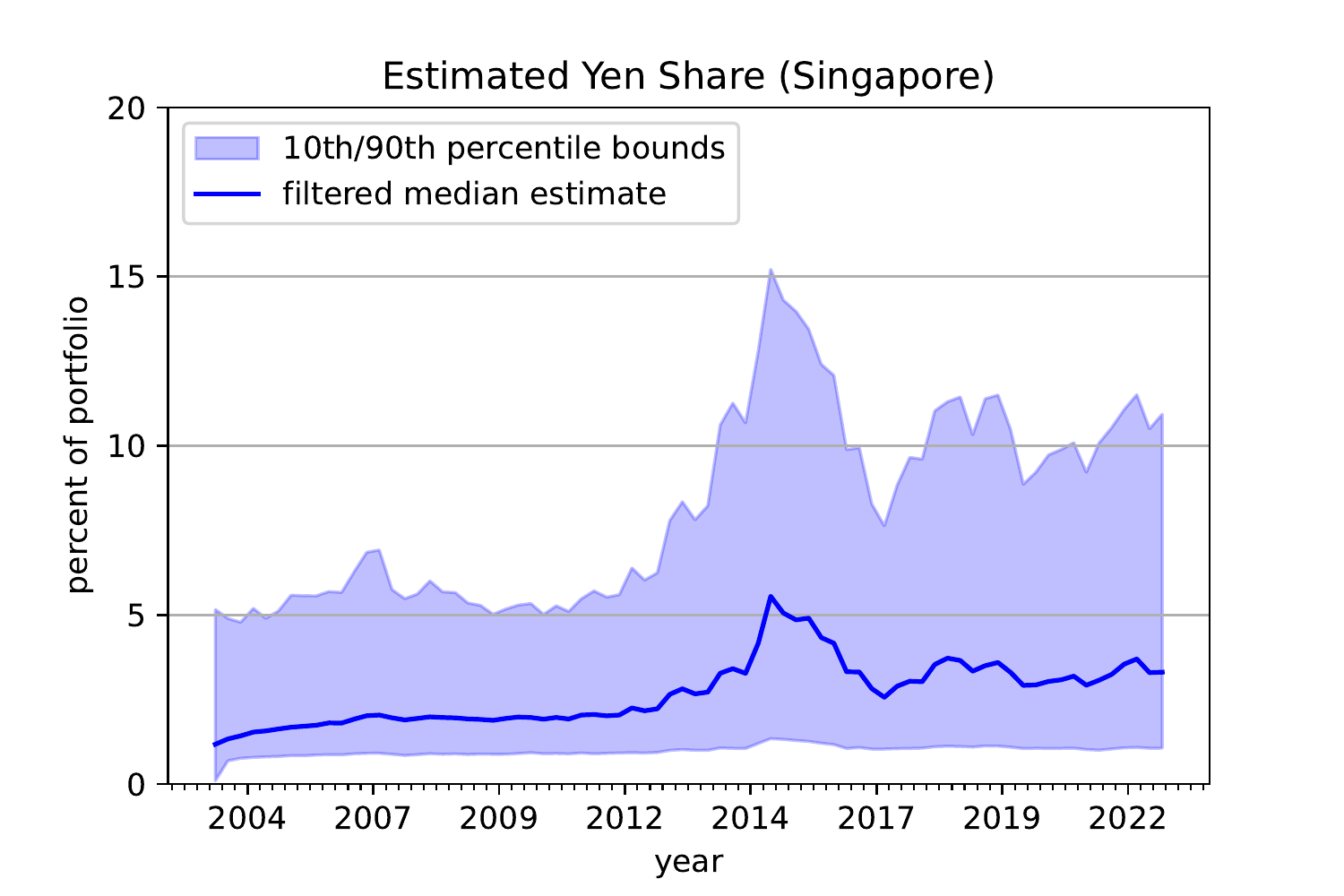}
\end{subfigure}\hspace*{\fill}

\caption{Filter estimates of Singapore's currency shares.}

\label{sgresults}

\end{figure}

\begin{figure}[h]\ContinuedFloat
\begin{subfigure}{1.0\textwidth}
\includegraphics[width=\linewidth]{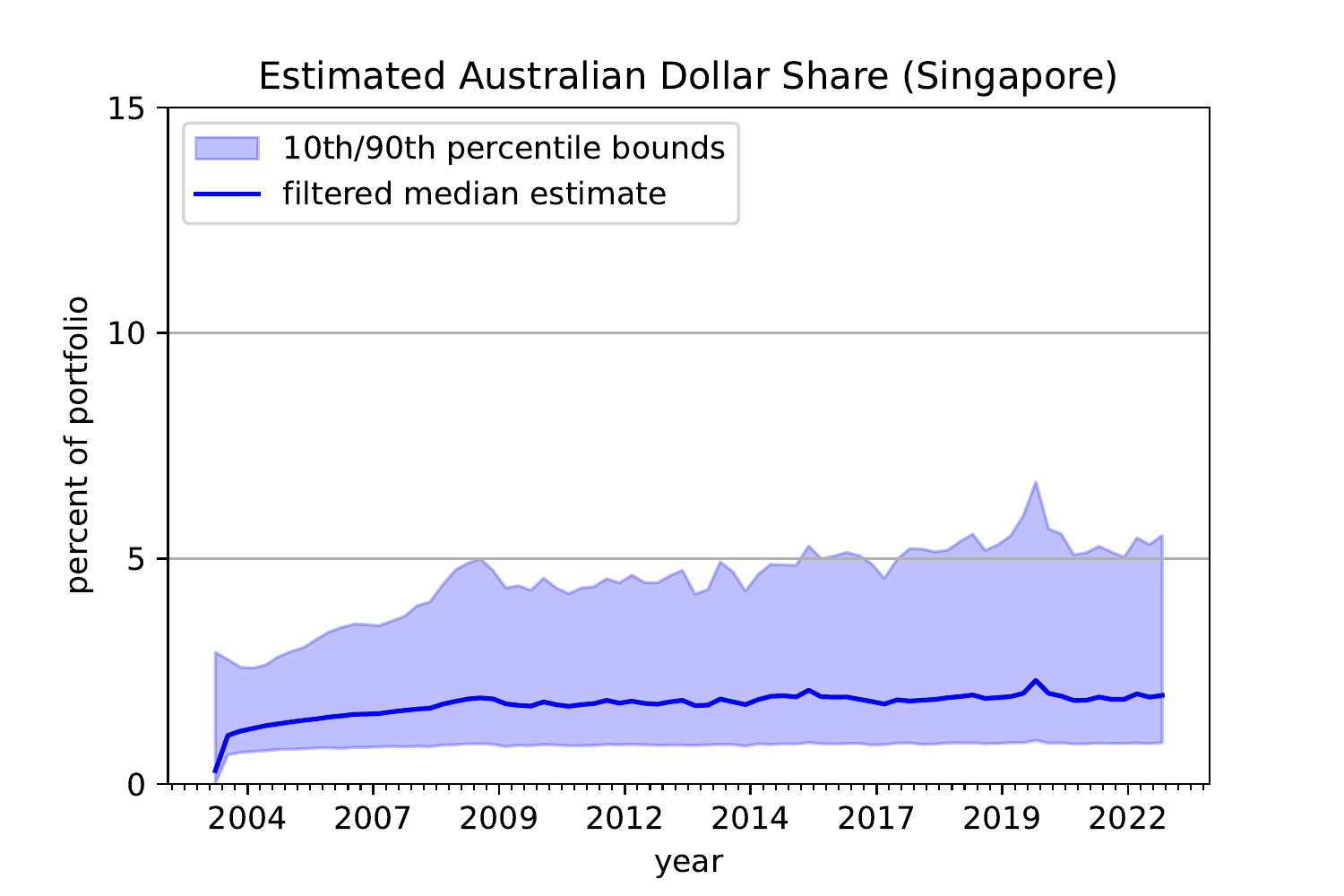}
\end{subfigure}\hspace*{\fill}

\begin{subfigure}{1.0\textwidth}
\includegraphics[width=\linewidth]{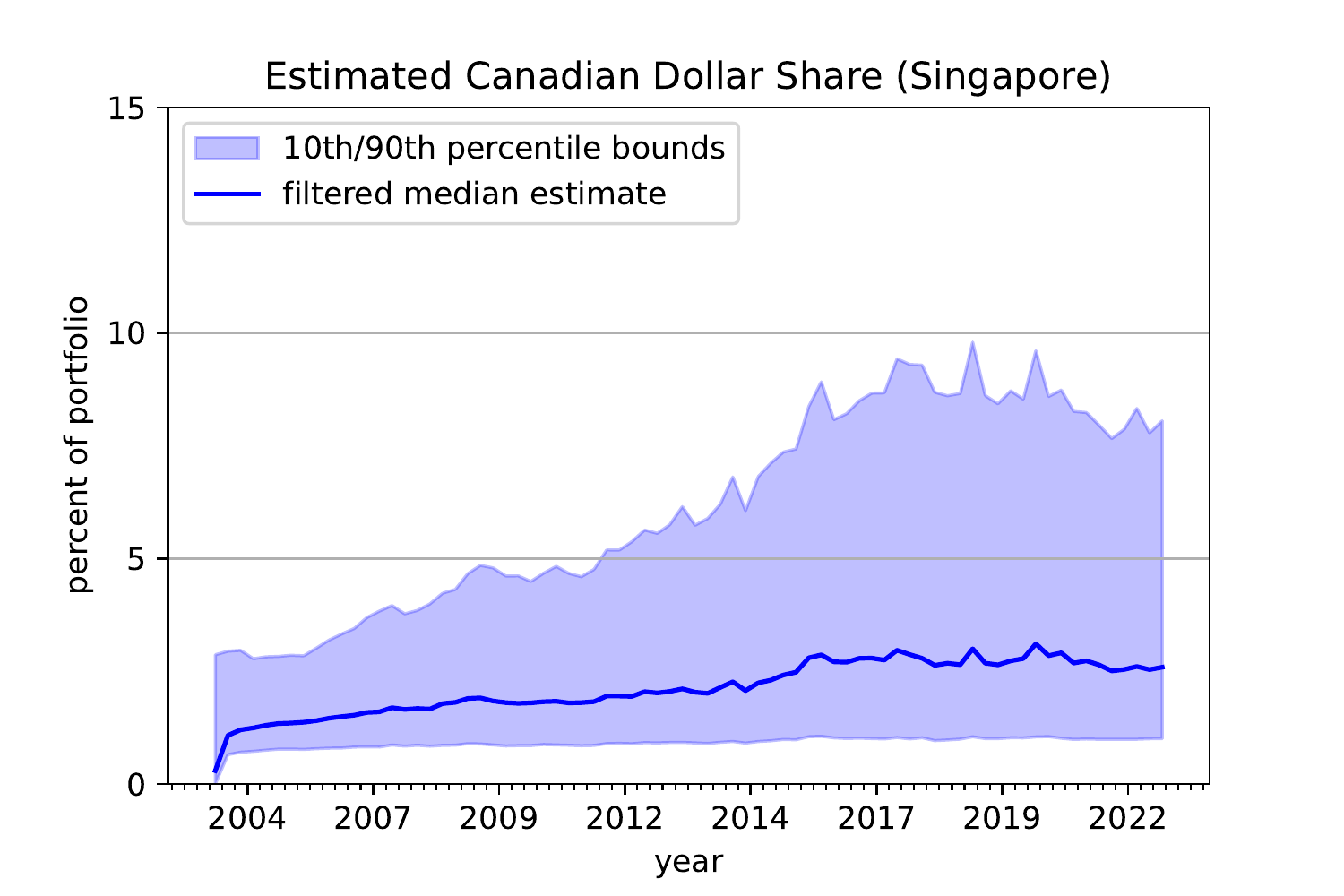}
\end{subfigure}\hspace*{\fill}

\caption{Filter estimates of Singapore's currency shares.}

\label{sgresults}

\end{figure}

\begin{figure}[h]\ContinuedFloat
\begin{subfigure}{1.0\textwidth}
\includegraphics[width=\linewidth]{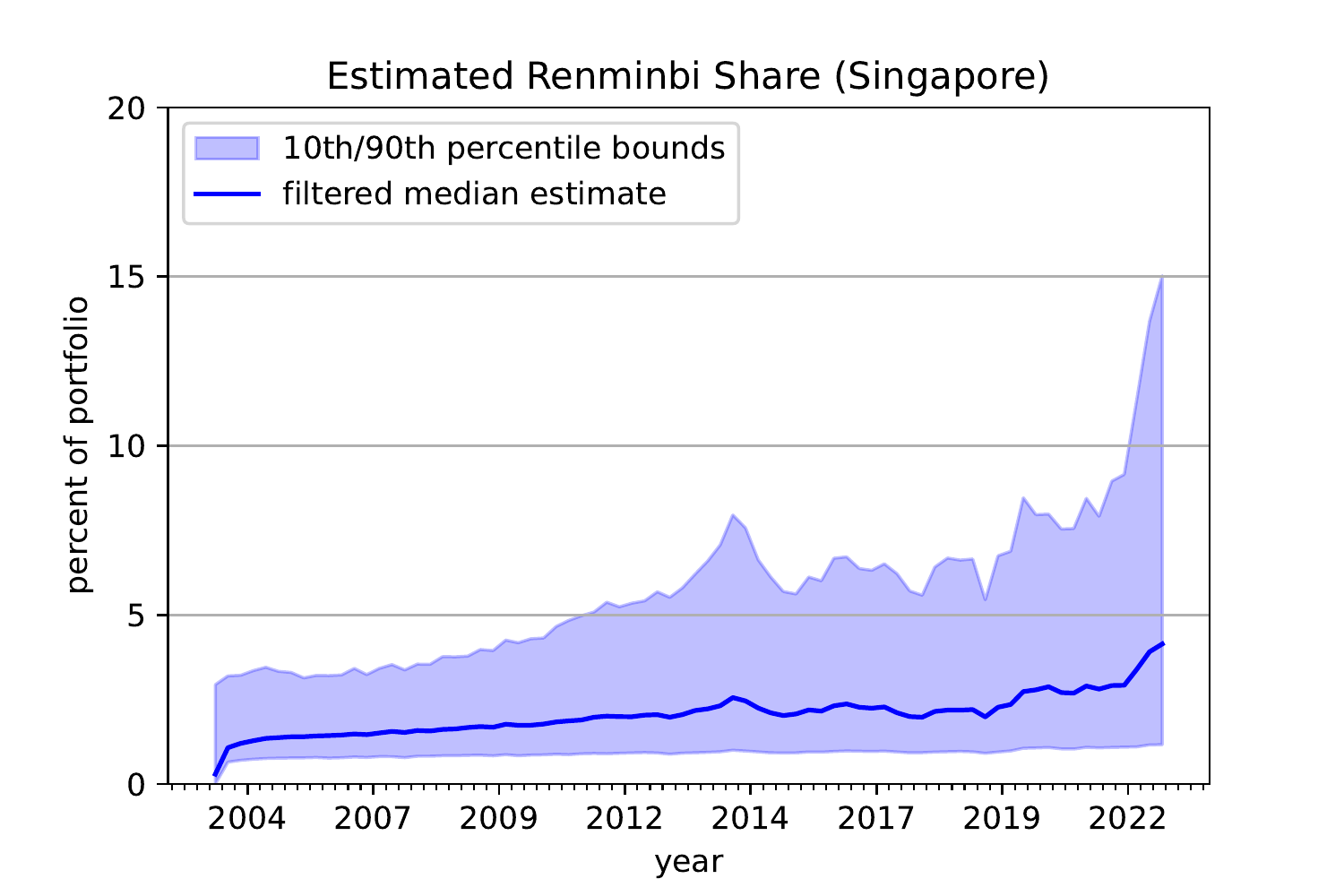}
\end{subfigure}\hspace*{\fill}

\caption{Filter estimates of Singapore's currency shares.}

\label{sgresults}

\end{figure}

\begin{figure}[h]
\begin{subfigure}{1.0\textwidth}
\includegraphics[width=\linewidth]{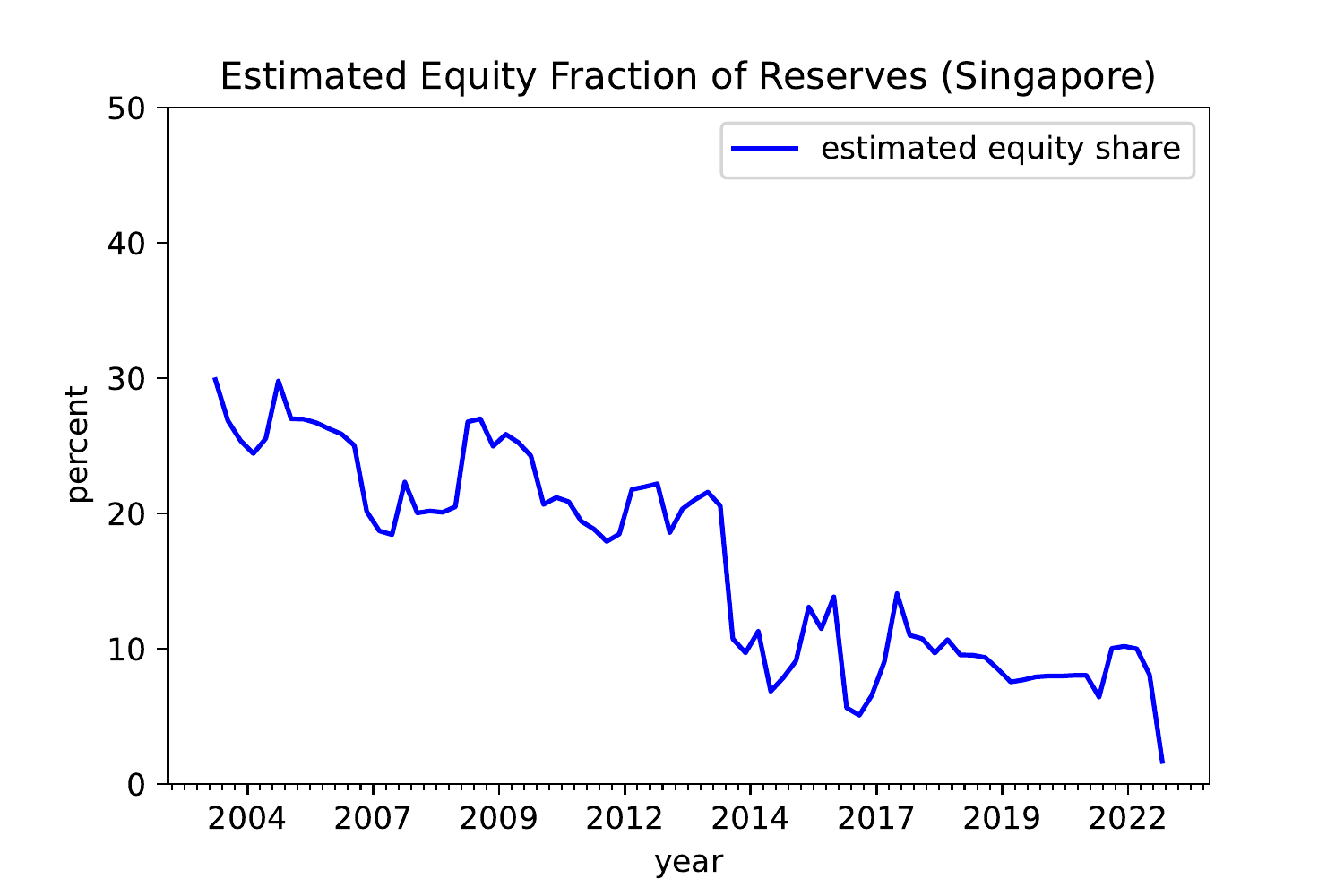}
\end{subfigure}\hspace*{\fill}

\caption{The estimated equity share of Singapore's reserves.}

\label{sgequity}

\end{figure}

\begin{figure}[h]
\begin{subfigure}{1.0\textwidth}
\includegraphics[width=\linewidth]{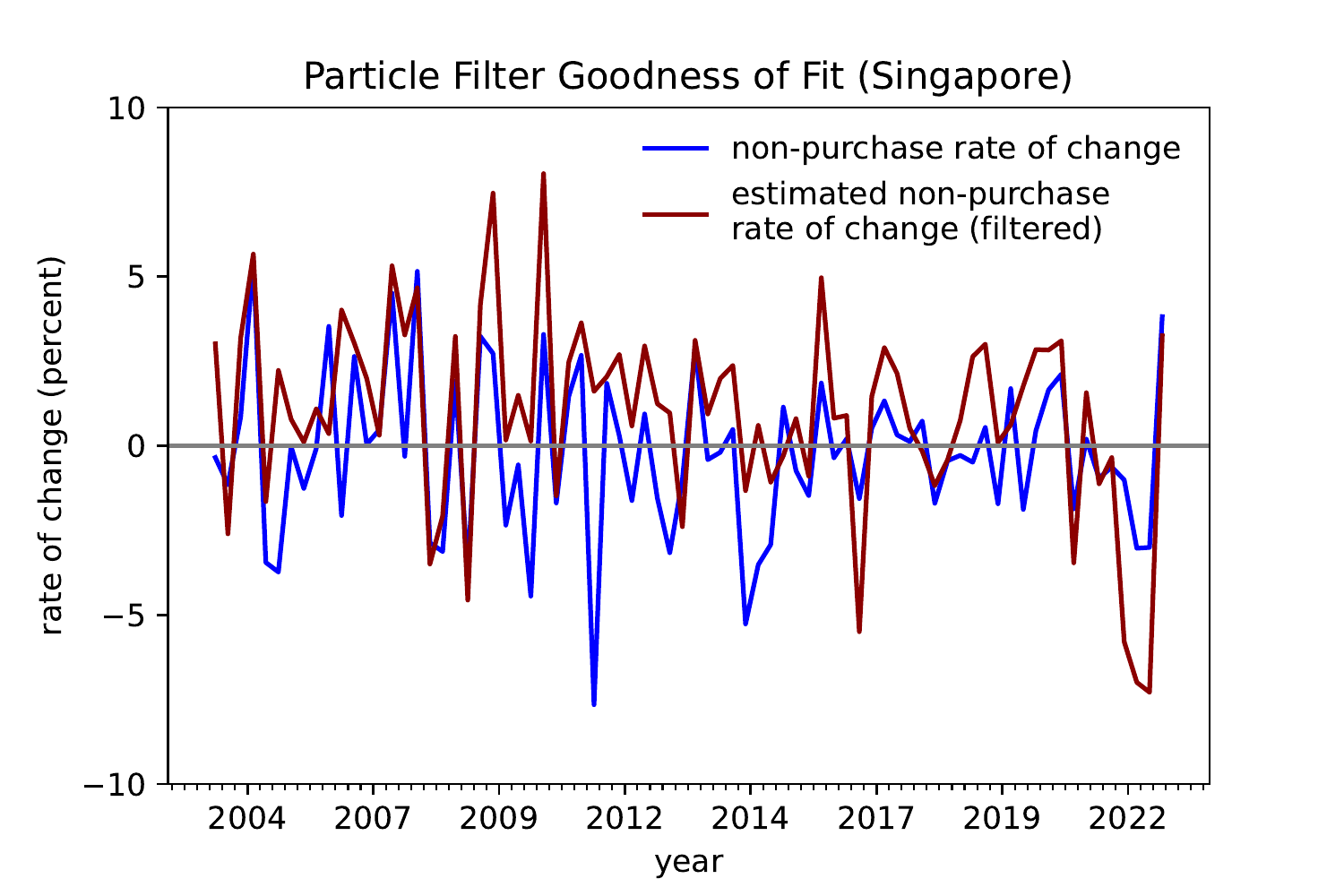}
\end{subfigure}\hspace*{\fill}

\caption{Median particle filter estimates, compared with the observed non-purchase rate of change of reserves.}

\label{sggoodness}

\end{figure}

\FloatBarrier
\section{Benchmarking}
\label{sec:benchmarking}

Because some countries do report the currency composition of their foreign exchange reserves, I can test the model by comparing its estimates to these countries' self-reported currency shares.

I use 5-year bonds for Switzerland, and 2-year bonds for Brazil, based on both countries' self-reported investment management practices.

Number of currencies for Brazil / Switzerland
"Other" Currencies for switzerland

\FloatBarrier
\subsection{Brazil}

Brazil possesses a foreign exchange reserve portfolio of about \$270 billion. Furthermore, Brazil invests primarily in high-quality, short-term debt, with an average duration ranging from 1.8 to 3.2 years. Therefore, the returns on Brazil's reserves should be similar to the returns on 2-year sovereign bonds. Brazil invests less than 1.5\% of its reserves in equities, and only began buying equities in 2018. The equity share optimization correctly identifies that the equity share of Brazil's reserves is very low.

Brazil poses a challenging test for the model because the variation in the US dollar and euro shares was very high between 2004 and 2007, when Brazil increased its US dollar share from 55\% to 90\% and reduced its euro share from 35\% to 10\%. A high US dollar allocation is typical of central banks in the Americas, based on the World Bank \cite{worldbanksurvey}, which surveyed 105 central banks with an 85\% response rate. Brazil implements a floating exchange rate regime. In 2019, Brazil added Chinese renminbi to its reserves. Therefore, I estimate the model based on seven reserve currencies.

Results are available in Figure \ref{brresults}. A comparison of the Laplace, Normal, and Cauchy distributions, illustrating the robustness of the Laplace approach, can be found in Appendix \ref{appendixcauchy}. The resolution limit of the filter is about 5 percentage points, so the estimated upper bound of the minor currencies is around 5 percentage points.

\begin{figure}[h]
 \centering
    \includegraphics[scale = 1.1]{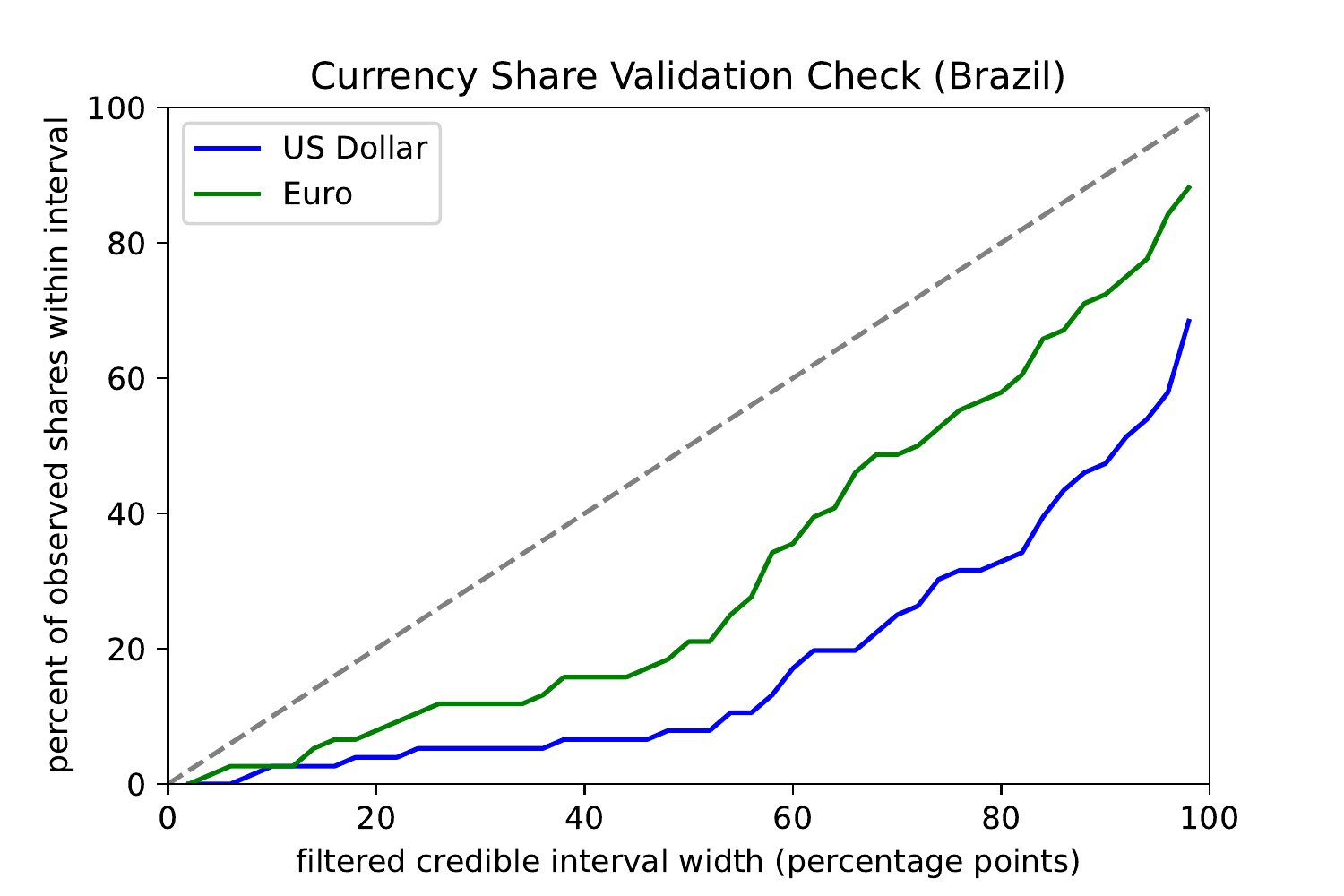}
    \vspace{-1em}
    \caption{The proportion of time periods in which the self-reported currency share falls within a credible interval centered around the estimated median. The closer to the diagonal line, the more precisely the filter captures the true currency shares at the specified level of confidence.}
    \label{brvalidation}
\end{figure}

Figure \ref{brvalidation} illustrates the extent of the agreement between Brazil's self-reported currency shares and the filter estimates. For Brazil, the uncertainty intervals around the filtered estimates are too narrow. For example, the credible interval ranging from the 10th to 90th percentile captures the correct self-reported currency share 33\% of the time for the US dollar share, and 58\% of the time for the euro share. The model's overconfidence derives from the rapid change in Brazil's currency shares between 2004 and 2007. The model assumes the quarterly standard deviation of the US dollar share is 1.5\%, so it takes a few years for the filter estimates to catch up to the new US dollar share. Although increasing the standard deviation parameter allows the estimated shares to adjust more quickly, setting the parameter too high results in implausible fluctuations in the estimated shares due to overfitting.

\begin{figure}[h]
\begin{subfigure}{1.0\textwidth}
\includegraphics[width=\linewidth]{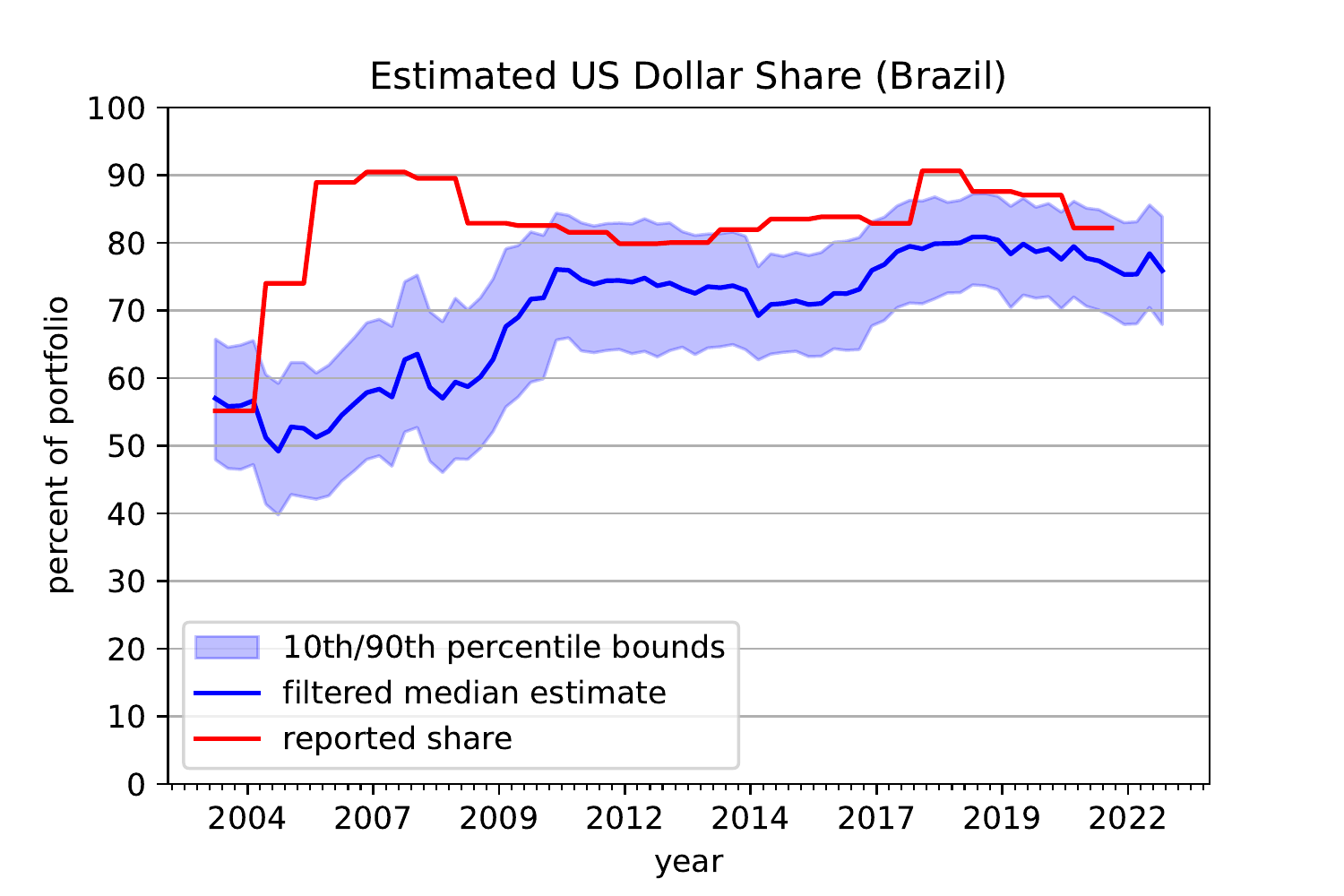}
\end{subfigure}\hspace*{\fill}

\begin{subfigure}{1.0\textwidth}
\includegraphics[width=\linewidth]{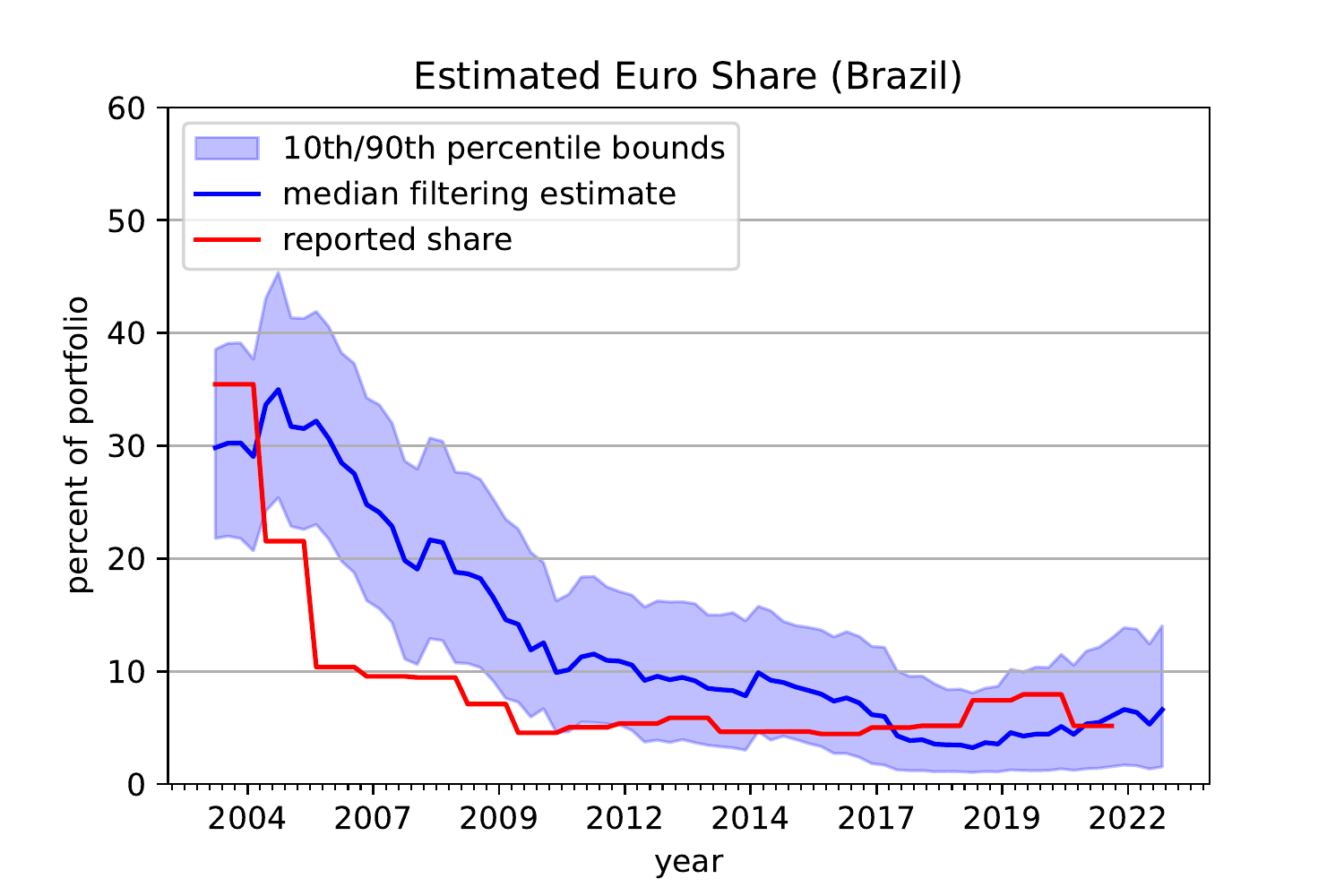}
\end{subfigure}\hspace*{\fill}

\caption{Filter estimates of Brazil's currency shares, compared with Brazil's self-reported shares.}

\label{brresults}
\end{figure}

\begin{figure}[h]\ContinuedFloat
\begin{subfigure}{1.0\textwidth}
\includegraphics[width=\linewidth]{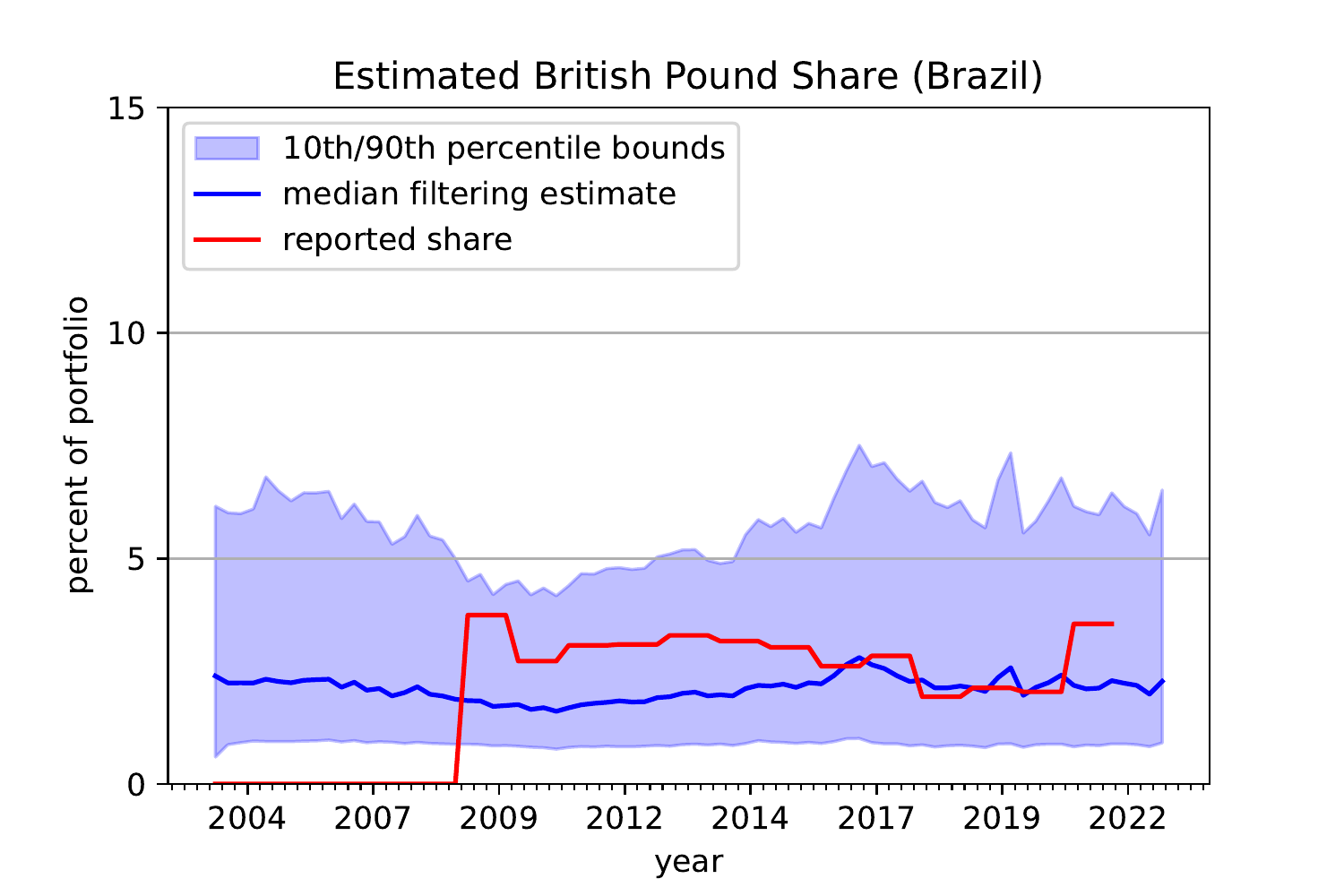}
\end{subfigure}\hspace*{\fill}

\begin{subfigure}{1.0\textwidth}
\includegraphics[width=\linewidth]{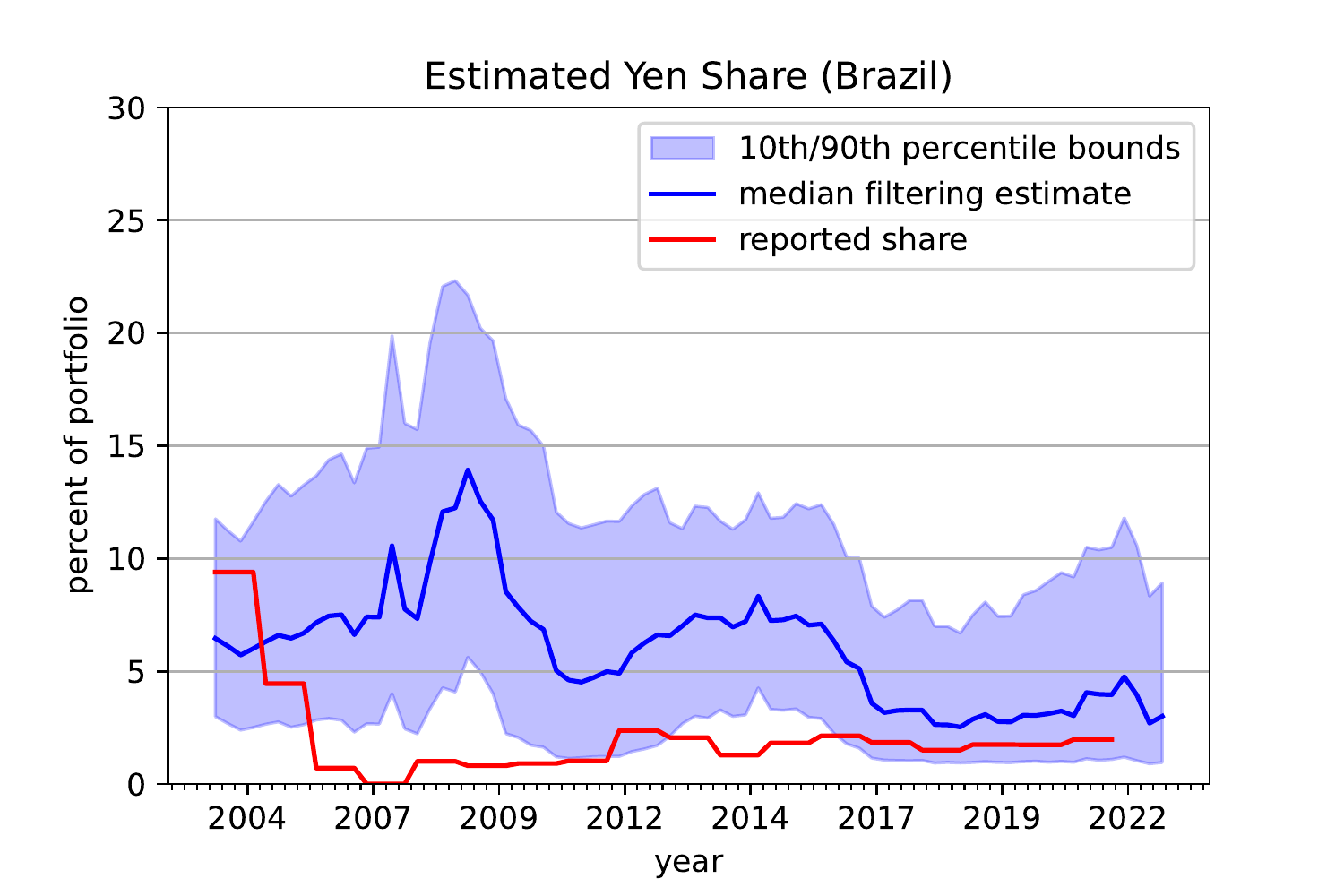}
\end{subfigure}\hspace*{\fill}

\caption{Filter estimates of Brazil's currency shares, compared with Brazil's self-reported shares.}

\label{brresults}

\end{figure}

\begin{figure}[h]\ContinuedFloat
\begin{subfigure}{1.0\textwidth}
\includegraphics[width=\linewidth]{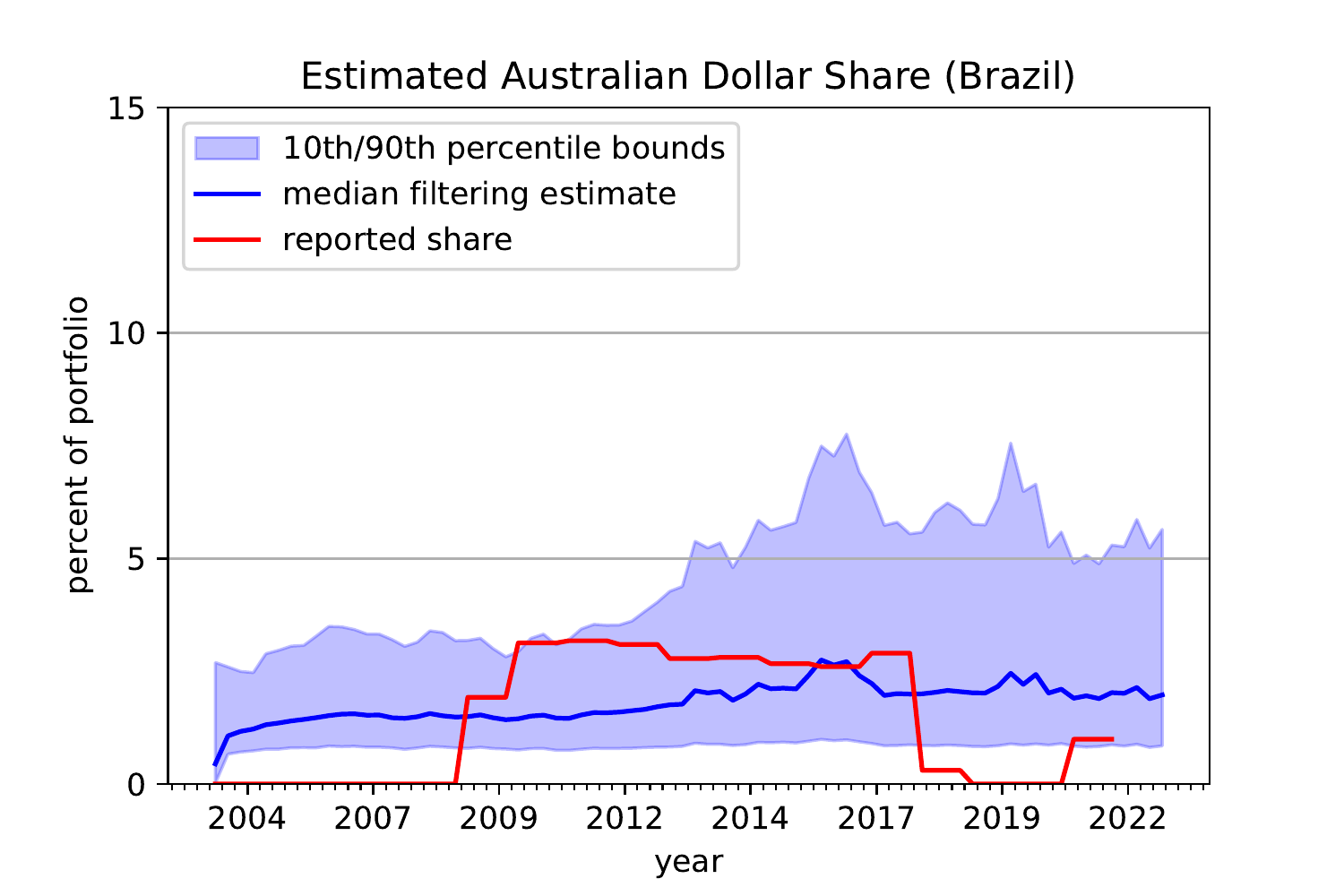}
\end{subfigure}\hspace*{\fill}

\begin{subfigure}{1.0\textwidth}
\includegraphics[width=\linewidth]{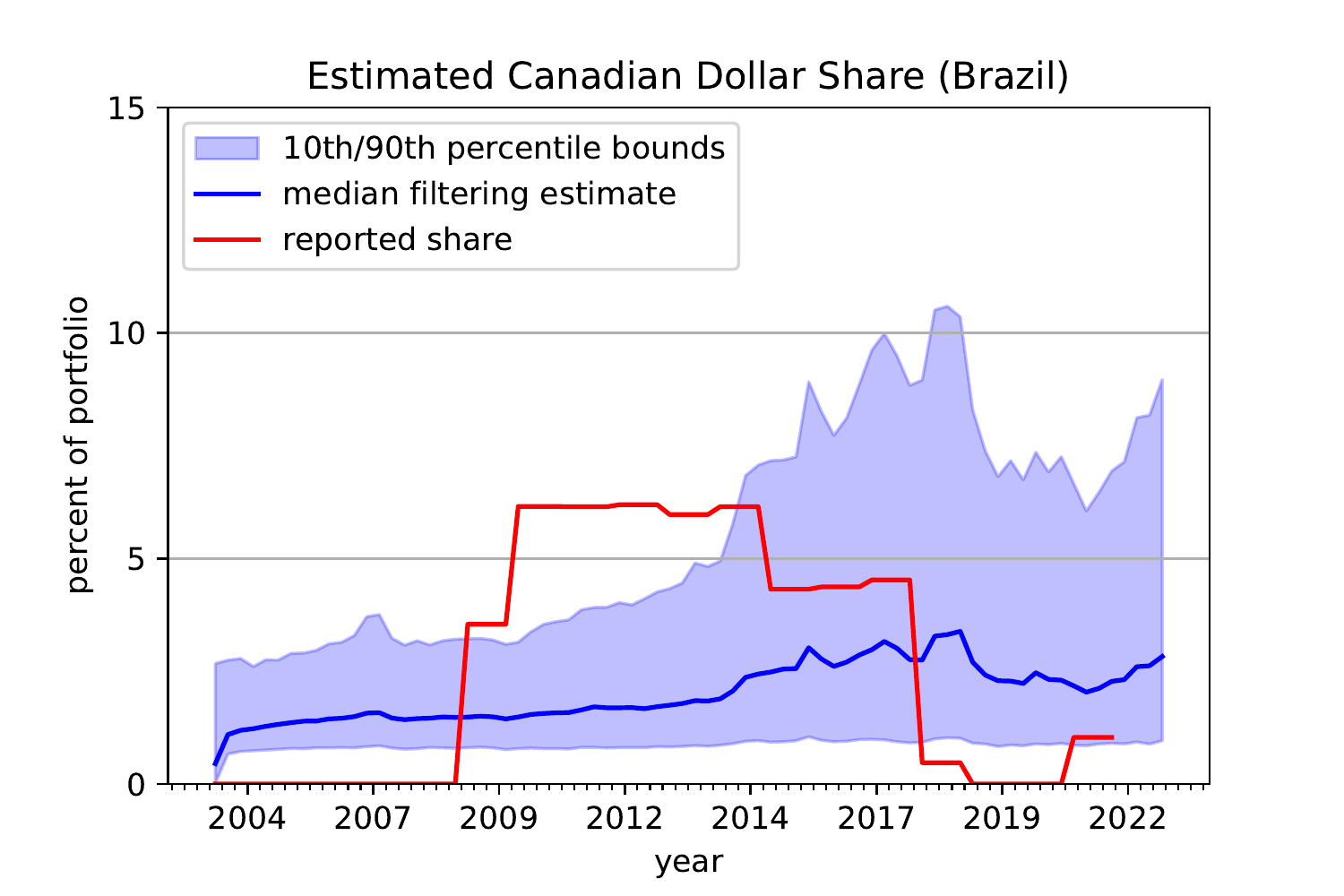}
\end{subfigure}\hspace*{\fill}

\caption{Filter estimates of Brazil's currency shares, compared with Brazil's self-reported shares.}

\label{brresults}

\end{figure}

\begin{figure}[h]\ContinuedFloat

\begin{subfigure}{1.0\textwidth}
\includegraphics[width=\linewidth]{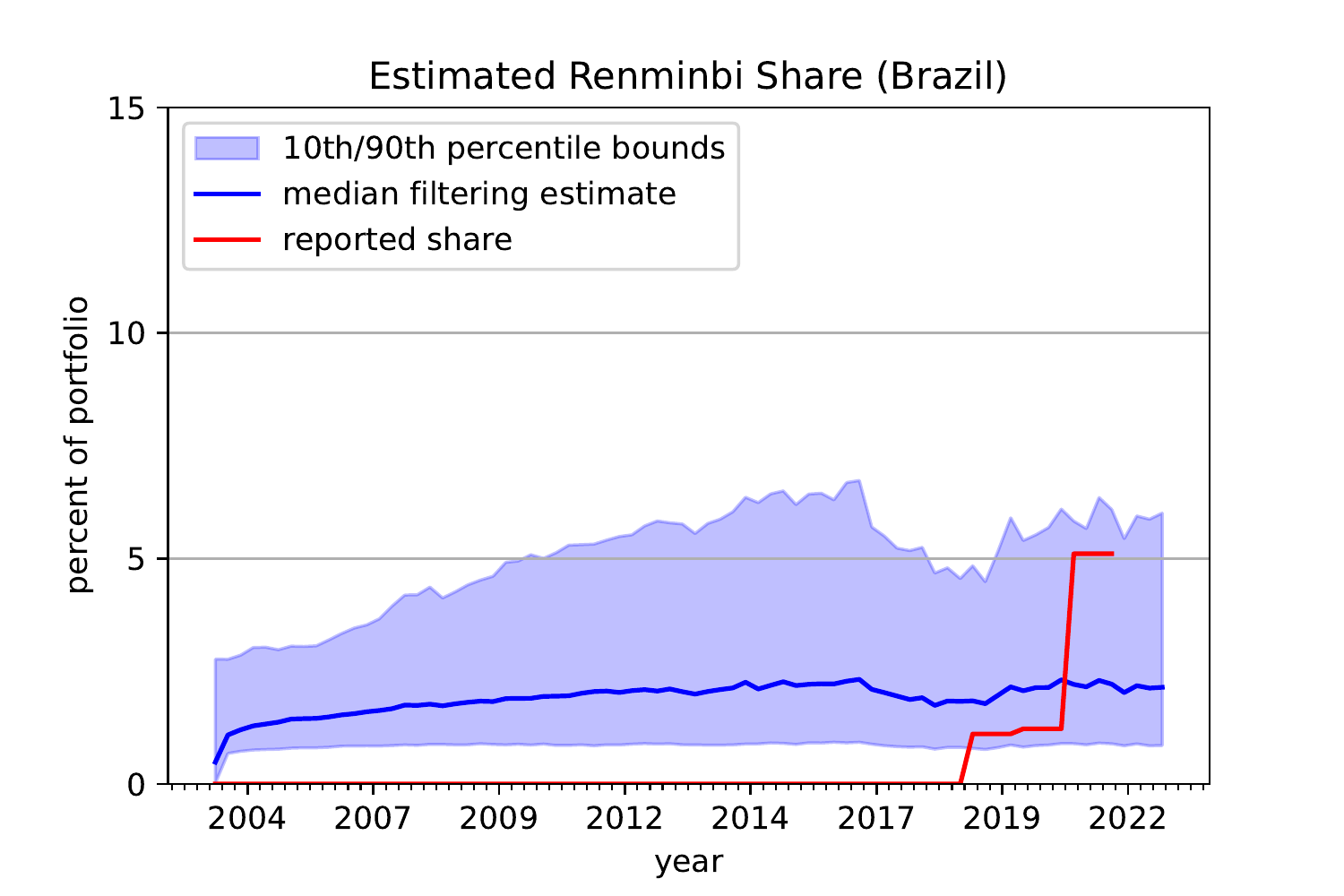}
\end{subfigure}\hspace*{\fill}

\caption{Filter estimates of Brazil's currency shares, compared with Brazil's self-reported shares.}

\label{brresults}

\end{figure}

\begin{figure}[h]
\begin{subfigure}{1.0\textwidth}
\includegraphics[width=\linewidth]{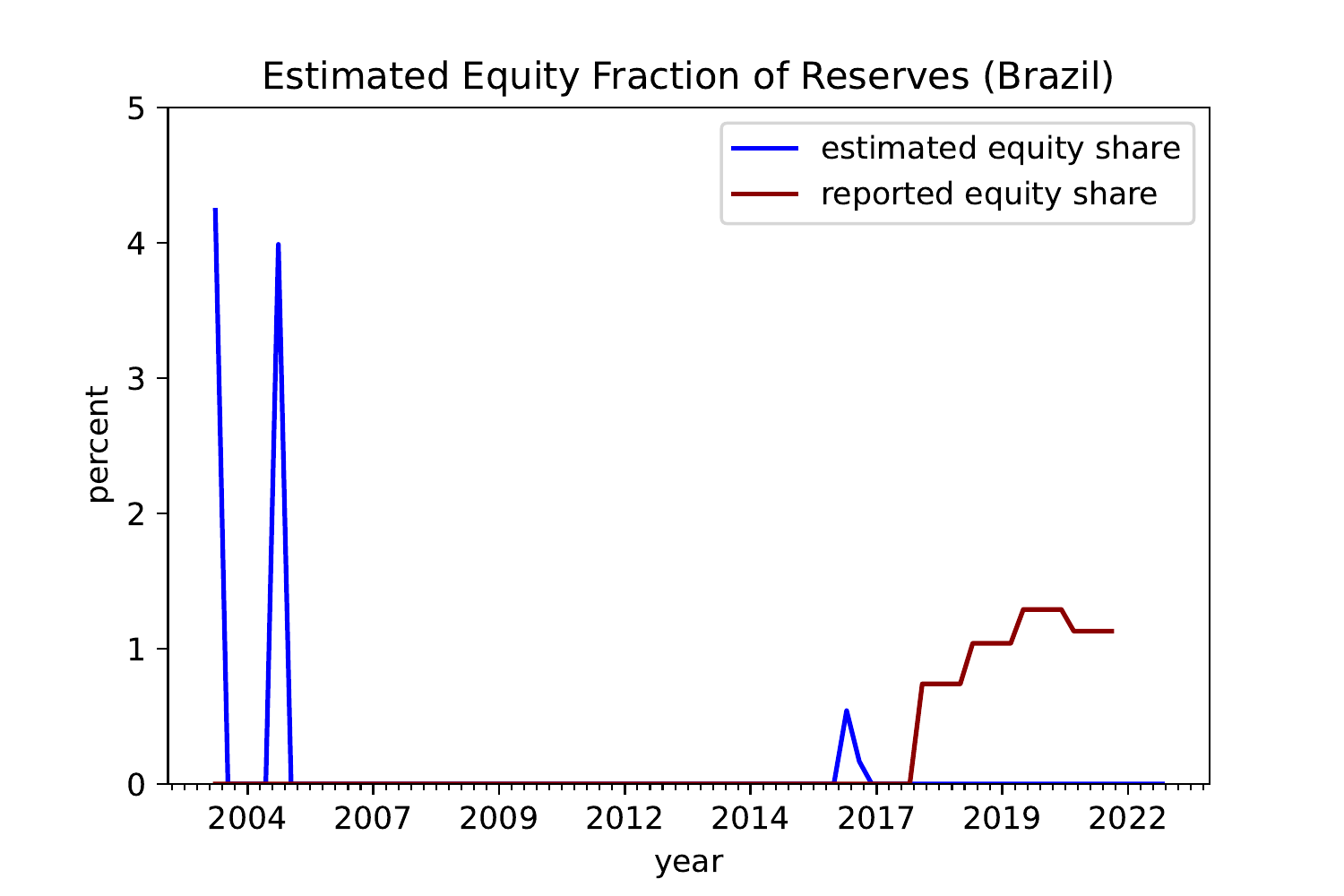}
\end{subfigure}\hspace*{\fill}

\caption{The estimated equity share of Brazil's reserves.}

\label{brequity}

\end{figure}

\begin{figure}[h]
\begin{subfigure}{1.0\textwidth}
\includegraphics[width=\linewidth]{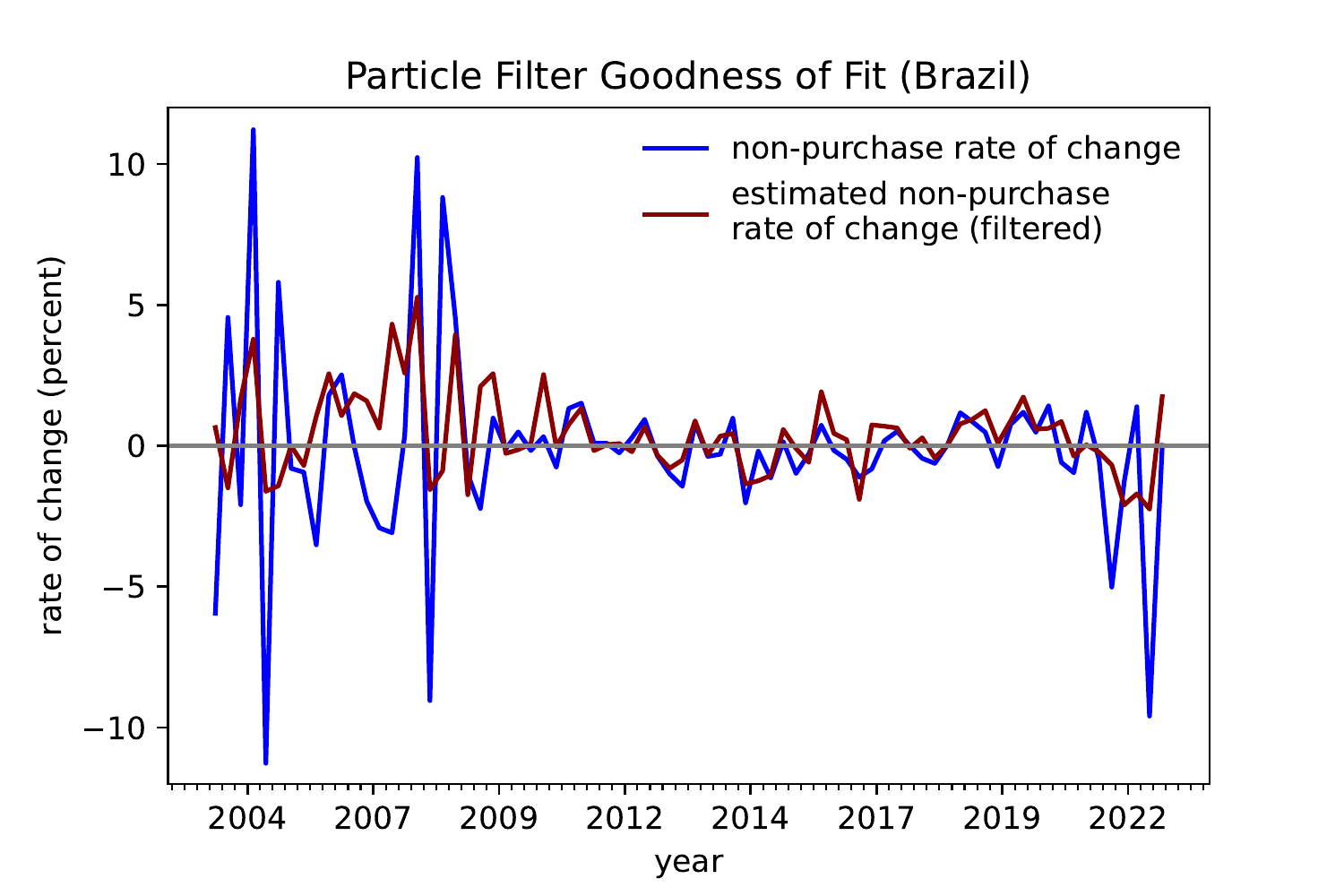}
\end{subfigure}\hspace*{\fill}

\caption{Median particle filter estimates, compared with the observed non-purchase rate of change of reserves.}

\label{brgoodness}
\end{figure}

\FloatBarrier
\subsection{Switzerland}

Switzerland's foreign exchange reserves total \$850 billion as of Q4 2022. Switzerland discloses that the average duration of its bond holdings is 4.6 years, so the returns on the fixed income portion of Switzerland's reserves should be similar to the returns on 5-year sovereign bonds. Unlike Brazil, Switzerland invests 25\% of its reserves in equities. The equity share optimization correctly identifies this proportion, although the optimization procedure suggests a higher equity share during the financial crisis, likely because the returns on some of Switzerland's investments were correlated with those of equities during that time.

Switzerland self-reports five currency shares: the US dollar, euro, pound, yen, and Canadian dollar. Switzerland reports that the remaining 0-10\% of its reserves is allocated to "other" currencies. Because I do not know the specific composition of this "other" category, I re-compute the five known currency shares excluding the "other" category, and run the model using only five currencies.

\begin{figure}[h]
\begin{subfigure}{1.0\textwidth}
\includegraphics[width=\linewidth]{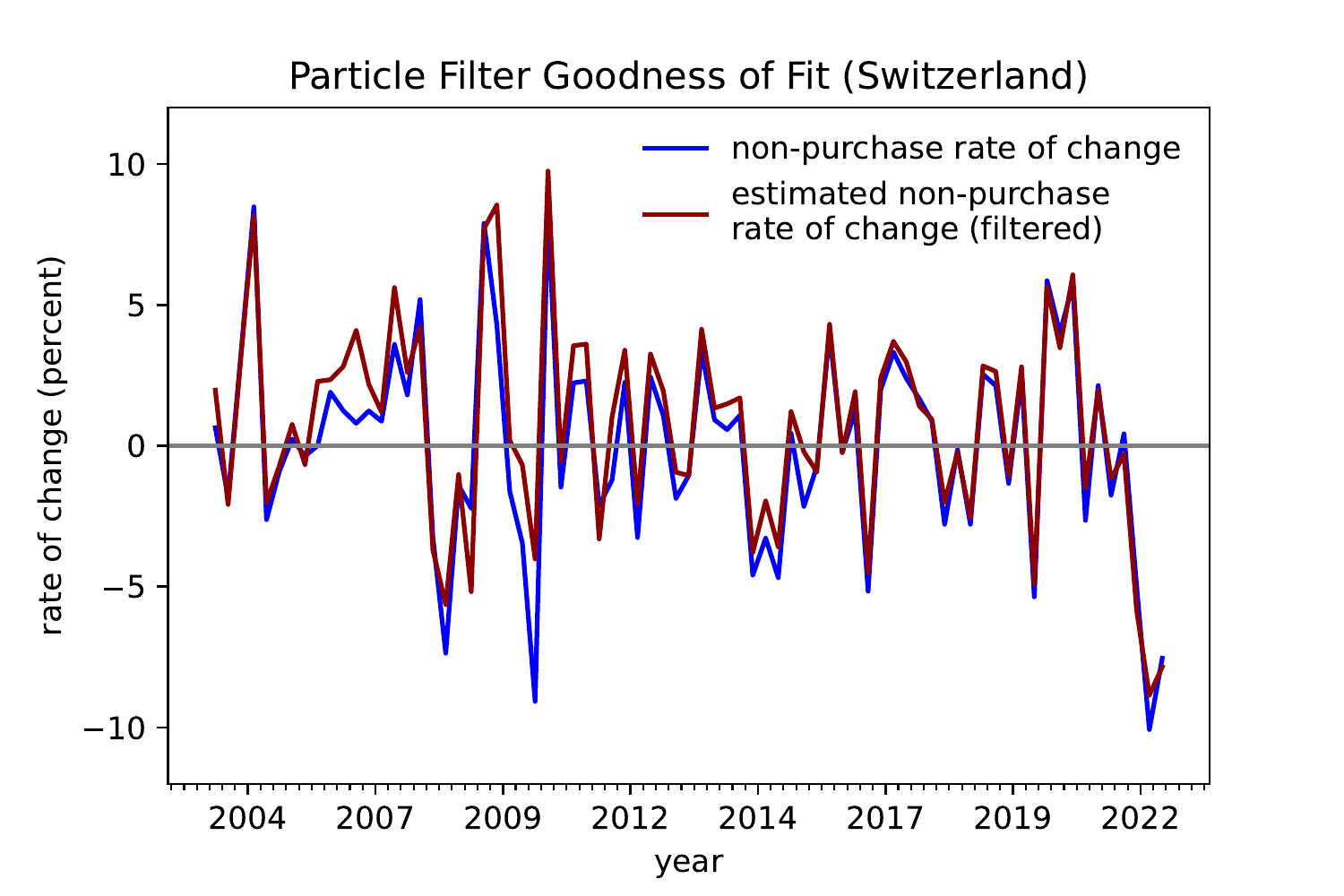}
\end{subfigure}\hspace*{\fill}

\caption{Median particle filter estimates, compared with the observed non-purchase rate of change of reserves.}

\label{swgoodness}
\end{figure}

Results are available in Figure \ref{swresults}. The filter matches the observed fluctuation in Switzerland's reserves very closely, as illustrated in Figure \ref{swgoodness}.

\begin{figure}[h]
\begin{subfigure}{1.0\textwidth}
\includegraphics[width=\linewidth]{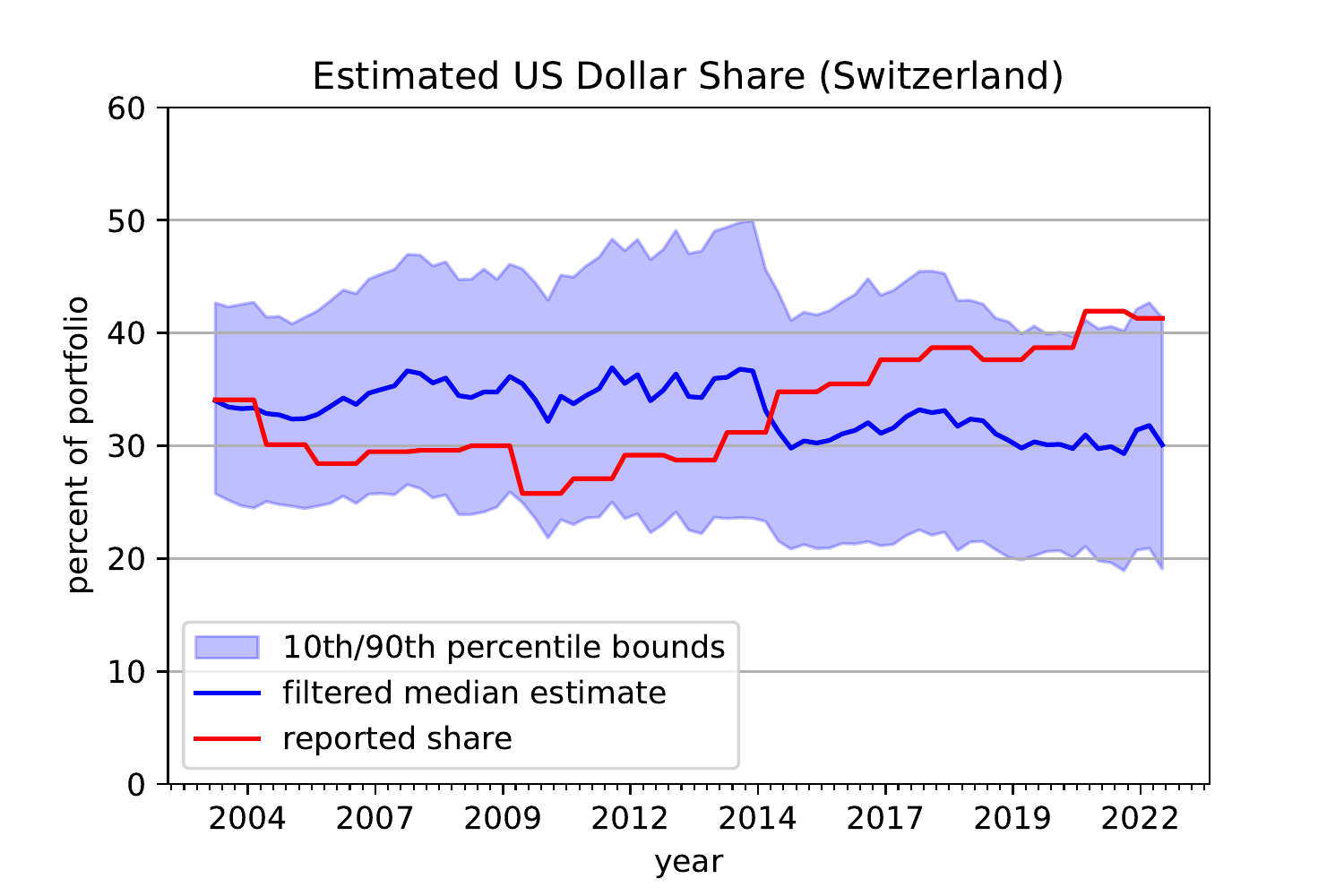}
\end{subfigure}\hspace*{\fill}

\begin{subfigure}{1.0\textwidth}
\includegraphics[width=\linewidth]{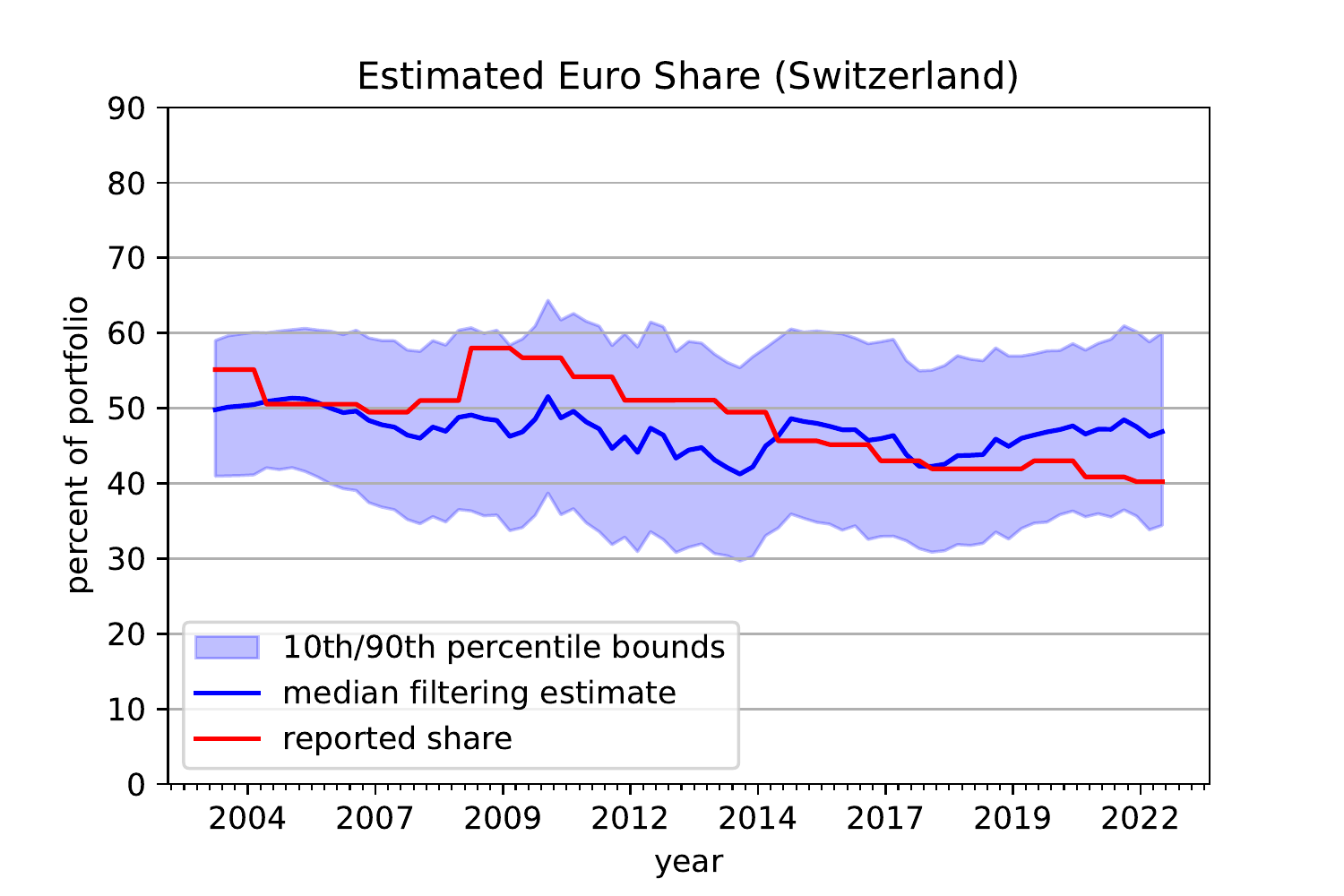}
\end{subfigure}\hspace*{\fill}

\caption{Filter estimates of Switzerland's currency shares, compared with Switzerland's self-reported shares.}

\label{swresults}
\end{figure}

\begin{figure}[h]\ContinuedFloat
\begin{subfigure}{1.0\textwidth}
\includegraphics[width=\linewidth]{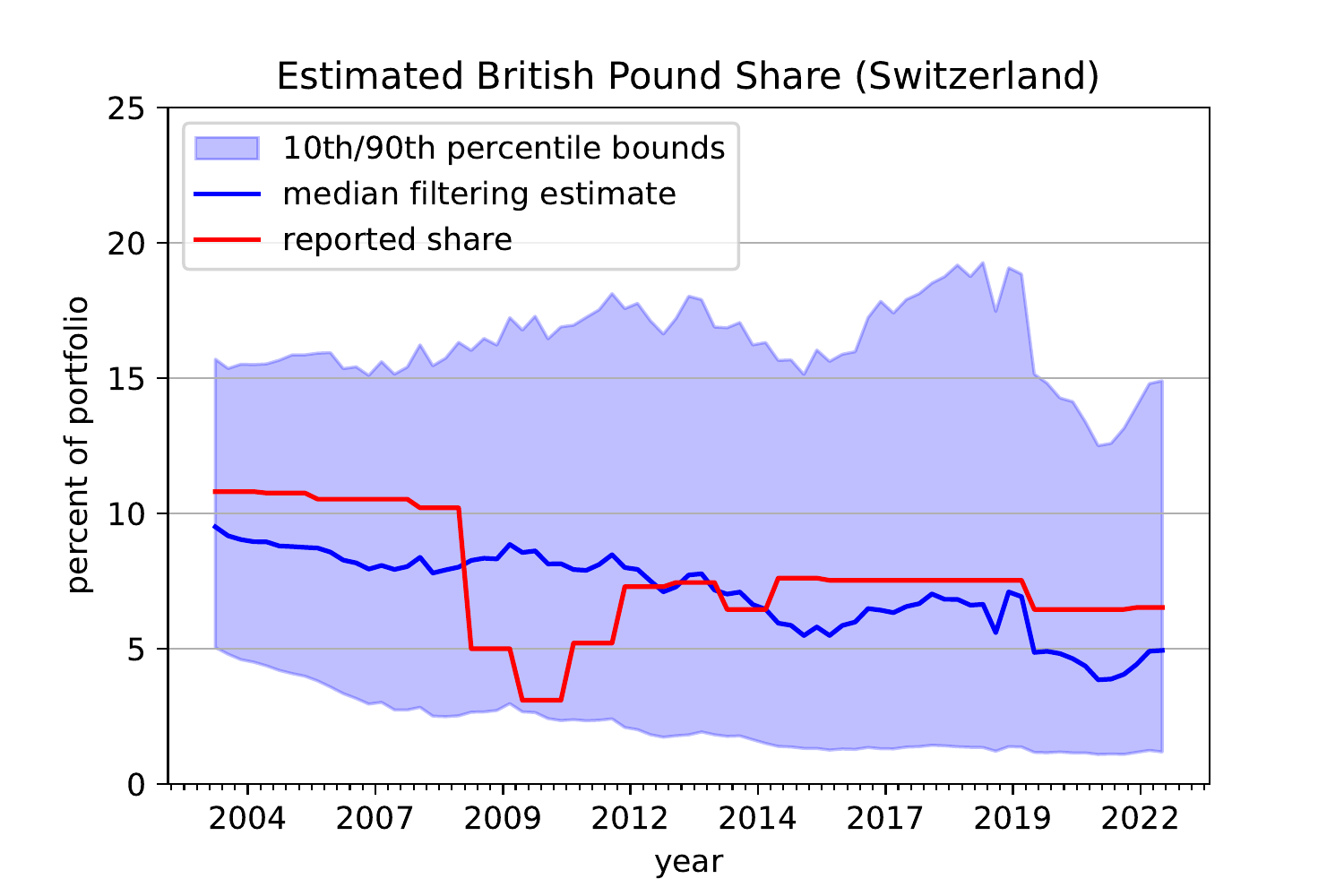}
\end{subfigure}\hspace*{\fill}

\begin{subfigure}{1.0\textwidth}
\includegraphics[width=\linewidth]{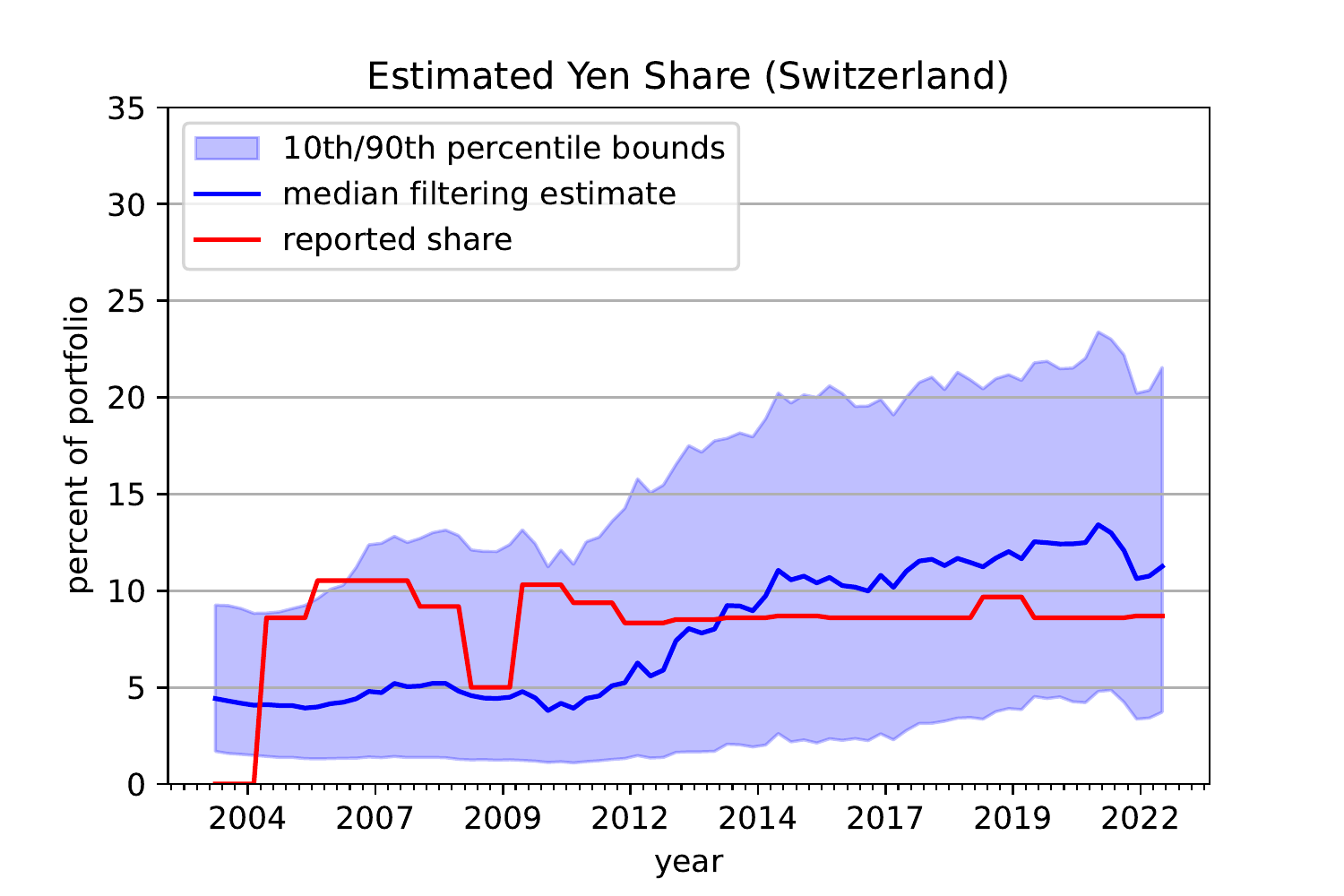}
\end{subfigure}\hspace*{\fill}

\caption{Filter estimates of Switzerland's currency shares, compared with Switzerland's self-reported shares.}

\label{swresults}

\end{figure}

\begin{figure}[h]\ContinuedFloat

\begin{subfigure}{1.0\textwidth}
\includegraphics[width=\linewidth]{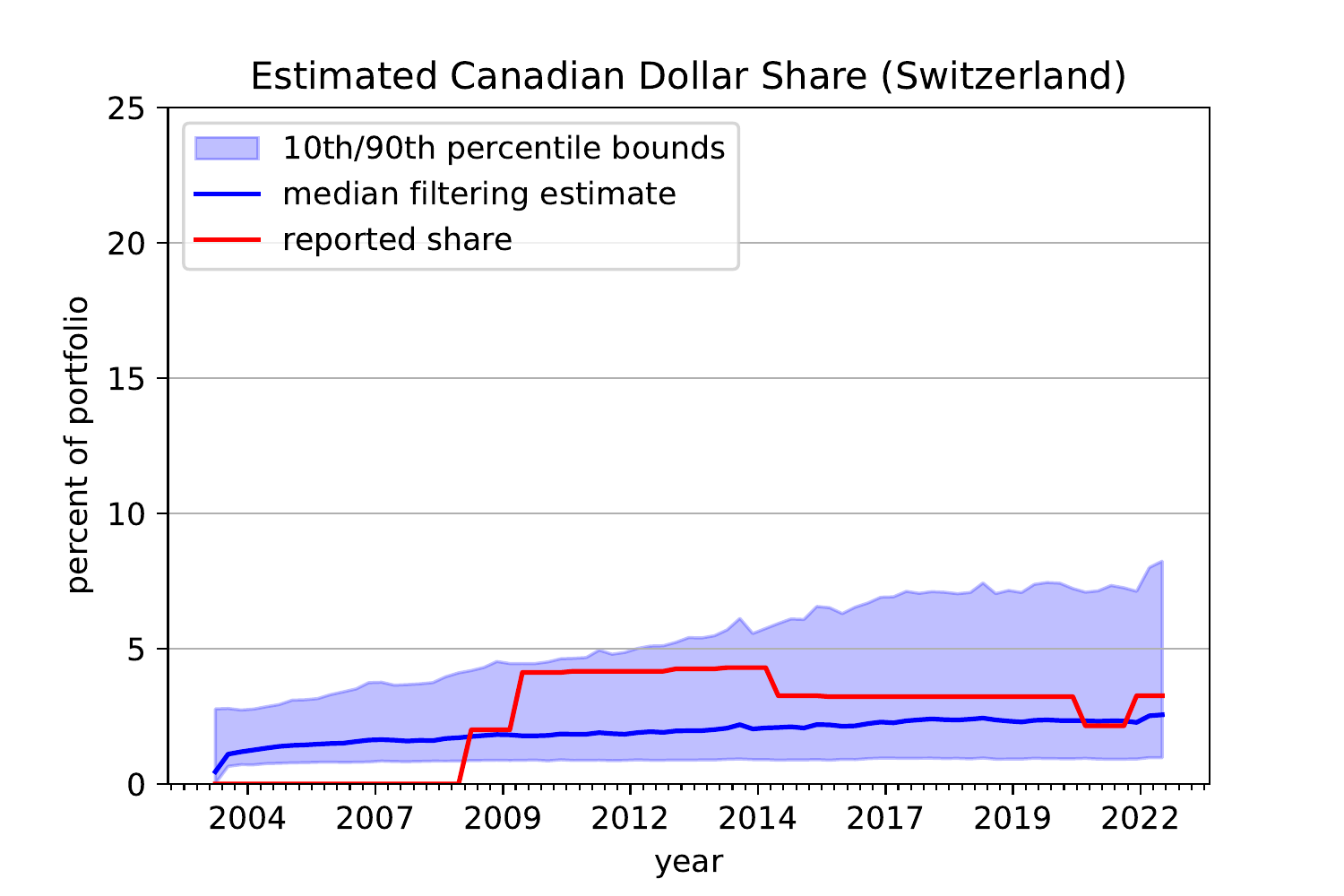}
\end{subfigure}\hspace*{\fill}

\caption{Filter estimates of Switzerland's currency shares, compared with Switzerland's self-reported shares.}

\label{swresults}

\end{figure}

\begin{figure}[h]
 \centering
    \includegraphics[scale = 1.1]{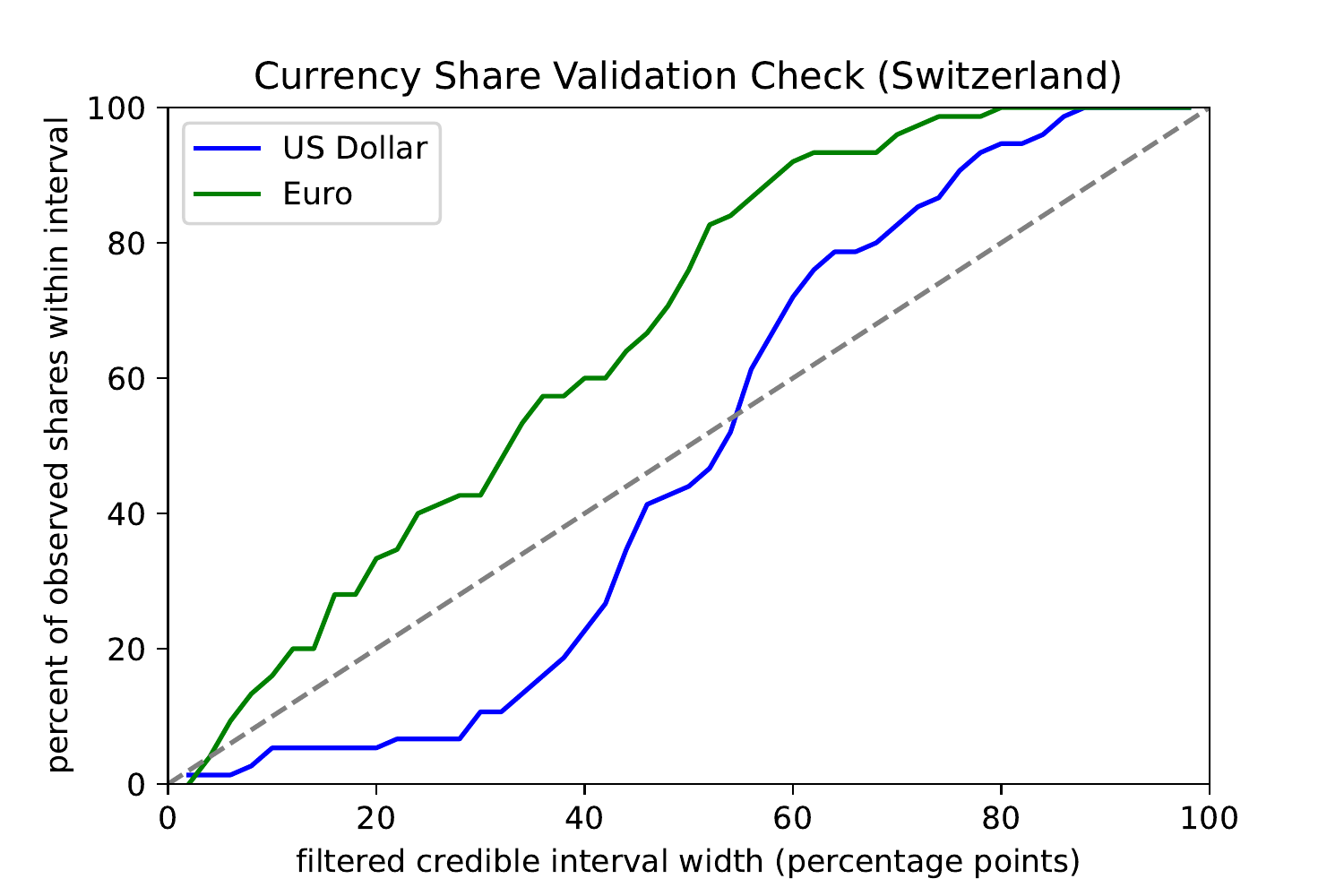}
    \vspace{-1em}
    \caption{The proportion of time periods in which the self-reported currency share falls within a credible interval centered around the estimated median. The closer to the diagonal line, the more precisely the filter captures the true currency shares at the specified level of confidence.}
    \label{swvalidation}
\end{figure}

\begin{figure}[h]
\begin{subfigure}{1.0\textwidth}
\includegraphics[width=\linewidth]{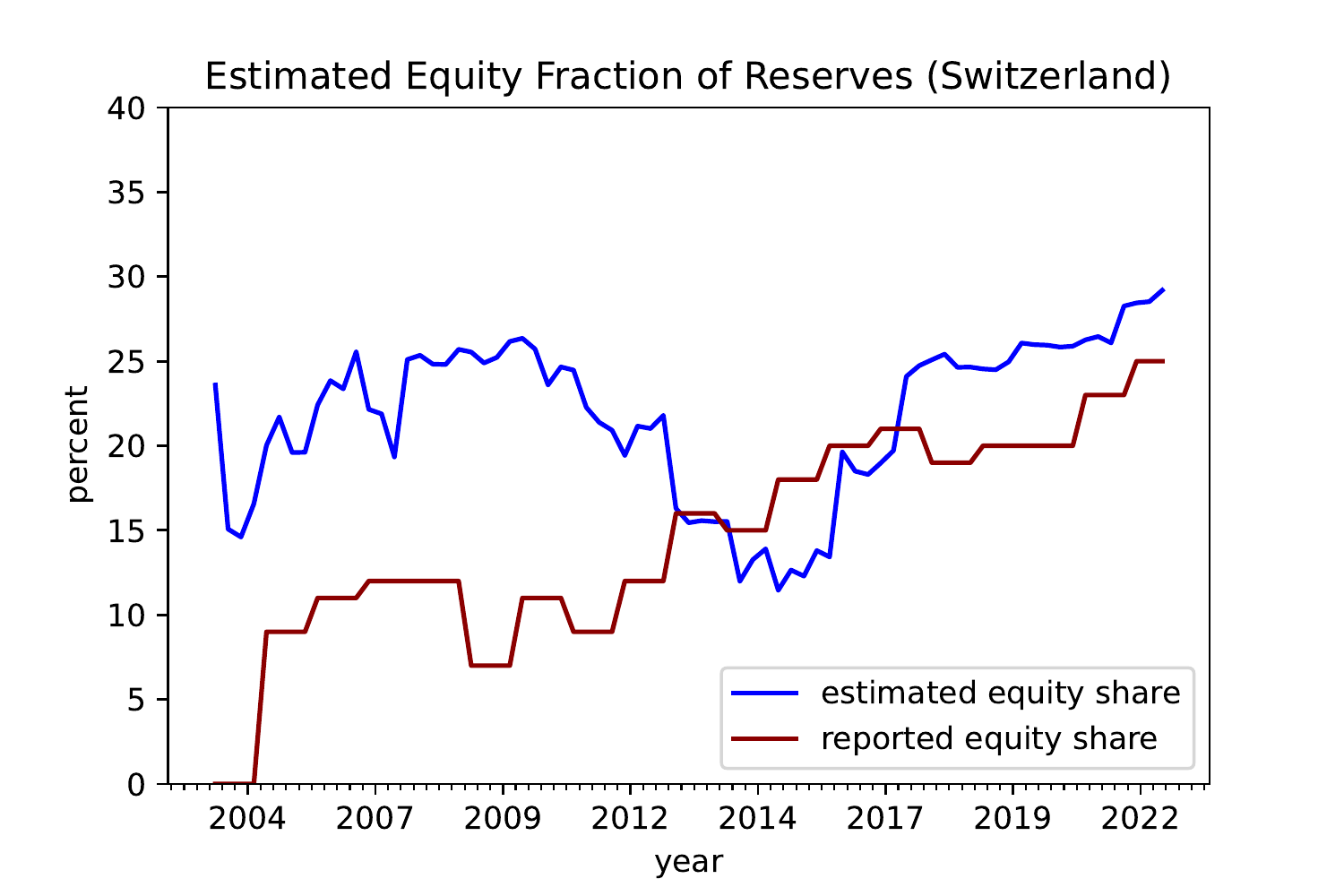}
\end{subfigure}\hspace*{\fill}

\caption{The estimated equity share of Switzerland's reserves.}

\label{swequity}

\end{figure}

\FloatBarrier
\section{Conclusion} \label{sec:conclusion}

Based on portfolio accounting identities, I estimated a Hidden Markov Model to identify the currency composition of the foreign exchange reserves of any country. To my knowledge, this paper outlines the first time series approach to estimating the currency composition of a country's foreign exchange reserves, regardless of the country's exchange rate regime. The model's structure, priors, and parameters provide sufficient flexibility to fit the circumstances of most countries. 

Despite speculation in the financial press about China reducing its US dollar holdings, the model suggests that the dollar share of China's foreign exchange reserves continues to track the global average closely. However, Singapore may hold fewer US dollars, more euros, and more British pounds than global averages.

Future research may use estimates from the Hidden Markov Model as inputs for regressions investigating factors that influence the currency composition of foreign exchange reserves. If this model is estimated simultaneously across many countries, it may be possible to pool information from multiple countries in order to infer the state space variances. Applying other filtering techniques, such as the particle learning approach of \cite{particlelearning}, could also assist in estimating time-invariant variance parameters. Lastly, similar types of Hidden Markov Models likely can be applied to other settings in economics in which researchers would like to estimate certain unknown time series. For example, this method may be applicable to infer the currency composition of a firm's assets or liabilities based on the firm's regulatory filings. Such estimates could augment the currency mismatch literature.

\clearpage
\printbibliography

\clearpage
\appendix

\setcounter{table}{0}


\section{Optimization Approach} \label{optimappendix}

Ignoring $\epsilon_t$, equation (\ref{observation}) can be estimated through a series of optimization problems. Given $N$ reserve currencies, a unique solution for $\beta_t$ can be found using a rolling window consisting of at least $N$ consecutive quarters.

\begin{mini}|l| 
  {\vec{\beta_t}}{\sum^{T+N}_{t=T} \left(y_t - \sum^N_{i=1} \beta^i_{t-1}(1+r^i_{t-1})\dt{e^i_t} + \sum^N_{i=1} \beta^i_{t-1}r^i_{t-1} \right)^2}{}{}
  {\label{optimization}}{}
  \addConstraint{\beta^i_t}{\geq 0,}{i=0,\ldots,N}
  \addConstraint{\sum_{i=1}^N \beta^i_t}{=1. \quad}
 \end{mini}

Although optimization provides some insight into a country's currency shares, this approach generally produces shares that vary to an implausible extent over time, because the optimization method is unduly influenced by observation error. I could penalize $\lVert \beta^i_{t+1} - \beta^i_t \rVert$ to reduce the fluctuation in the shares, or utilize a rolling window similar to the equity share optimization procedure described in (\ref{equityoptimization}). But the optimization method cannot estimate the uncertainty in the currency shares, because it does not account for $\epsilon_t$. A time series approach is necessary to model both observation error and fluctuation in the currency shares.

For purposes of comparison, I illustrate results from this optimization approach in the case of China. Although the optimization method produces currency shares that fluctuate to an implausible extent, the shares produced by the optimization method loosely oscillate around the filter estimates.

\FloatBarrier

\begin{figure}[h]
\begin{subfigure}{1.0\textwidth}
\includegraphics[width=\linewidth]{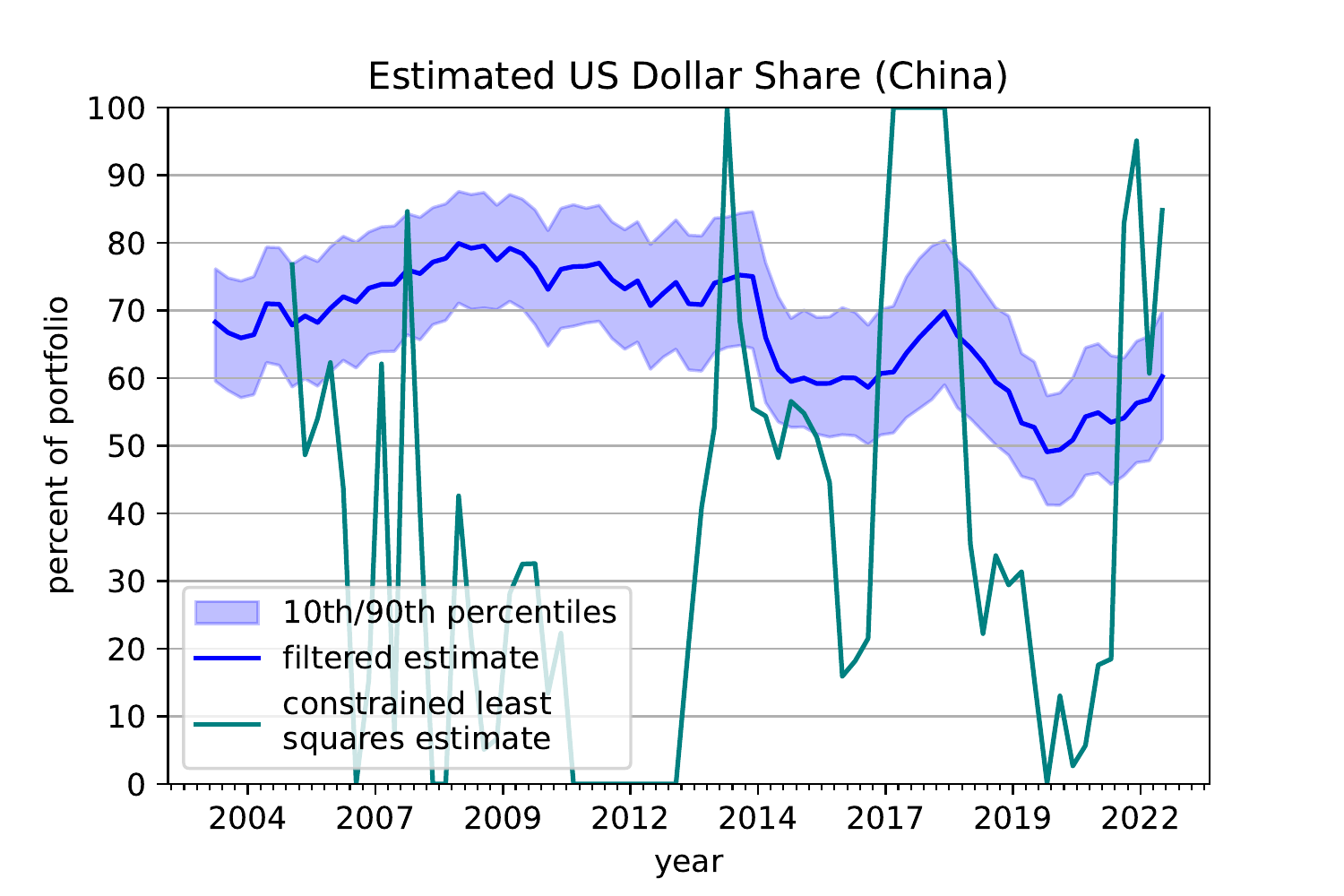}
\end{subfigure}\hspace*{\fill}

\begin{subfigure}{1.0\textwidth}
\includegraphics[width=\linewidth]{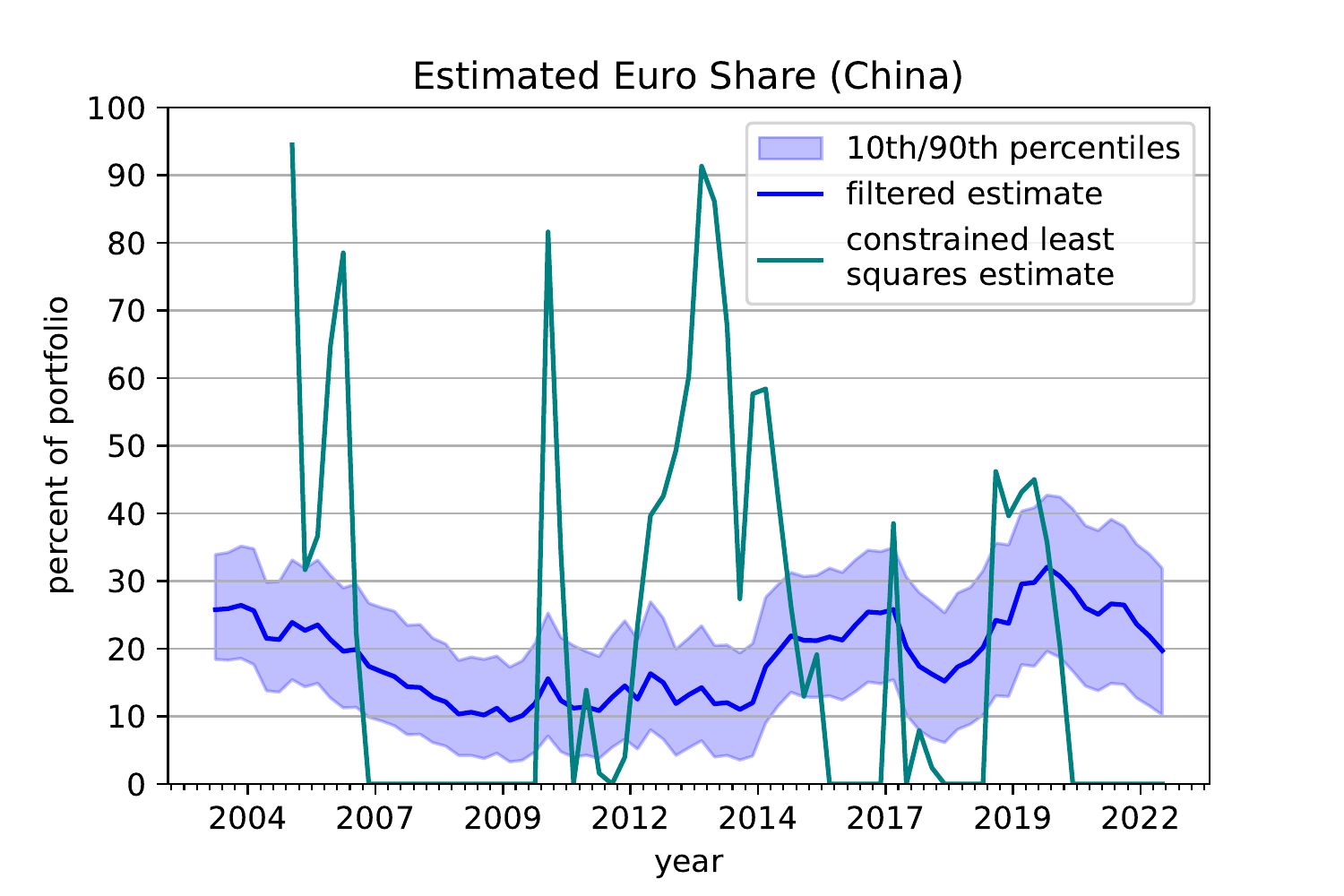}
\end{subfigure}\hspace*{\fill}

\caption{Optimization estimates of China's currency shares.}

\label{chinaoptim}
\end{figure}

\FloatBarrier

\section{Algorithms} \label{appendixalgorithms}

\subsection{Particle Filtering}

\begin{algorithm} 
\caption{Particle filtering}
\label{filteringalgorithm}
\begin{algorithmic} 
\STATE Sample $N_F$ initial particles from the Dirichlet prior: $\vec{\beta_0^i} \sim p(\vec{\beta_0})$ for $i=1, 2,...N_F$
\STATE Set weights $w^i_0 \leftarrow 1/N_F$ for $i=1, 2,...N_F$
\FOR{$t=1$ to $T-1$}
\FOR{$i=1$ to $N_F$}
\STATE Sample a $\vec{\beta^i_{t+1}} \sim p(\vec{\beta_{t+1}}|\vec{\beta_t^i})$ using a particle $\vec{\beta_t^i}$ and the state space equation
\STATE Compute a weight $w^i_{t+1} \leftarrow  w^i_t \cdot p(y_{t+1}|\vec{\beta^i_{t+1}})$ using $\vec{\beta^i_{t+1}}$ and the observation equation
\ENDFOR
\STATE Normalize the weights: $\widehat{w^i_{t+1}} \leftarrow w^i_{t+1}/\sum^{N_F}_{j=1} w^j_{t+1}$ for $i=1, 2,...N_F$
\FOR{$i=1$ to $N_F$}
\STATE Sample a filter particle $\vec{\beta^i_{t+1}} \sim p(\vec{\beta_{t+1}}|y_1, y_2, ... y_{t+1})$ from the categorical distribution defined by $(\vec{\beta^k_{t+1}}, \widehat{w^k_{t+1}})$, $k = 1, 2,...N_F$
\ENDFOR
\STATE Reset the weights: $w^i_{t+1} \leftarrow 1/N_F$ for $i=1, 2,...N_F$
\ENDFOR
\end{algorithmic}
\end{algorithm}

\clearpage
\FloatBarrier

\section{Model Parameters} \label{appendixparams}
\subsection{China}

\begin{table}[hbt!]
\caption{Dirichlet Prior Parameters} 
\centering 
\begin{tabular}{l *{6}{D..{-1}}} 
\toprule
 & \multicolumn{1}{c@{}}{USD} & \multicolumn{1}{c@{}}{EUR} & \multicolumn{1}{c@{}}{GBP} & \multicolumn{1}{c@{}}{JPY} & \multicolumn{1}{c@{}}{AUD} & \multicolumn{1}{c@{}}{CAD} \\ [0.5ex] 
\midrule
Parameter & 34.0 & 13.0 & 1.0 & 1.0 & 0.5 & 0.5 \\ [0.5ex]
Mean & 68.0\% & 26.0\% & 2.0\% & 2.0\% & 1.0\% & 1.0\% \\ [0.5ex]
Standard Deviation & 6.5\% & 6.1\% & 2.0\% & 2.0\% & 1.4\% & 1.4\% \\ [1ex] 
\bottomrule
\end{tabular}
\label{chinadirichlet} 
\end{table}

\subsection{Singapore}

\begin{table}[hbt!]
\caption{Dirichlet Prior Parameters} 
\centering 
\begin{tabular}{l *{7}{D..{-1}}} 
\toprule
 & \multicolumn{1}{c@{}}{USD} & \multicolumn{1}{c@{}}{EUR} & \multicolumn{1}{c@{}}{GBP} & \multicolumn{1}{c@{}}{JPY} & \multicolumn{1}{c@{}}{AUD} & \multicolumn{1}{c@{}}{CAD} & \multicolumn{1}{c@{}}{RMB} \\ [0.5ex] 
\midrule
Parameter & 22.3 & 8.7 & 0.7 & 0.7 & 0.3 & 0.3 & 0.3 \\ [0.5ex]
Mean & 67.0\% & 26.0\% & 2.0\% & 2.0\% & 1.0\% & 1.0\% & 1.0\% \\ [0.5ex]
Standard Deviation & 8.0\% & 7.5\% & 2.4\% & 2.4\% & 1.7\% & 1.7\% & 1.7\% \\ [1ex] 
\bottomrule
\end{tabular}
\label{sgdirichlet} 
\end{table}
\clearpage

\subsection{Brazil}

\begin{table}[hbt!]
\caption{Dirichlet Prior Parameters} 
\centering 
\begin{tabular}{l *{7}{D..{-1}}} 
\toprule
 & \multicolumn{1}{c@{}}{USD} & \multicolumn{1}{c@{}}{EUR} & \multicolumn{1}{c@{}}{GBP} & \multicolumn{1}{c@{}}{JPY} & \multicolumn{1}{c@{}}{AUD} & \multicolumn{1}{c@{}}{CAD} & \multicolumn{1}{c@{}}{RMB} \\
\midrule
Parameter & 28.5 & 15.0 & 1.5 & 3.5 & 0.5 & 0.5 & 0.5\\ [0.5ex]
Mean & 57.0\% & 30.0\% & 3.0\% & 7.0\% & 1.0\% & 1.0\% & 1.0\% \\ [0.5ex]
Standard Deviation & 6.9\% & 6.4\% & 2.4\% & 3.6\% & 1.4\% & 1.4\% & 1.4\% \\ [1ex] 
\bottomrule
\end{tabular}
\label{brdirichlet} 
\end{table}

\subsection{Switzerland}

\begin{table}[hbt!]
\caption{Dirichlet Prior Parameters} 
\centering 
\begin{tabular}{l *{5}{D..{-1}}} 
\toprule
 & \multicolumn{1}{c@{}}{USD} & \multicolumn{1}{c@{}}{EUR} & \multicolumn{1}{c@{}}{GBP} & \multicolumn{1}{c@{}}{JPY} & \multicolumn{1}{c@{}}{CAD} \\
\midrule
Parameter & 17.0 & 25.0 & 5.0 & 2.5 & 0.5 \\ [0.5ex]
Mean & 34.0\% & 50.0\% & 10.0\% & 5.0\% & 1.0\%  \\ [0.5ex]
Standard Deviation & 6.6\% & 7.0\% & 4.2\% & 3.1\% & 1.4\% \\ [1ex] 
\bottomrule
\end{tabular}
\label{swdirichlet} 
\end{table}

\clearpage

\section{Sensitivity Analysis: Influence of Prior} \label{appendixprior}

\subsection{China}

\begin{figure}[H]
\includegraphics[width=\linewidth]{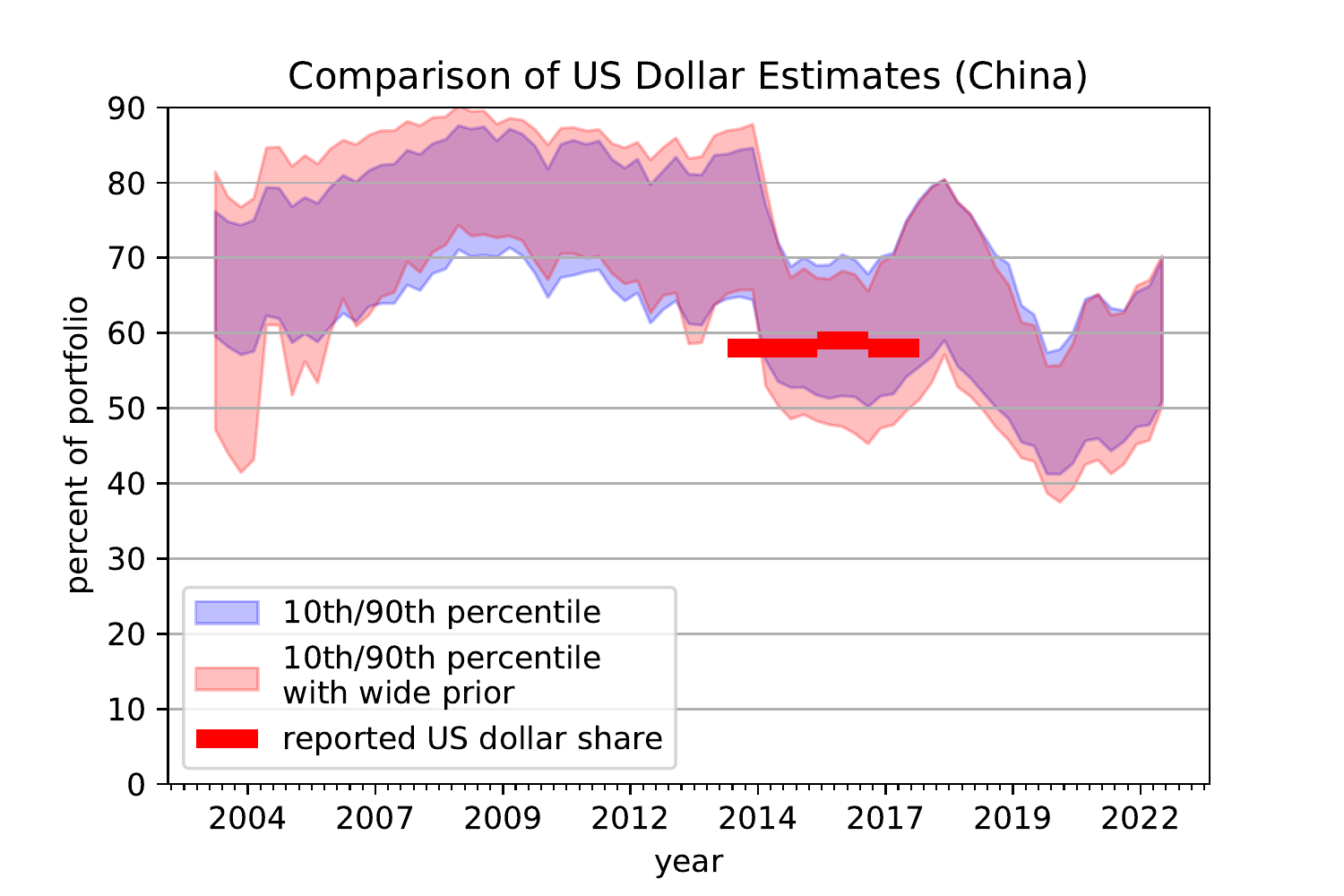}
\caption{The sensitivity of the China US dollar share estimates to the width of the prior.}
\end{figure}

\begin{figure}[H]
\includegraphics[width=\linewidth]{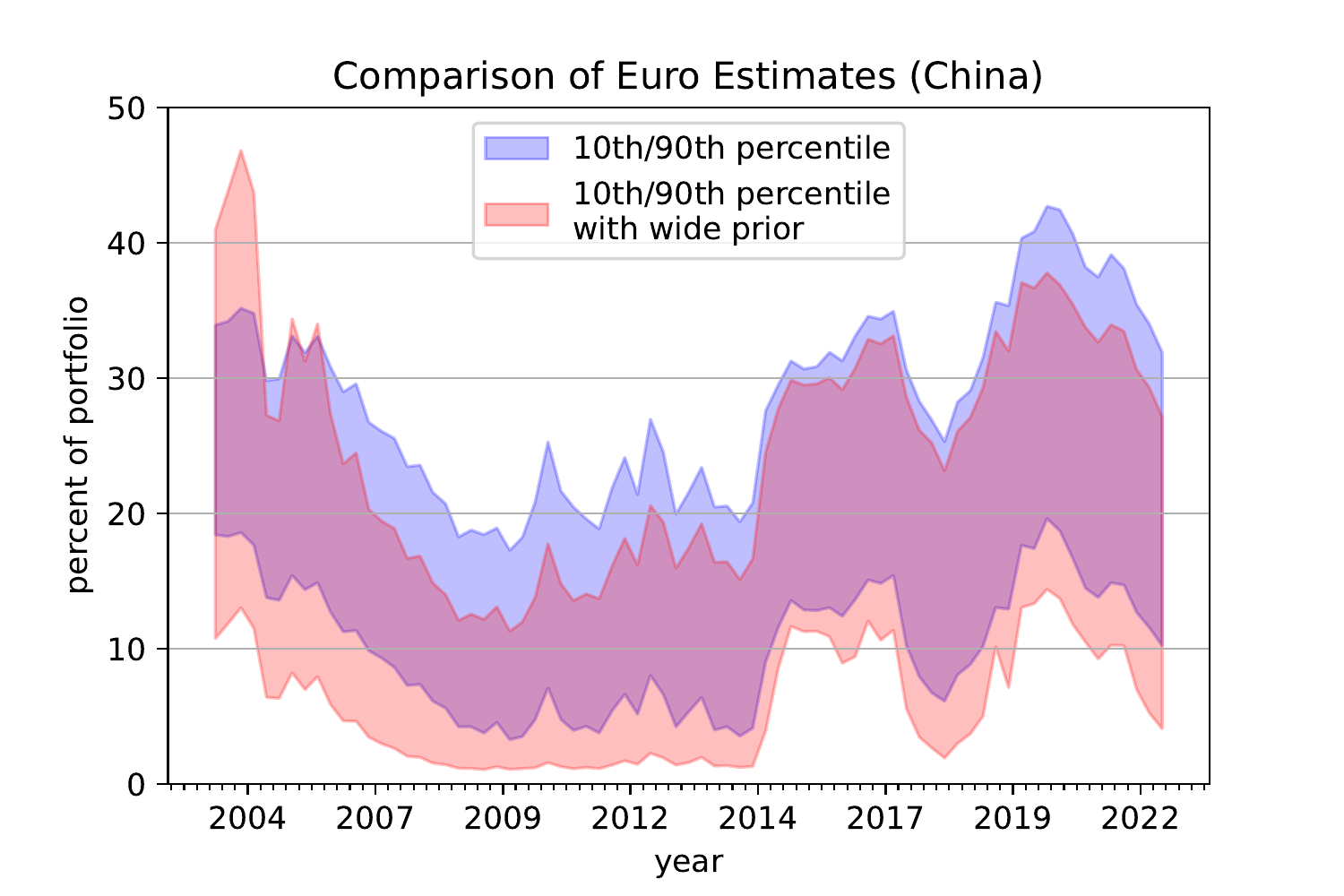}
\caption{The sensitivity of the China euro share estimates to the width of the prior.}
\end{figure}

\clearpage

\section{Sensitivity Analysis: Assumed Investment Returns} \label{appendixchinasensitivity}

\subsection{China}

\begin{figure}[H]
\includegraphics[width=\linewidth]{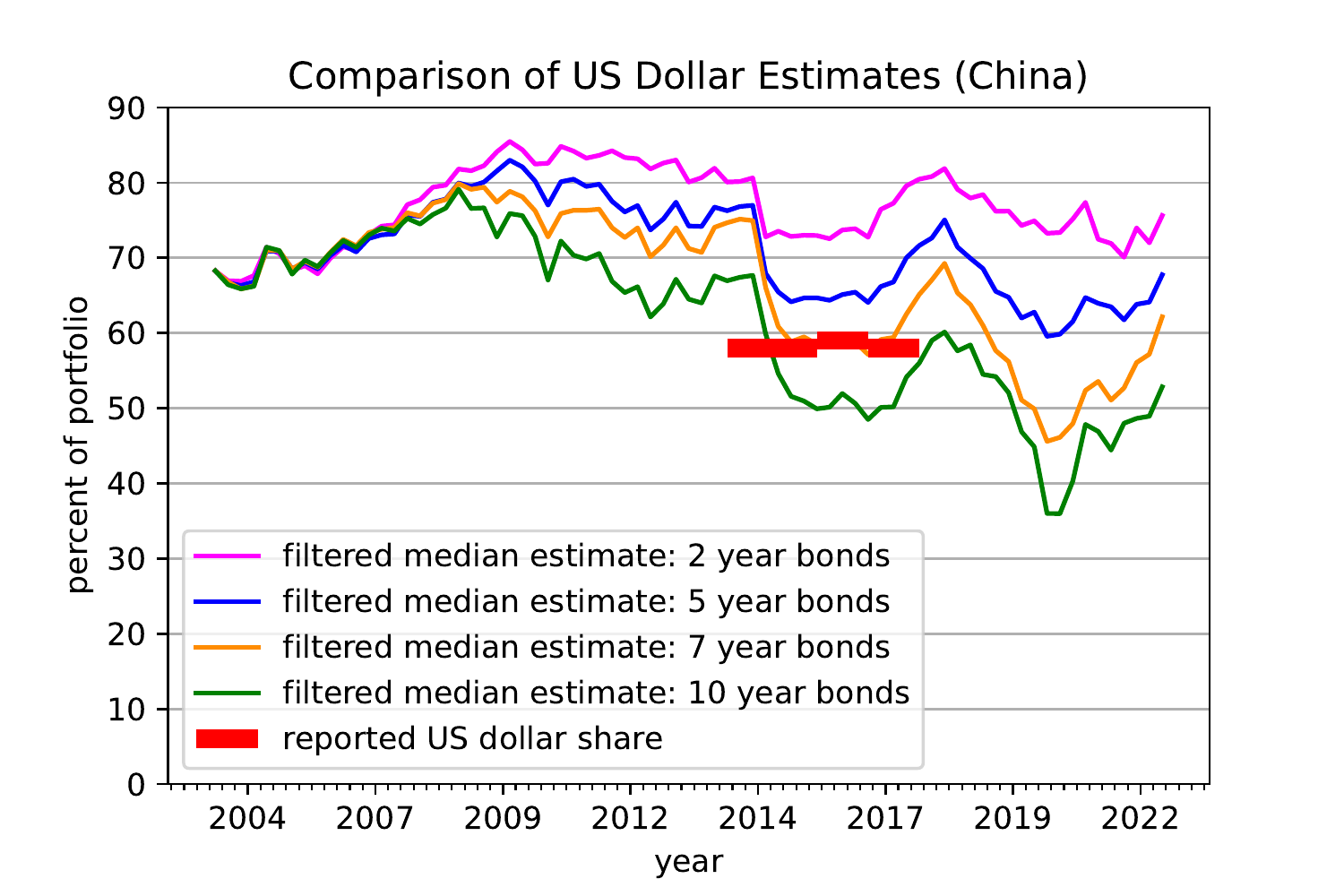}
\caption{The sensitivity of the China US dollar share estimates to the assumed duration of sovereign bonds in China's reserves.}
\end{figure}

\begin{figure}[H]
\includegraphics[width=\linewidth]{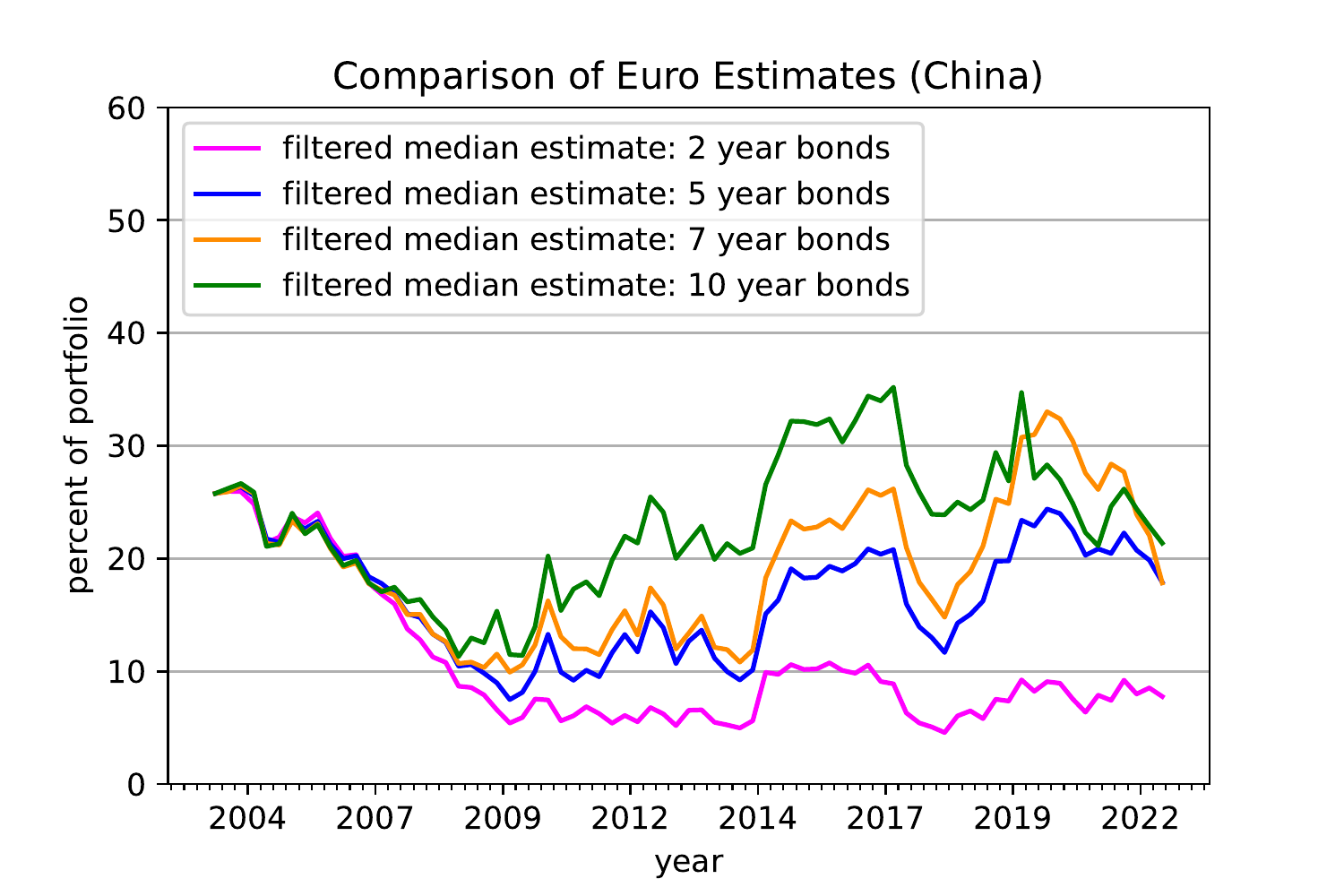}
\caption{The sensitivity of the China euro share estimates to the assumed duration of sovereign bonds in China's reserves.}
\end{figure}

\clearpage

\section{Comparison of Probability Distributions} \label{appendixcauchy}

The heavier tails of the Laplace distribution accommodate volatile net flows into reserve portfolios better than the Normal distribution, particularly during the mid-2000s when many emerging markets rapidly accumulated foreign exchange reserves. Also, the Laplace distribution is slightly more responsive to new observations than the Cauchy distribution.

\begin{figure}[h]
\begin{subfigure}{1.0\textwidth}
\includegraphics[width=\linewidth]{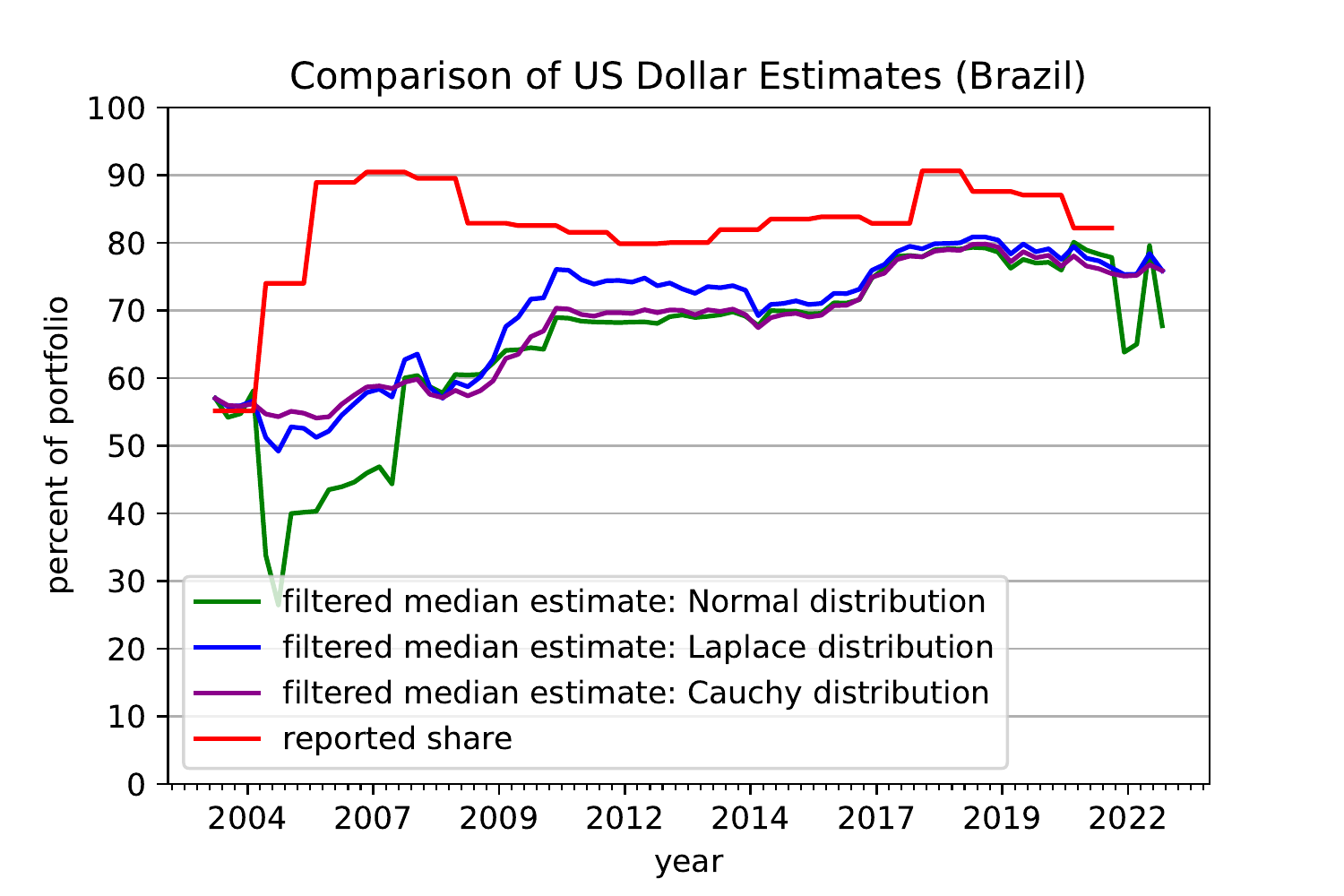}
\end{subfigure}\hspace*{\fill}

\begin{subfigure}{1.0\textwidth}
\includegraphics[width=\linewidth]{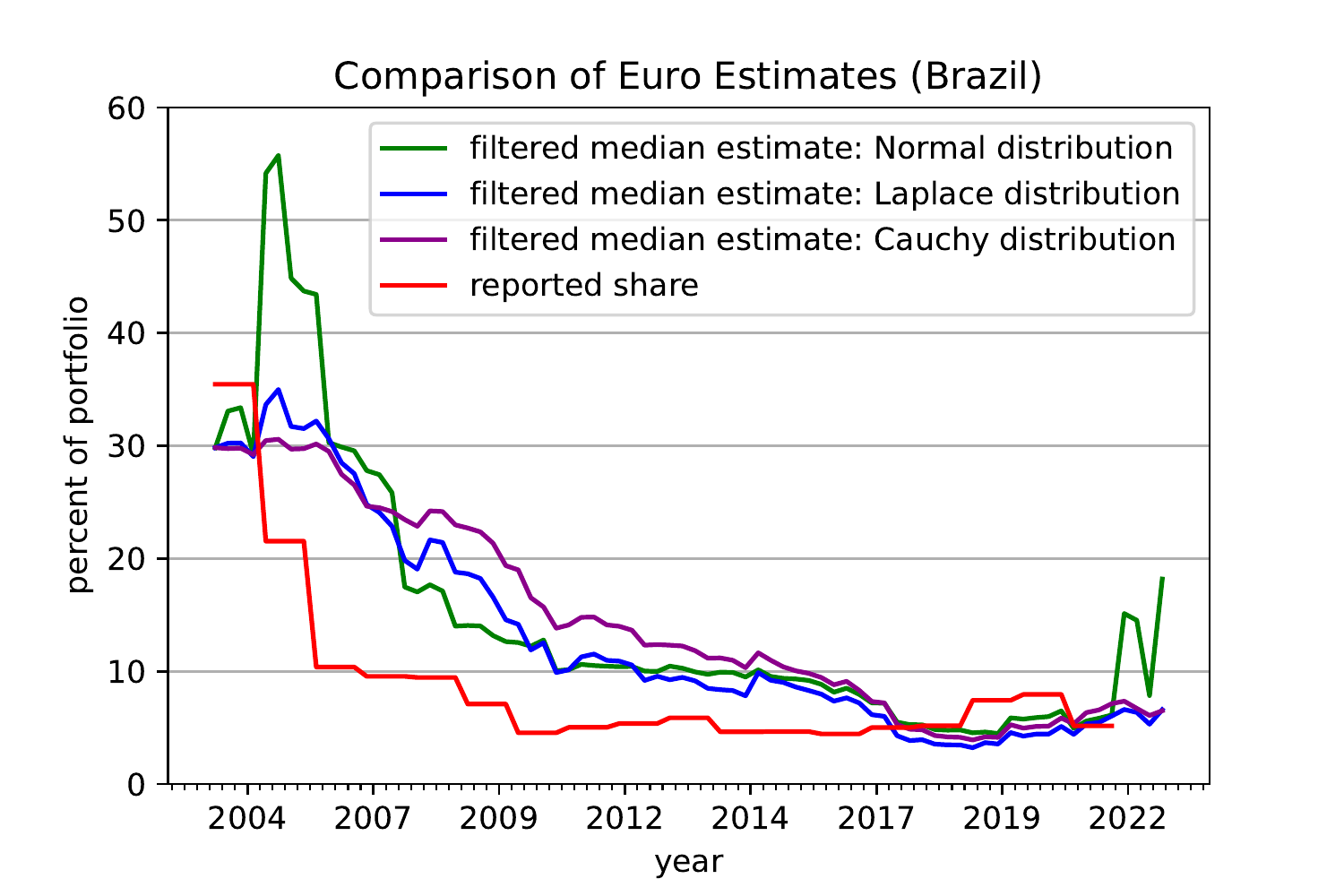}
\end{subfigure}\hspace*{\fill}

\caption{Laplace, Normal, and Cauchy estimates for Brazil's currency shares. The Normal distribution struggles to track the reported shares in the mid-2000s.}

\label{brazilcomparison}
\end{figure}


\end{document}